\documentclass[final,3p,times]{elsarticle}

\usepackage{amssymb}
\usepackage{caption}
\usepackage{subcaption}
\usepackage{todonotes}

\usepackage[utf8]{inputenc}
\usepackage[english]{babel}
\usepackage[T1]{fontenc}
\usepackage{amsmath}
\usepackage{amsthm}
\usepackage{amssymb}
\usepackage{bbm}
\usepackage{doi}
\usepackage{mathtools}

\usepackage{array}

\newtheorem*{theorem*}{Theorem}

\DeclareMathOperator{\tr}{\mathrm{tr}}
\def\identity{\leavevmode\hbox{\small1\kern-3.8pt\normalsize1}}

\newcommand{\ket}[1]{\left | #1 \right\rangle}
\newcommand{\bra}[1]{\left \langle#1 \right |}
\newcommand{\proj}[1]{\ket{#1}\bra{#1}}
\begin{document}

\begin{frontmatter}



\title{Analysing quantum systems with randomised measurements}


\author[1]{Pawe{\l} Cie\'{s}li\'{n}ski \fnref{fn1}}
\author[2,10]{Satoya Imai \fnref{fn1}}
\author[3,4,5]{Jan Dziewior}
\author[2]{Otfried G\"uhne}
\author[3,4,5]{Lukas Knips}
\author[1,6]{Wies{\l}aw Laskowski \corref{cor1}}
\ead{wieslaw.laskowski@ug.edu.pl}
\author[3,4,5,7]{Jasmin Meinecke}
\author[1,8]{Tomasz Paterek}
\author[9]{Tam\'as V\'ertesi}

\fntext[fn1]{These authors contributed equally as co-first authors}
\cortext[cor1]{Corresponding author}

\affiliation[1]{organization={Institute of Theoretical Physics and Astrophysics, University of Gda\'nsk},
           city={Gda\'nsk},
          postcode={80-308}, 
            country={Poland}}

\affiliation[2]{organization={Naturwissenschaftlich-Technische Fakult\"at, Universität Siegen},
            addressline={Walter-Flex-Stra{\ss}e 3}, city={Siegen},
            postcode={57068}, 
           country={Germany}}

\affiliation[10]{organization={QSTAR, INO-CNR and LENS},
          addressline={Largo Enrico Fermi 2}, 
           city={Firenze},
          postcode={50125}, 
          country={Italy}}

\affiliation[3]{organization={Max Planck Institute for Quantum Optics},
            city={Garching},
            postcode={85748}, 
            country={Germany}}
            
\affiliation[4]{organization={Faculty of Physics, Ludwig Maximilian University},
            city={Munich},
          postcode={80799}, 
          country={Germany}}
            
\affiliation[5]{organization={Munich Center for Quantum Science and Technology},
            city={Munich},
            postcode={80799}, 
            country={Germany}}

\affiliation[6]{organization={International Centre for Theory of Quantum Technologies, University of Gda\'nsk},
            city={Gda\'nsk},
            postcode={80-308}, 
            country={Poland}}

\affiliation[7]{organization={Institut f{\"u}r Festk{\"o}rperphysik, Technische Universit{\"a}t Berlin},
                city={Berlin},
                postcode={10623},
                country={Germany}}

\affiliation[8]{organization={School of Mathematics and Physics, Xiamen University Malaysia},
            city={Sepang},
            postcode={43900}, 
            country={Malaysia}}

\affiliation[9]{organization={
MTA ATOMKI Lend{\"u}let Quantum Correlations Research Group, Institute for Nuclear Research},
            city={Debrecen},
            postcode={4001}, 
            country={Hungary}}


\begin{abstract}
Measurements with randomly chosen settings determine many important properties of quantum states without the need for a shared reference frame or calibration.
They naturally emerge in the context of quantum communication and quantum computing when dealing with noisy environments, and allow the estimation of properties of complex quantum systems in an easy and efficient manner.
In this review, we present the advancements made in utilising randomised measurements in various scenarios of quantum information science.
We describe how to detect and characterise different forms of entanglement, including genuine multipartite entanglement and bound entanglement.
Bell inequalities are discussed to be typically violated even with randomised measurements, especially for a growing number of particles and settings.
Furthermore, we also present an overview on the estimation of non-linear functions of quantum states and shadow tomography from randomised measurements. 
Throughout the review, we complement the description of theoretical ideas by explaining key experiments.
\end{abstract}




\end{frontmatter}
\setcounter{tocdepth}{2}
\tableofcontents


\section{Introduction}

A key difference between classical and quantum information processing is the number of possible measurements one can perform.
While for a classical bit only a single type of measurement is possible, namely reading out its binary value, an infinite number of different measurements can be applied to even a single quantum bit.
Broadly speaking, this review explores the possibilities which arise when these measurements are chosen randomly.
While it was observed some time ago that randomised measurements are powerful tools for the analysis of quantum systems~\cite{munroe_photon_1995,  two-photon_beenakker_2009, liang_nonclassical_2010, laing_reference_2010,peeters_observation_2010, wallman_generating_2011, van_enk_measuring_2012, shadbolt_guaranteed_2012,
	laskowski_experimental_2012, palsson_experimentally_2012, laskowski_optimized_2013}, only in recent years a systematic approach was developed.
In particular, detailed strategies have been proposed which employ randomised measurements to detect and characterise entanglement, to determine certain invariant state properties and to certify Bell-type quantum correlations. These are the topics covered here.

\begin{figure*}[!t]
	\centering
	\includegraphics[width=0.65\textwidth]{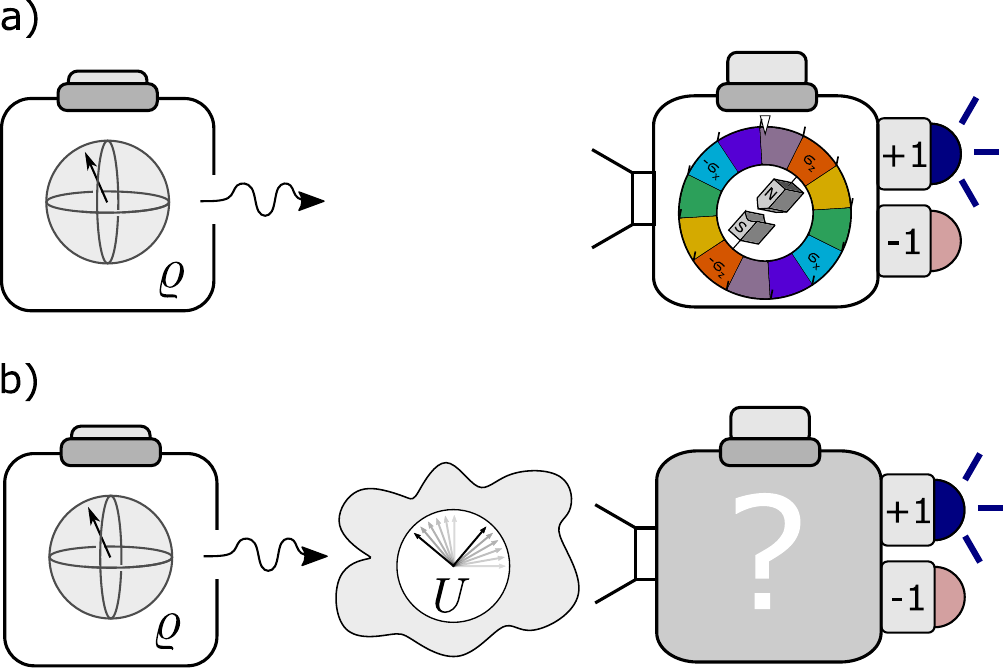}
	\caption{Different realisations of randomised measurements.
		a) Randomised measurements can be intentionally implemented in an experiment using a measurement device that applies random measurement settings, illustrated by the Stern-Gerlach magnet installed on the wheel of fortune.
		b) They may also arise inadvertently due to noisy environments.
		In this case, the measurement setting is not known to the experimenter even with full control over the measurement device.
	}
	\label{fig_concept1qubit_environment}
\end{figure*}

By construction, a randomised measurement strategy uses an apparatus where the measurement setting is chosen at random.
An example of a randomised measurement on a quantum system is given by a random unitary operation followed by a von Neumann measurement in some basis.
In general, the randomness can be introduced on purpose or could be a result of an unknown process, see Fig.~\ref{fig_concept1qubit_environment}.
There are foundational and practical motivations behind considering randomised measurements. 
On the fundamental side, it is intuitive that entangled quantum states are correlated in more bases than product states, and this prompts the question of whether correlations measured along random local directions are enough to capture the difference between entangled and product states.
The answer turns out to be affirmative.
Entanglement criteria in terms of randomised measurements provide necessary and sufficient conditions for entanglement in pure quantum states and in general give rise to witnesses capable of detecting genuine multipartite entanglement, bound entanglement or distinguishing various classes of entangled states, even for the case of mixed states.

A practical motivation comes from considering non-linear functions of quantum states.
Given a quantum state $\varrho$,  the outcome-statistics of von Neumann measurements are proportional to the first power of $\varrho$, i.e.~the probability of the $n$th result is governed by the Born rule $\tr(\varrho \proj{n})$. Therefore, a non-linear function of $\varrho$ is not directly measurable.
While it has been proposed to estimate non-linear functions with the help of multiple copies of $\varrho$, see e.g.~\cite{bruni_measuring_2004, mintert_observable_2007, walborn_experimental_2006}, randomised measurements provide an alternative which is still based on a single copy of a quantum state.
The ability to probe the non-linear properties has its origin in taking the non-linear functions of estimated correlations and averaging them over the randomly chosen measurement settings.
In this way, one gains access to a number of local unitary invariants such as purity, various entropies or many-body topological invariants.
Another practical motivation, which recently enjoyed increased attention, originates in the analysis of complex systems which in practice cannot be characterised by tomographic means.
Here, randomised measurements are used to gather a ``shadow’’ of the underlying complex system which turns out to contain considerable predictive power about the measurements that have not yet been performed.

In addition to the theoretical developments, this review describes a number of related experimental techniques in which randomised measurements are an attractive tool.
As mentioned above, they can either be implemented on purpose or can appear naturally due to noisy environments.
In a broad range of typically encountered environments, the transmission channel is modelled as a random unitary, as illustrated in Fig.~\ref{fig_concept1qubit_environment}b. For example, optical fibres rotate polarisation, changing phases affect the path degree of freedom, atmospheric turbulence acts on the modes of orbital angular momentum, and magnetic field fluctuations influence trapped ions.
If these fluctuations are slow enough, sufficient statistics in the computational basis measurement is attainable for every random unitary channel.
Note that in this case even the experimenter does not know which setting is actually measured.
Yet, meaningful statements about entanglement, Bell non-locality or non-linear functions of the state before the channel's action can be obtained.

In practice, dependent on the actual physical degree of freedom implementing the information processing, there are various ways in which randomised measurement can be engineered.
Typical realisations of such transformations are rotated waveplates and polarisers in optical polarisation measurements, or laser driven transitions in ion trap-based experiments. 
In the latter case, changing between the settings can be precisely controlled without determining exactly what the present setting is, which also naturally fits into the concept of randomised measurement.
In fact, the quantitative methods to analyse states with randomised measurements described here, rely on a uniform sampling of measurement settings.
In principle this can be very difficult or resource intensive to achieve since the amount of settings needed to sufficiently approximate uniformity is growing exponentially with the system size.
For such scenarios alternative techniques are available which allow to extract the same quantities via smaller sets of measurements, so called designs, that we also describe in detail.

As for any method, there are scenarios in which randomised measurements are especially useful but there are also scenarios where these tools are not optimal.
Randomised measurements are ideal tools when there is only a partial or no common reference frame between distant observers (or even local reference frames are missing) and when there is no prior knowledge about the measured quantum state.
Since the randomness could be introduced by a noisy environment, these methods are applicable in many natural scenarios of unitary noise as long as the rate with which the states are produced is high compared to the speed of fluctuations. 
Randomised measurements give access to non-linear functions of a quantum state without the need for the reconstruction of its density matrix and characterise many features of complex quantum systems through classical shadows.
Since a correlation estimated with even a single randomly chosen setting contains useful information about the quantum state, the method is attractive as a statistical entanglement witness in cases with very few detection events, e.g., in multi-photon optics.
The scenarios where randomised measurements would not be effective include estimation of parameters that are not invariant under local unitary operations, for example quantum coherence. The tools as presented here do not apply to systems where, say, subsequent emissions are not in the same state or contain correlations (not identical and independently distributed systems). More subtle properties of quantum states, such as bound entanglement, require the determination of more complex functions of measurement results which may have to be estimated with a greater number of experimental trials.

We stress that there is a recent excellent review article on the topic of randomised measurements\cite{elben_randomized_2022}, but in that article the focus lies mostly on scenarios where the random measurement setting is known. In contrast, we mainly consider types of randomness which is inevitably connected to a lack of information about the setting.
Still, in order to describe the full picture 
we also discuss the most relevant results for the case of known randomised measurements in this review.

The present review article is structured as follows.
In Sec.~\ref{SEC_NUTSHELL}, we offer an intuitive introduction to the concept of randomised measurements based on a two-qubit example and introduce basic mathematical tools.
In Sec.~\ref{SEC_ENTANGLEMENT} we start with a brief overview of entanglement theory and quantum designs.
Then, the heart of this section reviews various entanglement criteria in terms of correlations between randomised measurements.
They provide necessary and sufficient conditions for entanglement in pure states and certain mixed states, and in general give rise to witnesses capable of detecting genuine multipartite entanglement, bound entanglement or distinguishing various classes of entangled states.
In Sec.~\ref{SEC_FUNCTIONS} we describe estimations of local unitary invariants such as purity and give an overview of shadow tomography where randomised measurements play a crucial role.
In Sec.~\ref{SEC_BELL} we present different approaches to detect Bell non-local correlations between physical systems with randomised measurements.
We discuss quantifiers such as the probability of violation and strength of non-locality, present common definitions of genuine multipartite non-locality, and scenarios where even with randomised measurements a violation of a Bell-type inequality is certain.
We conclude in Sec.~\ref{SEC_CONCLUSIONS} and gather a list of interesting open problems encountered in the main body.
For a better overview, in Tab.~\ref{tab.randnomness} the various randomised measurement protocols are grouped systematically with respect to the available information and control of the respective settings.
Also links to the relevant sections in this review are provided.

\begin{table}
	\begin{center}
		\begin{tabular}{ m{0.25\textwidth}<{\centering\arraybackslash} | m{0.3\textwidth}<{\centering\arraybackslash} | m{0.3\textwidth}<{\centering\arraybackslash} }
			\hline
			\small{Partial Alignment} & \small{Only Local Alignment} & \small{No Alignment} \\
			\hline\hline
			\small{Bell Violations (Sec.~\ref{SEC_BELL})} & \small{Non-Product Observables (Secs.~\ref{SEC_ENTANGLEMENT}, \ref{SEC_FUNCTIONS}) \newline Bell Violations (Sec.~\ref{SEC_BELL})} & \small{Bell Violations (Sec.~\ref{SEC_BELL})}
			\newline
			\small{Product Observables (Secs.~\ref{SEC_ENTANGLEMENT}, \ref{SEC_FUNCTIONS}) \newline Average Correlation (Sec.~\ref{SEC_BELL})}\\ 
			\hline
		\end{tabular}
		\caption{Protocols that can be executed with varying degree of alignment between local reference frames. 
			``No alignment'' means that even locally the reference directions are not specified, i.e.~no information about the relative orientation of subsequent measurement settings is available, beyond e.g.~the fact that they are drawn from a specific distribution.
			``Only local alignment'' denotes the information and/or control over the relative orientation of local settings, but with pairwise random orientations between the observers.
			In ``Partial alignment'' different observers share certain common directions, e.g. for qubits the $z$ axes are the same, but $x$ and $y$ are randomly rotated.
		}
		\label{tab.randnomness}
		
	\end{center}
\end{table}


\section{Randomised measurements in a nutshell}
\label{SEC_NUTSHELL}


In this section, we explain the basic formalism and give a brief overview of the possible applications of randomised measurements to the study of quantum systems.
As an illustrative example, we start with the case of two qubits and introduce the distributions of random correlation values as fundamental tools to investigate state properties.
We identify the moments of these distributions as quantities which are straightforwardly accessible by randomised measurements and illustrate how they can be used in various criteria.
The section includes several relevant concepts, such as quantum designs, sector lengths, local unitary invariants and PT moments.
We conclude this section with a discussion of CHSH inequalities and the usefulness of Bell-type inequalities in the context of randomised measurements. 

\subsection{Distributions of correlation functions}
\label{sec:example}

To understand the general concept of randomised measurements and the difference in observations for entangled and separable states, let us first consider a simple two-qubit example with generalisations and extensions being discussed in subsequent sections.
Assume that we are given two states, a pure product state $|{\psi_{\mathrm{prod}}}\rangle \equiv |{\phi^A}\rangle \otimes |{\phi^B}\rangle$, that we will choose as $\ket{00}$, and  an ideal Bell state, say a singlet state $\ket{\psi^-}=(\ket{01}-\ket{10})/\sqrt{2}$, where we denote $\ket{0} = (1 \ 0)^\top$, $\ket{1} = (0 \ 1)^\top$, and abbreviate the tensor product as $\ket{xy} = \ket{x} \otimes \ket{y}$.
The state $\ket{\psi^-}$ does not admit a product form.
In fact, this property is a general definition of entanglement for any bipartite pure state.
The two introduced states are now subjected to local measurements with random settings.
More precisely, many local projective measurements along randomly chosen measurement directions are performed such that a set of \textit{correlation values} of the outcomes is obtained. 

The \textit{correlation function} is a statistical parameter characterising the statistical dependence of the results and is given by the mean of their product.
In a two-qubit experiment, where the $j$-th particle (for $j=1,2$) is measured in a setting represented by the normalised vector $\mathbf{u}_j$ on the Bloch sphere, with a binary measurement outcome $r_j = \pm 1$, the correlation function reads
\begin{equation}
	E(\mathbf{u}_1, \mathbf{u}_2) = \langle r_1 r_2 \rangle
	= \tr\big[\varrho \; (\mathbf{u}_1 \cdot \sigma \otimes \mathbf{u}_2 \cdot \sigma)\big].
	\label{eq:2-qubit_correlation_function}
\end{equation}
The average is denoted here by $\langle \cdots \rangle$ and is in practice estimated by repeating the experiment sufficiently many times. 
The last expression in Eq.~(\ref{eq:2-qubit_correlation_function}) represents the quantum mechanical prediction for the average measured given the system's state $\varrho$.
We use the short notation $\sigma$ for the vector of Pauli matrices, $\sigma = (\sigma_x,\sigma_y,\sigma_z)$, which will also be conveniently enumerated as $(\sigma_1, \sigma_2, \sigma_3)$, such that $\mathbf{u}_j \cdot \sigma$ is an arbitrary dichotomic observable, with outcomes $\pm 1$, of the $j$-th qubit. 
Such defined correlation functions are well known and feature, among other important applications, in violations of Bell inequalities or in solid state physics.

\begin{figure}[b!]
	\begin{center}
		\includegraphics[width=0.8\textwidth]{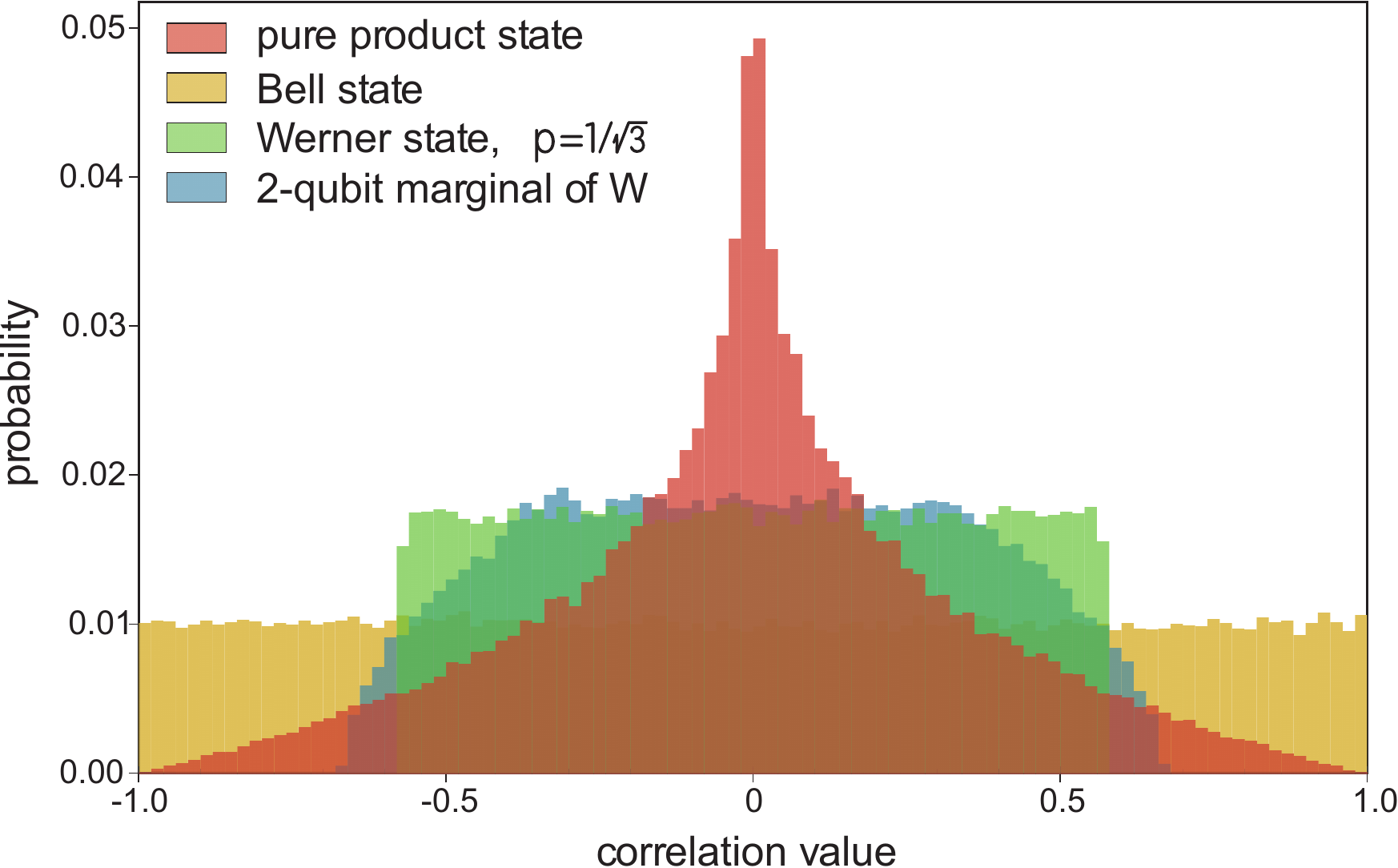}
		\caption{Numerically simulated probability distributions over values of the correlation function obtained under Haar-randomly chosen measurement directions. The results are shown for four different states indicated on the label. While the distributions are clearly distinct, states equivalent up to local unitary transformations result in equivalent distributions. The figure is taken from Ref.~\cite{knips_moment_2020}.}
		\label{2qubitExample}
	\end{center}
\end{figure}

Consider now a scenario with a large amount of different directions that are distributed uniformly on a Bloch sphere.
This is an example of a Haar random distribution, the general properties of which we describe in Sec.~\ref{sec:general_definitions}.
For each randomly chosen set of measurement directions, the experiment is repeated sufficiently many times to obtain a correlation value arbitrarily close to the quantum prediction. 
Fig.~\ref{2qubitExample} shows histograms of correlation values for different two-qubit states.
A very different behaviour is observed for the Bell state (yellow) and the product state (red).
For comparison, this figure also includes distributions for two mixed entangled states, i.e.~states which cannot be written in a separable form given by
\begin{equation}
	\varrho^{AB}_{\text{sep}}=\sum_i p_i \varrho^A_i \otimes \varrho^B_i,
\end{equation}
where $p_i>0$ and $\sum_i p_i=1$.
The two presented classes of mixed states are Werner states $\varrho_{\mathrm{Werner}}=(1-p)/4~{\mathbbm{1}}_4  + p\proj{\psi^-}$, and a two-qubit marginal of a $W_3$ state, 
that is $\varrho^{\mathrm{W_3}}_2 = \tr_3(\ket{W_3}\bra{W_3})$ with $\ket{W_3} = \left(\ket{001}+\ket{010}+\ket{100}\right)/\sqrt{3}.$
The data presented in Fig.~\ref{2qubitExample} shows that the knowledge of the probability distribution of correlations contains valuable information characterising the state.
Although this numerical simulation uses the ideal correlation values as described in Eq.~(\ref{eq:2-qubit_correlation_function}), a finite amount of different measurement directions has been chosen, leading to, e.g.~the deviations of the yellow distribution from a perfect uniform distribution.

It should be noted that, due to the nature of random measurements, states equivalent under local unitary (LU) transformations cannot be distinguished.
For example, any maximally entangled two-qubit state gives rise to the same distribution of outcomes as the singlet state and every pure product state is indistinguishable from the product state used to compute the distribution in Fig.~\ref{2qubitExample}. This is not surprising since all entanglement properties are by definition LU invariant. 
We elaborate more on this in Sec.~\ref{subsection:LUinva}.

\subsection{Moments of probability distributions}
\label{sec:moments}
A glance at Fig.~\ref{2qubitExample} shows that the distributions for the Bell state
$\ket{\psi^-}$ and the pure product state $|{\psi_{\mathrm{prod}}}\rangle$ have different variances.
This immediately raises the question of whether a larger variance in general implies entanglement.
In the following, we expand on this intuition and formulate the corresponding entanglement criterion.

To proceed, let us define the \textit{moments} of the probability distribution for the values of the correlation function as
\begin{align}\label{eq:momenttwoqubit}
	\mathcal{R}^{(t)}(\varrho)
	= N_t \int \mathrm{d}\mathbf{u}_1
	\int \mathrm{d}\mathbf{u}_2 \,
	[E(\mathbf{u}_1, \mathbf{u}_2)]^t,
\end{align}
where
$\mathrm{d}\mathbf{u}_j = \sin{\theta_j} \mathrm{d}\theta_j \mathrm{d}\phi_j$
denotes the uniform measure on the unit sphere which is also the Haar measure, see Refs.~\cite{tran_quantum_2015, tran_correlations_2016, ketterer_characterizing_2019} and Sec.~\ref{sec:general_definitions}. 
Here, $N_{t}$ is a normalisation constant, which is chosen differently throughout the literature.
Since we are interested in comparing moments of separable and entangled states, $N_{t}$ can be chosen such that $\mathcal{R}^{(t)}$ conveys scale-independent information in its probability distribution. Usually, the choice is such that the moments are directly given by other well-known quantities.
However, the choice of $N_t$ does not matter as long as it remains consistent across comparisons of $\mathcal{R}^{(t)}$ with different $N_t$.

Importantly, moments are invariant under any local unitary transformations of the state such that $\mathcal{R}^{(t)}(\varrho) = \mathcal{R}^{(t)}(V_1\otimes V_2 \varrho V_1^\dagger \otimes V_2^\dagger)$ for any single-qubit unitaries $V_1, V_2$. Since the amount of entanglement does not change under local unitaries, the moments thus seem well suited to capture the essential correlation properties of $\varrho$.

Note that $\mathcal{R}^{(t)}$ vanishes for odd $t$ since the sign of the correlation function is flipped under $\mathbf{u}_j \to -\mathbf{u}_j$.
Indeed, as seen in Fig.~\ref{2qubitExample}, the expectation ($t=1$) is zero for all states.
Thus, the first nontrivial result appears for $t=2$.
As discussed in more detail in Sec.~\ref{sec:n_qubit_entanglement}, the second moment gives rise to the following entanglement criterion:
\begin{equation}
	\textrm{If } \mathcal{R}^{(2)}(\varrho) > 1, \textrm{ then } \varrho \textrm{ is entangled.}
	\label{EQ_SIMPLE_R2}
\end{equation}
The elegant unit bound is obtained for the normalisation $N_2 = (3/4\pi)^2$, which will be further justified in the next section.
This entanglement criterion has been originally derived in Ref.~\cite{tran_quantum_2015} and generalised in Ref.~\cite{tran_correlations_2016}.
Equivalent statement in terms of sector length, see Sec.~\ref{sec:sector_lengths}, appeared in~\cite{hassan_separability_2008, hassan_experimentally_2008, hassan_geometric_2009} though note that the main proof is incomplete~\cite{tran_quantum_2015}.
The entanglement criterion is illustrated in Fig.~\ref{2qubitExample}, where the Bell state has clearly a larger variance than the pure product state.

We should stress that the criterion (\ref{EQ_SIMPLE_R2}) is only sufficient but not necessary for mixed two-qubit entangled states. 
That is, there exist mixed entangled states that this criterion does not detect.
Examples are the Werner states $\varrho_{\mathrm{Werner}}$ for $1/3 < p \le 1/\sqrt{3}$, and the two-qubit marginal of the $W_3$, state $\varrho^{\mathrm{W_3}}_2$.
To distinguish between these states and detect a broader range of entanglement, we need to use higher moments ($t>2$) to construct more refined entanglement criteria.
The details will be discussed in Secs.~\ref{sec:mixed_states} and \ref{sec:bipartite_higher_dimensions}.

\subsection{Quantum designs}
\label{sec:designs}
In order to evaluate the uniform average over the sphere in Eq.~(\ref{eq:momenttwoqubit}), the concept of \textit{designs} is very helpful.
Consider a polynomial function $f_t (\mathbf{x})$ in $n$ variables with degree $t$.
We call a set $X = \{\mathbf{x}_j \in S_n\}_{j=1, \ldots, K}$ \textit{spherical $t$-design} if
\begin{align}\label{eq:tdesignex}
	\frac{1}{K}
	\sum_{\mathbf{x}_j \in X} f_t (\mathbf{x}_j)
	= \int \mathrm{d}\mathbf{x} \, f_t (\mathbf{x}),
\end{align}
for any polynomial of at most degree $t$, where $\mathrm{d}\mathbf{x}$ is the spherical measure on the $n$-dimensional unit sphere $S_n$ with $\int \mathrm{d}\mathbf{x}=1$~\cite{delsarte_spherical_1991,colbourn_crc_2010}.
That is, the integral for the continuous polynomial $f$ can be evaluated by knowing its value at $K$ discrete points $\mathbf{x}_j$ of the spherical $t$-design set $X$.
By definition, the integral on the right-hand side in Eq.~(\ref{eq:tdesignex}) is invariant under any rotation on the sphere, so the evaluated expression on the left-hand side is also invariant.
In general, if the allowed degree $t$ or the dimension $n$ increases, then a larger set $X$ is required.
Further details are discussed in Sec.~\ref{sec:quantum_tdesigns}.

To give a concrete example, let us evaluate the second moment $\mathcal{R}^{(2)}$ using the idea of spherical designs.
This corresponds to the case $t=n=2$.
It is well known that a set of $K=6$ unit vectors on orthogonal antipodals, $\{\mathbf{x}_j = \pm \mathbf{e}_j: j=x,y,z\}$, where $\mathbf{e}_j$ are the Cartesian axes, is a spherical $2$-design (and also $3$-design)~\cite{ketterer_characterizing_2019, wyderka_learning_2020, seymour_averaging_1984}.
Using this spherical design, we rewrite each of the two integrals in $\mathcal{R}^{(2)}$ over a two-dimensional unit sphere as the average over the set of six points on the sphere:
\begin{equation}
	\mathcal{R}^{(2)}(\varrho)
	=3^2 \frac{1}{6^2}
	\sum_{j,k=1}^6
	[E(\mathbf{e}_j, \mathbf{e}_k)]^2
	=
	\sum_{j,k=x,y,z}
	[\tr(\varrho \; \sigma_j
	\otimes \sigma_k)]^2,
	\label{second_moment_tools}
\end{equation}
where we choose the normalisation $N_2=(3/4 \pi)^2$ and use the fact that the even function $[E(\mathbf{u}_1, \mathbf{u}_2)]^2$ does not change under the sign flip.
As a result, the integral over the entire spheres $\mathbf{u}_1, \mathbf{u}_2$ is replaced by a sum of nine (squared) correlation functions computed along orthogonal directions on local Bloch spheres.
Note that higher moments may be found in a similar manner, using designs for larger $t$.
Recalling that $\mathcal{R}^{(2)}$ is LU invariant and a convex function of a state, the separability bound can be found, without loss of generality, by considering the pure product state $|{\psi_{\mathrm{prod}}}\rangle = \ket{00}$.
We therefore arrive at the criterion discussed in the last section,
$\mathcal{R}^{(2)}(\varrho_{\text{sep}}) \leq 1$ for any two-qubit separable state $\varrho_{\text{sep}}$.

\subsection{Bloch decomposition of multipartite quantum states}
\label{sub_sec:bloch_decomposition}

Any single-qubit state $\varrho_A$ can be expressed using the operator basis of Pauli matrices as
\begin{align}
	\varrho_A = \frac{1}{2}\left(\mathbbm{1}_2 + a_x \sigma_x + a_y \sigma_y + a_z \sigma_z\right).
	\label{bloch_single_qubit}
\end{align}
Since Paulis are traceless the overall factor of $1/2$ follows from normalisation $\tr(\varrho_A) = 1$.
The positive semidefiniteness of the state $\varrho_A$ is equivalent to the constraint $\sum_{j=x,y,z} a_j^2 \leq 1$ \cite{nielsen_quantum_2011}.
The parametrisation in Eq.~(\ref{bloch_single_qubit}) enables us to visualise the state as a point within a unit sphere in a three-dimensional space with coordinates ${\textbf{a}} = (a_x, a_y, a_z)$.
It is called the Bloch sphere and has the property that a pure state corresponds to a point on the surface of the sphere, while a mixed state corresponds to a point inside.
It is essential to note that the \textit{length} of its radius, denoted as $L(\varrho) = \sum_{j=x,y,z} a_j^2$, corresponds to the purity $\mathcal{P}(\varrho)=\tr(\varrho^2)$, which remains invariant under unitary rotations.
That is, $0 \leq L(\varrho) = L(U\varrho U^\dagger)  \leq 1$, where the first and second inequalities are respectively saturated by the completely mixed state and pure states.

The decomposition (\ref{bloch_single_qubit}), in terms of Pauli operators $\{\mathbbm{1}_2, \sigma_x, \sigma_y,\sigma_z\}$, is based on the orthogonality relation $\tr(\sigma_{\mu} \sigma_{\nu}) = 2 \delta_{\mu\nu}$ for $\mu,\nu = 0,1,2,3$.
In the same way, a tensor product of Pauli operators forms a basis for composite quantum states.
For example, we can represent a two-qubit state $\varrho_{AB}$ in the Bloch form
\begin{align}
	\varrho_{AB} = \frac{1}{4}\left(
	\mathbbm{1}_2^A \otimes \mathbbm{1}_2^B
	+ \sum_{j=x,y,z} a_j \sigma_j^A \otimes \mathbbm{1}_2^B
	+ \sum_{j=x,y,z} b_j \mathbbm{1}_2^A \otimes \sigma_j^B
	+ \sum_{j,k=x,y,z} T_{jk} \sigma_i^A \otimes \sigma_k^B
	\right),
	\label{bloch_two_qubit}
\end{align}
with
$\sigma_j^{A} \in \mathcal{H}_2^{A}$ and $\sigma_j^{B} \in \mathcal{H}_2^{B}$ for $j = x,y,z$.
The coefficients $a_j$ and $b_j$ describe the reduced states, whereas the so-called correlation tensor $T = (T_{jk})$ captures two-body quantum correlations. 
Here, the positivity of $\varrho_{AB}$ implies several non-trivial constraints on the possible values of $\{a_j, b_j, T_{jk}\}$.
Thus in general it is difficult to find the complete set of values satisfying these constraints.
For example, all two-qubit states obey the condition
$\sum_{j=x,y,z}
(a_j^2 + b_j^2) +\sum_{j,k=x,y,z} T_{jk}^2 \leq 3$, following from the purity condition
$\tr(\varrho_{AB}^2) \leq 1$, but this is not sufficient to characterise the positivity, for details see Refs.~\cite{kurzyski_correlation_2011,gamel_entangled_2016,wyderka_characterizing_2020,morelli_correlation_2023}.
Notice that Eq.~(\ref{second_moment_tools}) can be written as
$\mathcal{R}^{(2)}(\varrho_{AB}) = \sum_{j,k=x,y,z} T_{jk}^2$.
Thus the sufficient criterion for two-qubit entanglement reads:
if $\sum_{j,k=x,y,z} T_{jk}^2 > 1$, then the state is entangled.

The Bloch decomposition can be generalised to $n$-particle $d$-dimensional quantum states ($n$ qudits) with
\begin{align} \label{eq-nqudits}
	\varrho = \frac{1}{d^{n}} \sum_{j_{1}, \cdots, j_{n}=0}^{d^2-1} T_{j_{1} \cdots j_{n}}
	\lambda_{j_{1}} \otimes \cdots \otimes \lambda_{j_{n}}, 
\end{align}
where $\lambda_{0}$ is the identity $\mathbbm{1}_d$,
and $\lambda_{j}$
are the generalised Gell-Mann matrices, such that
$\lambda_j=\lambda_j^\dagger$,
$\mathrm{tr}[\lambda_{j} \lambda_{k}]=d \delta_{j,k}$,
and $\mathrm{tr}[\lambda_{j}]=0$ for $j > 0$
\cite{kimura_bloch_2003,gell-mann_symmetries_1962}.
For $d=3$, one usually calls them simply the Gell-Mann matrices.
Note that some references use different normalisation of the $\lambda_j$ or yet different bases such as, for example, the Heisenberg-Weyl matrices~\cite{bertlmann_bloch_2008,asadian_heisenberg-weyl_2016}.
The correlation tensor $T_{j_{1} \cdots j_{n}}$ was first considered by Schlienz and Mahler in Ref.~\cite{schlienz_description_1995}.
We remark that the $k$-fold tensor $T_{j_{1} \cdots j_{n}}$ for $1 \leq k \leq n$, i.e.~the entries, for which $k$ indices are non-zero, characterises the $k$-body correlations of the (reduced) state.

\subsection{Sector lengths}
\label{sec:sector_lengths}
The Bloch representation directly leads to the notion of so-called sector lengths.
As mentioned, the length of the one-party Bloch vector quantifies the degree of mixing of the state.
Accordingly, the length encodes information about the state that can be obtained in a basis-independent way.
The sector lengths are its direct extension to multipartite quantum systems.

The sector lengths $S_k$ are defined based on the generalised Bloch decomposition of a $n$-qudit state in Eq.~(\ref{eq-nqudits}) as
\begin{align} \label{eq:sectorlengths}
	S_k(\varrho)
	= \sum_{k \textrm{ non-zero indices}}
	T_{j_{1} \cdots j_{n}}^2,
\end{align}
where
$S_0=\lambda_{0 \cdots 0}=1$ due to the normalisation condition $\tr(\varrho)=1$.
The sector lengths quantify the amount of $k$-body correlations in the state $\varrho$.
For example, in the case of the three-qubit GHZ state $\ket{\mathrm{GHZ}_3} = (\ket{000}+\ket{111})/\sqrt{2}$, one obtains $(S_1, S_2, S_3) = (0,3,4)$.
Additionally, note that Eq.~(\ref{EQ_SIMPLE_R2}) along with (\ref{second_moment_tools}) give an entanglement criterion in terms of sector length.

Sector lengths have several useful properties.
(i) The sector lengths are invariant under any local unitary transformation.
That is, for a local unitary $V_{1} \otimes \cdots \otimes V_{N}$, it holds that $S_k(\varrho)=S_k\big(V_{1} \otimes \cdots \otimes V_{n} \varrho V_{1}^{\dagger} \otimes \cdots \otimes V_{n}^{\dagger}\big)$.
(ii) The sector lengths are convex on quantum states.
That is, for the mixed quantum state $\varrho=\sum_i p_i \ket{\psi_i}\!\bra{\psi_i}$, it holds that $S_k \left(\sum_i p_i \ket{\psi_i}\!\bra{\psi_i}\right) \leq \sum_i p_i S_k(\ket{\psi_i})$.
(iii) The sector lengths have a convolution property:
For a $n$-particle product state $\varrho_P \otimes \varrho_Q$, where $\varrho_P$ and $\varrho_Q$ are, respectively, $j$-particle and $(n-j)$-particle states, we have
$S_k(\varrho_P \otimes \varrho_Q) = \sum_{i=0}^{k} S_i(\varrho_P) S_{k-i}(\varrho_Q)$ \cite{wyderka_characterizing_2020}.
(iv) The sector lengths are directly associated with the purity of $\varrho$, namely
\begin{align} 
	\mathcal{P}(\varrho)
	= \frac{1}{d^{n}} \sum_{k=0}^{n}S_k(\varrho).
	\label{sectorandpurity}
\end{align}
That is, the purity can be decomposed into the sector lengths of different orders.
Using this relation, the sector lengths can be always represented as the purities of reduced states of $\varrho$, and vice versa.
(v) The $n$-body (often called full-body) sector length $S_n$ for all $n$-qubit states has been shown to be always maximised by the $n$-qubit GHZ state, denoted by
$\ket{\text{GHZ}_n}=(\ket{0}^{\otimes n} + \ket{1}^{\otimes n})/\sqrt{2}$.
Its maximal value is given by
$S_n(\text{GHZ}_n)
= 2^{n-1} + \delta_{n, \, \text{even}}$
\cite{tran_correlations_2016,eltschka_maximum_2020, miller_small_2019}.
However, this is not always true in higher dimensions, i.e.~quantum states that are not of the GHZ form can attain the maximal $S_n$ value~\cite{eltschka_maximum_2020}.
Even more interestingly, it has been demonstrated that there exist multipartite entangled states with zero $S_n$ for an odd number of qubits~\cite{kaszlikowski_quantum_2008,laskowski_incompatible_2012,schwemmer_genuine_2015,tran_genuine_2017,klobus_higher_2019}.

Finally and importantly, the sector lengths can be directly obtained from the randomised measurement scheme.
In fact, the $k$-body sector lengths $S_k$ can be represented as averages over all second-order moments of random correlations in $k$-particle subsystems.
Entanglement criteria using the sector lengths are therefore accessible with randomised measurements, for details see Sec.~\ref{sec:multipartite_entanglement}.

\subsection{Local unitary invariants}\label{subsection:LUinva}
Moments of random correlations and sector lengths are special cases of a broader class of functions of the quantum state which are invariant under local unitary transformations.
In general, such \textit{local unitary (LU) invariants} $q(\varrho)$ are functions of the quantum state $\varrho$ for which
\begin{align}
	q\big(V_{1} \otimes \cdots \otimes V_{n}
	\varrho V_{1}^{\dagger} \otimes \cdots \otimes V_{n}^{\dagger}\big)
	= q(\varrho),
\end{align}
for any $V_{1} \otimes \cdots \otimes V_{n}$ with 
$V_i$ defined in a $d$-dimensional unitary group.
Since the purity of the global state $\mathcal{P}(\varrho) =  \tr(\varrho^2)$ is invariant under any global unitary, one can interpret the relation in Eq.~(\ref{sectorandpurity}) as a decomposition of a global unitary invariant into LU invariants.
Moreover, for two qubits $\tr(\varrho^3)$, can be expressed with the help of the determinant $\det{(T)}$ of the correlation tensor introduced in Eq.~(\ref{bloch_two_qubit}).
This is one of the so-called Makhlin LU invariants~\cite{makhlin_nonlocal_2002}.

Here a nontrivial question arises:
How can we access certain LU invariants from randomised measurements?
Since LU invariants include detailed information about quantum correlations in the state, addressing this question can be related to the improvement of entanglement detection and can reveal many other important properties of the state.
In Secs.~\ref{sec:PPT_criterion} and \ref{sec:makhlin_invariants}, we discuss how LU invariants for two qubits can be characterised by randomised measurements.

Another example of LU invariant is the Rényi entropy of order $\alpha$, defined as
\begin{equation}
	H_\alpha\left(\varrho_M \right)=
	\frac{1}{1-\alpha}
	\log\left[\tr
	\left( \varrho_M^\alpha\right)\right],
	\label{eq:Rényientropy}
\end{equation}
where
$\alpha\in\mathbb{R}, \alpha\neq0, \alpha\neq1$,
the reduced state is defined as $\varrho_M = \tr_{\Bar{M}}(\varrho)$
for any $M=\{1,2,\ldots, n\}$ and the trace is taken over the complement $\Bar{M}$.
In particular, the second-order Rényi entropy $H_2$ is often used to analyse entanglement~\cite{horodecki_quantumentropy_1996}. This quantity is accessible through randomised measurements, see Secs.~\ref{sec:second_moments} and \ref{sec:general_determination_uni_inv}.

\subsection{PT moments}
\label{subsec:pt_moments}

A different route to witnessing entanglement via randomised measurements is based on the Peres-Horodecki separability criterion~\cite{peres_separability_1996, horodecki_separability_1996}.
It states that if a bipartite state $\varrho_{AB}$ is separable, then the partially transposed density matrix $\varrho^{\Gamma_B}_{AB}:=(\text{id} \otimes T)(\varrho_{AB})$ is positive semi-definite, where $\text{id}$ is the identity map and $T$ is the transposition map.
States with this property are called PPT states, as they have a positive semidefinite partial transpose.
Contrary, if $\varrho^{\Gamma_B}_{AB}$ has negative eigenvalues, the state is called NPT and must be entangled.
Importantly, for systems consisting of two qubits or a qubit and a qutrit a positive semi-definite $\varrho^{\Gamma_B}_{AB}$ is also a sufficient criterion for separability. 
In general, however, there exist entangled states that the PPT criterion can never detect.
The criterion can also be used to quantify entanglement and the corresponding entanglement monotone is provided by the logarithmic 
negativity defined as~\cite{zyczkowski_volume_1998, zyczkowski_volume_1999,vidal_computable_2002,plenio_logarithmic_2005, Lee_partial_2000}
\begin{equation}
	\mathrm{E}_N(\varrho_{AB})=\log ||\varrho_{AB}^{\Gamma_B}||_1=\log\sum_{i} |\lambda_i|,
\end{equation}
with $|| \cdot ||_1$ being the trace norm and $\lambda_i$ the eigenvalues of the partially transposed density matrix.

In order to make the PPT criterion accessible by randomised measurements one considers the so-called PT (or negativity) moments. The $k$-th PT moment is defined as
\begin{equation}
	p_k(\varrho_{AB}) = \tr \left[ \left( \varrho_{AB}^{\Gamma_B} \right)^k \right].
	\label{eq:pt_moments}
\end{equation}
These quantities are LU invariant for any order $k$, since the eigenvalues of the partially transposed matrix are LU invariant.
Similarly to the moments of the density matrix [see their use in Eq.~(\ref{eq:Rényientropy})], these moments can be determined by randomised measurements \cite{zhou_single-copies_2020, elben_mixed-state_2020}, as described in Sec.~\ref{sec:pt_estimation}. 

Furthermore, it is a well-established mathematical fact that the coefficients of the characteristic polynomial of a matrix can be expressed in terms of traces of the power of this matrix~\cite{roman_advanced_2005}, so knowledge of the moments $p_k$ for any $k$ allows to evaluate the PPT criterion. In practice, however, only a few of these moments can be measured and the question arises: Is this data compatible with a PPT density matrix or not?
This question is similar to security analysis in entanglement-based quantum key distribution, where the protocol is insecure if the measured data is compatible with a separable state~\cite{curty_entanglement_2004}.

In Ref.~\cite{gray_machine-learning-assisted_2018} it was shown via a machine-learning approach that the logarithmic negativity can be 
estimated using $p_3$.
As an analytical result, the following moment-based entanglement criterion was introduced in Ref.~\cite{elben_mixed-state_2020},
\begin{equation}
	p_3 < p_2^2 \implies \mbox{$\varrho_{AB}$ is NPT and hence entangled.}
	\label{p3pptc}
\end{equation}
This so-called $p_3$-PPT criterion was utilised to detect entanglement in the experimental data from Ref.~\cite{brydges_probing_2019}. It is worth noting that the PT-moment approach, even with lower orders, can detect the Werner state in a necessary and sufficient manner for any dimension, for more details see the Appendix in Ref.~\cite{elben_mixed-state_2020}.

Still, the $p_3$-PPT criterion is not the optimal way to extract information from the moments $p_2$ and $p_3.$
This problem can be solved with a family of optimal criteria ($p_n$-OPPT) derived in Ref.~\cite{yu_optimal_2021}, see also Ref.~\cite{neven_symmetry-resolved_2021}.
For the special case of $n=3$, the necessary and sufficient $p_3$-OPPT condition for compatibility of the PT moments with a PPT state is given by
\begin{equation}
	p_3 \geq \alpha x^3+(1-\alpha x)^3,
	\label{p30ppt}
\end{equation}
where $\alpha=\lfloor1/p_2 \rfloor$ and $x=[\alpha+(\alpha [p_2(\alpha+1)-1])^{1/2}]/[\alpha (\alpha +1)]$.
Note that the above expression does not depend on the Hilbert space dimension.
Also, if the PT moments are compatible with the spectrum of a PPT state, they are compatible with a separable state, since one can directly write down a separable state (diagonal in the computational/product basis) for a given nonnegative spectrum of the partial transpose. 
Finally, the $p_n$-OPPT criteria are defined for all $n \geq 3$ and demonstratively stronger than their $p_n$-PPT counterparts as shown with numerical simulations \cite{yu_optimal_2021}.

\subsection{Bell-type inequalities}
\label{sec.nutshell.bell}
Another topic where randomised measurements are highly useful tools is testing of so-called non-local correlations in quantum systems.
One of the most fundamental properties of quantum mechanics is that measurement results at spatially separated measurement sites exhibit correlations that do not permit a classical description.
As shown by Bell's theorem~\cite{bell_einstein_1964,bell_theory_1976} such correlations can only be explained if certain fundamental assumptions about the physical world are given up. These include relativistic causality, the possibility to choose measurement settings independently of the experimental results or the ability to causally explain the occurrence of the outcomes altogether, sometimes also referred to as giving up ``realism'', for a careful analysis see~\cite{wiseman_twobell_2014}.
Apart from such basic questions, this class of correlations is also an important resource used in numerous quantum information processing protocols, in particular in quantum key distribution~\cite{acin_device-independent_2007}, in the certified generation of unpredictable randomness~\cite{acin_certified_2016}, and in reducing the communication complexity of computation~\cite{buhrman_nonlocality_2010}. 
These unique properties of quantum systems are sometimes called ``quantum nonlocality'' or ``Bell non-locality'' in the literature~\cite{brunner_bell_2014,scarani_bell_2019}.

Whether a given state produces Bell non-locality is usually tested via inequalities which give bounds on functions of expectation values for joint measurements at spatially separated sites sharing an entangled state~\cite{scarani_bell_2019, brunner_bell_2014}.
The simplest of these, the Clauser-Horne-Shimony-Holt (CHSH) inequality~\cite{clauser_proposed_1969} applies to the scenario of two observers who share an entangled state of two qubits.
They perform dichotomic measurements, with the first observer choosing between two alternative observables $\mathbf{u}_1 \cdot \sigma$ and $\mathbf{u}_1' \cdot \sigma$, and the second observer between $\mathbf{u}_2 \cdot \sigma$ and $\mathbf{u}_2' \cdot \sigma$.
It can be proven that any Bell-local model~\cite{fine_hidden_1982}, i.e.~any model respecting all assumptions of Bell's theorem, satisfies the inequality
\begin{equation}
	S \equiv |E(\mathbf{u}_1,\mathbf{u}_2)+E(\mathbf{u}_1,\mathbf{u}_2')+E(\mathbf{u}_1',\mathbf{u}_2)-E(\mathbf{u}_1',\mathbf{u}_2')| \leq 2.
	\label{CHSH}
\end{equation}
For a suitable maximally entangled state and an optimal choice of observables, as for example
$\mathbf{u}_1 = \mathbf{x}$, $\mathbf{u}_1' = \mathbf{z}$, and $\mathbf{u}_2 = (\mathbf{x} + \mathbf{z})/\sqrt{2}$, $\mathbf{u}_2' = (\mathbf{x} - \mathbf{z})/\sqrt{2}$,
we obtain the largest value of the left-hand side of the inequality $S_{max}=2\sqrt{2} >2$ and hence the maximal violation of the inequality. 
The quantum violations of similar inequalities have been observed in precisely dedicated experiments~\cite{hensen_loophole_2015,giustina_significant_2015,shalm_strong_2015,rosenfeld_event_2017}.

One can also ask if always a Bell-local model exists whenever the CHSH inequality is not violated.
The answer is negative and a necessary and sufficient condition for the existence of such a model is given by a set of inequalities (not just one of them) which describe the facets of so-called Bell-Pitovsky polytope~\cite{pitovsky_quantum_1989}.
These polytopes are different for scenarios with different numbers of observers, measurement settings and outcomes. It turns out that for two parties, each choosing between two dichotomic measurements, it is sufficient to permute the observables in the CHSH inequality to generate the complete set of 16 CHSH inequalities describing the Bell-Pitovsky polytope.
For more complex scenarios the corresponding polytopes have been fully characterised analytically only for special cases~\cite{werner_bell_2001, zukowski_all_2002, sliwa_symmetries_2003,
	bancal_looking_2010,pironio_all_2014, deza_enumeration_2015}.

The maximum value $S$ of the CHSH expression that a given state $\varrho$ can achieve, optimised over the choice of measurements, is given by $S(\varrho) = 2 \sqrt{\lambda_1 + \lambda_2}$, where $\lambda$'s are the two largest eigenvalues of the matrix $T^\top T$~\cite{horodecki_violating_1995} with $T$ defined in Eq.~(\ref{bloch_two_qubit}).
The corresponding state violates the CHSH inequality if and only if $\sqrt{\lambda_1 + \lambda_2} > 1$.
This condition can also be expressed directly in terms of correlation matrix elements \cite{zukowski_bells_2002} as $\sum_{i,j=x,y} T_{ij}^2 > 1$, where the axes $x$ and $y$ define the plane in which the optimal settings for the inequality lie.

While, in general, a particular choice of settings is crucial to obtain the violation of Bell-type inequalities such as the CHSH inequality, it is interesting to investigate whether Bell non-local correlations can also be witnessed in the scenario of randomised measurements.
It turns out that with suitable states a violation can still be guaranteed, even if certain fixed random rotations are added between the reference frames of the two observers or even with randomness in the local frames.
A detailed discussion of this topic is presented in~ Secs.~\ref{sec:makhlin_invariants} and \ref{sec:non_classicasl_correlations}.

\section{Entanglement} 
\label{SEC_ENTANGLEMENT}

In this section, we review several results on detecting entanglement using randomised measurements.
We begin with an introduction to the theory of entanglement and the general framework of randomised measurements focusing on the $t$-th moment of the distributions of correlation values in multipartite high-dimensional systems.
We also provide an overview of quantum $t$-designs as a powerful tool for the computation of integrals over Haar randomly distributed unitaries.
Subsequently, applications of these tools to detect and characterise entanglement in a broad range of scenarios are discussed.
The section concludes with an analysis of the effects of statistical noise due to limited data in experimental situations and a proposal of proper strategies to account for this noise.

\subsection{Multipartite entanglement}
\label{sec:introduction_entanglement_detection}

In the previous section, basic intuitions behind the structure of entanglement and its detection with randomised measurements have been introduced.
Here, we discuss entanglement beyond the two-qubit scenario to include systems with an arbitrary dimension and number of parties.
The interested reader can find more details about the field of multipartite entanglement in several in-depth review articles \cite{guhne_entanglement_2009,horodecki_quantum_2009,friis_entanglement_2018,plenio_introduction_2007,eltschka_quantifying_2014,vicente_further_2008,noauthor_lectures_2006}.

An $n$-partite $d$-dimensional quantum state ($n$-qudit) defined in the Hilbert space $\mathcal{H}_d^{\otimes n}$ is \textit{fully separable} if it can be written as
\begin{align}
	\varrho_{\text{fs}}
	= \sum_i p_i 
	\varrho_i^1 \otimes 
	\varrho_i^2 \otimes \cdots \otimes
	\varrho_i^n, 
\end{align}
where $\varrho_i^j$ are quantum states and the $p_i$ form a 
probability distribution, i.e.~$p_i \geq 0$ and $\sum_i p_i = 1$.
We say that an $n$-particle state contains entanglement if it is not fully separable.
Note that this does not imply anything about the structure of the entanglement, as for example whether all parties are entangled with each other.
One option to intuitively understand different types of entanglement is to consider how states are prepared.
For instance, $\varrho_{\text{fs}}$ can be prepared from a product state by local operations and classical communication (LOCC) by operating on each particle separately.
One can also consider states which can be prepared from a product state by LOCC where the operations are performed jointly on groups of particles (not just on one particle).
For instance, a state is called \textit{biseparable} with respect to a bipartition $M|\Bar{M}$, for a 
subset $M \subset \{1,2,\ldots, n\}$,
if it can be written as
\begin{align}
	\varrho_{M|\Bar{M}}
	= \sum_i
	q_i^M
	\varrho_i^M \otimes 
	\varrho_i^{\Bar{M}},
\end{align}
where the $q_i^M$ form a probability distribution,
$\Bar{M}$ is the complement of $M$ and $\varrho_i^M$ is a quantum state of particles in set $M$. 
In order to prepare state $\varrho_{M|\Bar{M}}$ via LOCC one needs to operate jointly on subsystems in the set $M$ and in the set $\Bar{M}$.
Moreover, one can consider mixtures of biseparable states for all bipartitions,
\begin{align}
	\varrho_{\text{bs}}
	= \sum_{M} p_M \varrho_{M|\Bar{M}},
	\label{EQ_GME}
\end{align}
where $p_M$ are probabilities and the summation includes at most $2^{n-1}-1$ terms.
Such a general state is simply called biseparable (without reference to any particular bipartition).
A quantum state which cannot be written in the form (\ref{EQ_GME}) is called \textit{genuine $n$-particle entangled} and involves entanglement between all subsystems.

For example, a three-particle state is called biseparable for a bipartition $A|BC$ if
\begin{align}
	\varrho_{A|BC}
	=\sum_i q_i^A
	\varrho_i^A \otimes
	\varrho_i^{BC},
\end{align}
where $\varrho_i^{BC}$ may be entangled.
We can furthermore construct mixtures of biseparable states with respect to different partitions, i.e.~states of the form
\begin{align}
	\varrho_{\text{bs}}
	= p_A\varrho_{A|BC}
	+ p_B\varrho_{B|CA}
	+ p_C\varrho_{C|AB},
	\label{eq:biseparablethreepartite}
\end{align}
where the $p_A, p_B, p_C$ are probabilities.
In contrast, a typical example of a genuine $n$-qudit entangled state is the generalised Greenberger–Horne–Zeilinger (GHZ) state given by
\begin{align}
	\ket{\text{GHZ}(n,d)}
	= \frac{1}{\sqrt{d}}
	\sum_{i=0}^{d-1}
	\ket{i}^{\otimes n}.
\end{align}
In particular, in two-qudit systems (that is, $n=2$),
this state is the maximally entangled two-particle state.
Other examples of genuine $n$-partite entangled states include W states~\cite{dur_three_2000}, Dicke states~\cite{toth_detection_2007}, cluster states~\cite{briegel_persistent_2001}, graph states~\cite{hein_multiparty_2004}, and absolutely maximally entangled (AME) states~\cite{ helwig_absolute_2012,huber_absolutely_2017,Klobus_k-uniform_2019}.

The question of whether a given quantum state is separable or entangled is known as the \textit{separability problem} and is central for quantum information theory.
It has several aspects:
\begin{itemize}
	\item[(a)] \textit{Complexity}:
	Even if the density matrix is completely known, in general it remains a complicated mathematical problem to determine whether a state is entangled,
	known to belong to the NP-hard class of computational complexity~\cite{gharibian_strong_2010}.
	Following the Choi-Jamiolkowski isomorphism, connecting quantum states and channels~\cite{depillis_linear_1967,choi_completely_1975,jamiolkowski_linear_1972,lewenstein_optimization_2000}, the separability problem is equivalent to the problem of distinguishing positive and completely positive maps which is as yet unsolved.
	
	\item[(b)] \textit{Practical issues and limited control}:
	In experiments, sometimes only partial information about the state is accessible.
	If some \textit{a priori} information about the state is available, e.g.~that an experiment is aimed at producing a certain entangled state, then so-called entanglement witnesses may allow for the efficient detection using directly measurable observables~\cite{lewenstein_optimization_2000,bruss_reflections_2002,guhne_entanglement_2009}.    
	In other situations, where one cannot be sure about the appropriate description of measurements and cannot trust the underlying quantum devices, it is still possible to certify entanglement in a device-independent manner~\cite{acin_from_2006}, using, e.g.~Bell-type inequalities, based only on the measurement data observed from input-output statistics~\cite{bancal_device-independent_2011,pal_device_2014}.
	Moreover, when considering ensembles of quantum particles, such as cold atoms, individual control over local subsystems may be lost, but entanglement can still be characterised by measuring macroscopic quantities and applying, e.g. spin squeezing inequalities~\cite{sorensen_many_2001,toth_optimal_2007,ma_quantum_2011,wieniak_magnetic_2005,wieniak_heat_2008}.
	
	\item[(c)] \textit{Meaningful entanglement}:
	Addressing the separability problem highlights distinctions between quantum physics and classical physics in terms of correlations.
	The features of entanglement, such as the negativity of conditional entropy~\cite{cerf_negative_1997,horodecki_partial_2005}, monogamy of entanglement~\cite{coffman_distributed_2000,osborne_general_2006}, or the presence of bound entanglement~\cite{horodecki_mixedstate_1998,divincenzo_evidence_2000}, are associated with entanglement conditions from fundamental and operational viewpoints.
	In fact, whether a given entangled state is useful or not (can be used to outperform classical solutions), is decided by certain thresholds in terms of several quantum communication protocols~\cite{bennett_teleporting_1993,curty_entanglement_2004,horodecki_quantum_2006} or quantum metrology~\cite{pezze_entanglement_2009,toth_quantum_2014,pezze_quantum_2018}.
	
	\item[(d)] \textit{Generalizations}:
	As a generalisation of the separability problem, one can ask, for example, how many partitions are separated in a multipartite state based on the concept of $k$-separability~\cite{dur_separability_1999,dur_classification_2000} (see Sec.~\ref{sec:k_separability_drom_SL}), or how many particles are entangled based on the concept of $k$-producibility~\cite{guhne_multipartite_2005,guhne_energy_2006,hyllus_fisher_2012,sorensen_entanglement_2001}.
	Other interesting concepts are given by $k$-stretchability~\cite{szalay_kstretchability_2019,toth_stretching_2020,ren_metrological_2021}, tensor rank~\cite{eisert_schmidt_2001}, and the bipartite and multipartite dimensionality~\cite{spengler_examining_2013,huber_structure_2013,kraft_characterizing_2018} of entanglement.
	Genuine multipartite entanglement can in turn again be classified into several types, such as the W class or GHZ class of states, for details see Sec.~\ref{sec:W_class}.
	More recently, also different notions of network entanglement came into the focus of attention~\cite{navascues_genuine_2020,tavakoli_bell_2022,hansenne_symmetries_2022}.
\end{itemize}

\subsection{Randomised measurements on multipartite systems}
\label{sec:general_definitions}

While in Section \ref{SEC_NUTSHELL} we have mainly discussed the second moment of correlations obtained via randomised measurements on two qubits, in the following we generalise this scheme to $t$-th moments in $n$-particle $d$-dimensional quantum systems.
Using this formulation, we review several systematic methods to detect various types of entanglement.

When measuring an observable $\mathcal{M}$ on a state with $n$ parties, $\varrho \in \mathcal{H}_d^{\otimes n}$, such that each party rotates their measurement direction in an arbitrary manner according to a randomly chosen unitary matrix $U_i$, the corresponding correlation function reads
\begin{align}
	E(\mathcal{M}, U_1 \otimes U_2 \otimes \cdots \otimes U_n)
	= \tr\left[\varrho
	(U_1 \otimes U_2 \otimes \cdots \otimes U_n)^\dagger
	\mathcal{M}
	(U_1 \otimes U_2 \otimes \cdots \otimes U_n)
	\right].
	\label{eq:correlation_function_general}
\end{align}
By sampling random unitaries uniformly from the unitary group, the resulting distribution of the correlation functions can be characterised by its moments with
\begin{align}
	\mathcal{R}^{(t)}_{\mathcal{M}}(\varrho)
	=N_{n,d,t}
	\int \mathrm{d}\mu(U_1) \int \mathrm{d}\mu(U_2) \cdots \int \mathrm{d}\mu(U_n) \,
	[E(\mathcal{M}, U_1 \otimes U_2 \otimes \cdots \otimes U_n)]^t,
	\label{eq:generalmoment}
\end{align}
where the integral is taken according to the Haar measure $\mathrm{d}\mu(U)$. 
Here, we denote $N_{n,d,t}$ as a suitable normalisation constant, which again is defined differently throughout the literature.
For the case of $n=d=2$ we arrive at the form of Eq.~(\ref{eq:momenttwoqubit}) independently of which Pauli product observable $\mathcal{M} = \sigma_i \otimes \sigma_j$ with $i,j=x,y,z$ is chosen.
In the same manner, without loss of generality, the observable can be assumed to be $\sigma_z \otimes \sigma_z \otimes \ldots \sigma_z$ for larger $n$.

The notion of the Haar measure used in the definition of a moment is defined as follows. Let $\mathcal{U}(d)$ be the group of all $d \times d$ unitaries and $f(U)$ be a function on $\mathcal{U}(d)$.
Then, the Haar measure $\mathrm{d}\mu(U)$ used in the
integral of $f(U)$ over the unitary group $\mathcal{U}(d)$ is defined by its invariance properties.
By this one means the left and right invariance under shifts via multiplication by an arbitrary unitary $V \in \mathcal{U}(d)$, which is respectively given by
\begin{align}
	\int \mathrm{d}\mu(U)\, f(U)
	= \int \mathrm{d}\mu(U)\, f(VU)
	= \int \mathrm{d}\mu(U)\, f(UV),
\end{align}
see Refs.~\cite{collins_integration_2006,puchala_symbolic_2017,spengler_composite_2012,zhang_matrix_2015,kostenberger_weingarten_2021,collins_weingarten_2022,mele_2023} for further details.
A general parametrization of the unitary group $\mathcal{U}(d)$ and the associated Haar measure are known~\cite{spengler_composite_2012,tilma_generalized_2002}.
For instance, any single-qubit unitary ($d=2$) can be written in the Euler angle representation~\cite{sakurai_modern_1995} as
$U(\alpha, \beta, \gamma)
= U_z(\alpha)U_y(\beta)U_z(\gamma)$,
where
$U_i(\theta) = e^{-i\theta \sigma_i/2}$ for $i=y,z$
and the Haar measure 
$\mathrm{d}\mu(U) = \sin(\beta) \mathrm{d}\alpha \mathrm{d}\beta \mathrm{d}\gamma$. 
For the qubit case, $d=2$, the Haar unitary integrals can be replaced by integrals with respect to the uniform measure on the Bloch sphere $S^2$,
\begin{align}
	\int \mathrm{d}\mu(U) \to \frac{1}{4\pi} \int_{S^2} \mathrm{d}\mathbf{u}, 
\end{align}
where
$\mathrm{d}\mathbf{u}= \sin(\theta) \mathrm{d}\theta \mathrm{d}\phi$.
With the help of quantum designs, one can simplify such integrals to a finite sum over certain directions on the Bloch sphere.

In many applications, it is useful to know the specific properties and issues associated with moments. The ones which are especially relevant for the discussed subjects are: 
\begin{itemize}
	\item[(a)] \textit{Local unitary invariance}: 
	By their definition, the moments are invariant under any local unitary transformation.
	More precisely, since the Haar measure is invariant under left and right translation,
	it holds that
	\begin{align} 
		\mathcal{R}^{(t)}_{\mathcal{M}}(\varrho)
		= \mathcal{R}^{(t)}_{\mathcal{M}}
		(V_{1} \otimes \cdots \otimes V_{n} \; \varrho \;  V_{1}^{\dagger} \otimes \cdots \otimes V_{n}^{\dagger}),
	\end{align}
	for any local unitary $V_{1} \otimes \cdots \otimes V_{n}$.
	Thus, we can characterise the state $\varrho$ with the moments $\mathcal{R}^{(t)}(\varrho)$ independently of the choice of local bases, that is, independent of reference frames between parties or unknown local unitary transformations.
	This invariance is one of the most important properties of randomised measurements and suggests that the moments of the measured distributions contain essential information about the entanglement of the corresponding quantum states.
	
	\item[(b)] \textit{Choice of observables}:
	In general, the observable $\mathcal{M}$ does not necessarily have to be a product observable of the form $\mathcal{M}_{\text{P}} = M_1 \otimes M_2 \otimes \cdots \otimes M_n$,
	but can be of the more general form $\mathcal{M}_{\text{NP}} = \sum_i m_i M_1^i \otimes M_2^i \otimes \cdots \otimes M_n^i$, with real coefficients $m_i$ \cite{horodecki_quantum_2008,wyderka_complete_2022}.
	The measurement of non-product observables requires a certain restriction of randomness, where the unitary cannot change significantly while the various observers switch between the particular local observables $M^i_k$ in a synchronised manner.
	However, as a tradeoff, it enables the extraction of additional information not accessible via product observables as discussed in 
	Sec.~\ref{sec:mixed_states} and also Secs.~\ref{sec:makhlin_invariants} and \ref{sec:pt_estimation}.
	
	\item[(c)] \textit{Marginal moments}:
	By discarding the measurements of some parties, one can obtain the marginal moments of the reduced states of $\varrho$.
	For illustration, let us consider a three-particle state $\varrho_{ABC}$ and discard the measurements of the parties $B$ and $C$,
	that is, $M_B=M_C=\mathbbm{1}$.
	This yields the corresponding one-body marginal moments $\mathcal{R}^{(t)}_A(\varrho_A)$ of the party $A$,
	while on the other hand, the case of $M_C=\mathbbm{1}$ yields the two-body marginal moments $\mathcal{R}^{(t)}_{AB}(\varrho_{AB})$ of the parties $A$ and $B$.
	Here, $\varrho_A, \varrho_{AB}$ are the one and two-body reduced states of $\varrho_{ABC}$, respectively.
	In general all $k$-body moments for $k \in [1, n]$ can be accessed by measuring the full $n$-body moments and discarding the corresponding measurements of $(n-k)$ parties.
	In particular, the averaging over all second-order $k$-body moments with product observables yields the $k$-body sector length $S_k$, discussed in Sec.~\ref{sec:sector_lengths}.
	
	\item[(d)] \textit{Challenging issues}: 
	When higher-order moments are considered, additional information may be extracted, allowing more powerful entanglement detection schemes.
	On a more technical level, however, this requires at least two additional steps.
	The first step is to evaluate the Haar integrals in the moments and obtain analytically tractable expressions such as simple symmetric polynomials of the correlation tensor $T_{j_1\ldots j_n}$ in Eq.~(\ref{eq-nqudits}). Since the moments depend on the choice of observable $\mathcal{M}$ in general, finding suitable families of observables may not be straightforward. For instance, in the case with $t=2$, the moments are independent of the choice of measurement observables as long as the observables are traceless~\cite{tran_correlations_2016,imai_bound_2021}, which, in general, is not the case~\cite{imai_bound_2021,wyderka_probing_2023}.
	The next step is to find entanglement criteria using the evaluated higher-order moments.
	Intuitively, one can power up entanglement detection by combining, e.g.~$\mathcal{R}^{(2)}$ and $\mathcal{R}^{(4)}$, rather than using solely $\mathcal{R}^{(2)}$.
	For this purpose, one should systematically search for the most effective combination of such nonlinear functions.
	Addressing the above questions is nontrivial and is considered in more detail in Secs.~\ref{sec:mixed_states} and \ref{sec:bipartite_higher_dimensions}.
\end{itemize}

\subsection{Quantum $t$-designs}
\label{sec:quantum_tdesigns}

In general, quantities which are at least approximately accessible by randomised measurements correspond to integrals over the space of unitary rotations.
This has two potential drawbacks.
For once, they require a large amount of sampled measurement directions to be approximated well, see e.g.~\cite{ohliger_efficient_2012} and secondly, the integral form is cumbersome for analytical derivations and proofs. 
Quantum designs represent a powerful tool to address both issues by replacing the integration over the full space with the average over several particular points only.
In the following, we will give an overview of the concept of designs both from a mathematical and a physical perspective and show how they can be applied in the context of randomised measurements.

\subsubsection{Spherical $t$-designs}
\label{sec:spherical_designs}

Historically, quantum $t$-designs were discussed by analogy with classical $t$-designs in combinatorial mathematics. Their basic 
idea is the following. Let us consider a real quadratic function $f_2(x)$ for a variable $x$ and take an integral in the interval from $a$ to $b$.
According to the rule found by Thomas Simpson in the $18$th century, it holds that the integral for the quadratic function can be exactly evaluated as a simple expression using only three points, namely
\begin{align}
	\int_a^b \mathrm{d}x\, f_2(x)
	=
	\frac{b-a}{6}
	\left[
	f_2(a)+4f_2\left(\frac{a+b}{2}\right)+f_2(b)
	\right].
\end{align}
An extension of Simpson's rule to a greater number of points is possible and can be found under the name of Gauss-Christoffel quadrature rule.

A spherical $t$-design can be seen as a generalisation of Simpson's rule for the efficient computation of integrals of certain polynomials over spheres~\cite{delsarte_spherical_1991,bannai_survey_2009}.
In fact, a spherical $t$-design has already been used in Sec.~\ref{sec:designs} to simply evaluate the moments $\mathcal{R}^{(2)}$.

Let $\mathcal{S}_{n-1}$ be the $n$-dimensional real unit sphere and let $X=\{\textbf{x}: \textbf{x} \in \mathcal{S}_{n-1}\}$ be a finite set of points on it with the number of elements $K = |X|$.
We call this set a \textit{spherical $t$-design} if
\begin{align}
	\frac{1}{K}\sum_{\textbf{x}\in X}f_t(\textbf{x})
	= \int \mathrm{d}\textbf{x} \, f_t(\textbf{x}),
	\label{spherical_tdesign}
\end{align}
for any homogeneous polynomial function $f_t(\textbf{x})$ in 
$n$ variables with degree $t$, where $\mathrm{d}\textbf{x}$ is the spherical measure in $n$ dimensions. The spherical design property ensures that integrals over the entire sphere can be efficiently computed by taking the average over the set of only $K$ different points.

Clearly, any spherical $t$-design is also a spherical $(t-1)$-design and it can be shown that spherical $t$-designs 
exist for any positive integer $t$ and $n$~\cite{seymour_averaging_1984}, although they may be difficult to construct explicitly~\cite{hardin_mclaren_1996}.
Furthermore, as expected, if a design for a higher degree $t$ is considered, then a larger number of points $K$ is needed.

\subsubsection{Complex projective $t$-designs}
\label{sec:complexprojective_designs}

Complex projective $t$-designs (or quantum spherical $t$-designs) are a generalisation of spherical designs to a complex vector space~\cite{renes_symmetric_2004, ambainis_quantum_2007}.
As such they allow, for example, to evaluate expressions based 
on a random sampling of quantum states.
A finite set of unit vectors $D = \{\ket{\psi_i}: \ket{\psi_i} \in C\mathcal{S}_{d-1} \}_{i=1}^K$ defined on a $d$-dimensional sphere $C\mathcal{S}_{d-1}$ in the complex vector space, forms a \textit{complex projective $t$-design} if
\begin{align}
	\frac{1}{K}\sum_{\ket{\psi_i} \in D} P_t({\psi_i})
	= \int \mathrm{d}\mu({\psi}) \, P_t({\psi}),
	\label{complex_projective_tdesign1}
\end{align}
for any homogeneous polynomial function $P_t$ in $2d$ variables with degree $t$ (that is, $d$ variables with degree $t$ and their complex conjugates with degree $t$),
where $\mathrm{d}\mu({\psi})$ is the spherical measure on the complex unit sphere $C\mathcal{S}_{d-1}$.
Here it is important to note that $C\mathcal{S}_{d-1}$ is isomorphic to the $d$-dimensional projective Hilbert space denoted as $P(\mathcal{H}_{d})$, where complex unit vectors $\ket{x}, \ket{y} \in P(\mathcal{H}_d)$ are identified iff $\ket{x} = e^{i\phi} \ket{y}$ with a real $\phi$~\cite{bengtsson_geometry_2017}.
For example, the Bloch sphere is known as $P(\mathcal{H}_{2})$, as a point on its surface corresponds, up to a global phase, to a pure single-qubit state.

Since polynomials of degree $t$ can be written as linear functions on $t$ copies of a state,
the definition of complex projective $t$-designs is equivalent to requiring
\begin{align}
	\frac{1}{K}\sum_{\ket{\psi_i} \in D}
	\left(\ket{\psi_i}\!\bra{\psi_i}\right)^{\otimes t}
	= \int \mathrm{d}\mu({\psi}) \, \left(\ket{\psi}\!\bra{\psi}\right)^{\otimes t}.
	\label{complex_projective_tdesign2}
\end{align}
This form is called the \textit{quantum state $t$-design} and involves an ensemble of states that is indistinguishable from a uniform random ensemble over all states, if one considers $t$-fold copies of quantum states.
Since the integral on the right-hand side of  Eq.~(\ref{complex_projective_tdesign2}) is proportional to
the projector onto the symmetric subspace~\cite{barenco_stabilization_1997,harrow_church_2013,brandao_mathematics_2016} (or see Lemma 2.2.2. in Ref.~\cite{low_pseudo_2010}), one can simplify this to
\begin{align}
	\int \mathrm{d}\mu({\psi}) \, \left(\ket{\psi}\!\bra{\psi}\right)^{\otimes t}
	=
	\frac{P_{\text{sym}}^{(t)}}{d_{\text{sym}}^{(t)}},
	\label{complex_projective_tdesign3}
\end{align}
where $P_{\text{sym}}^{(t)}$ is the projector onto the permutation-symmetric subspace and $d_{\text{sym}}^{(t)} = \binom{d+t-1}{t}$ is its dimension. In particular, for multi-qubit systems 
($d=2$), the symmetric subspace is spanned by the Dicke 
states $\{|{D_{t,m}}\rangle \}_{m=0}^t$ given by
\begin{align}
	|{D_{t,m}}\rangle
	= \frac{1}{\sqrt{\binom{t}{m}}}\sum_k
	{\pi}_k \left(\ket{1}^{\otimes m}\otimes \ket{0}^{\otimes (t-m)}\right),
\end{align}
where the summation in $\sum_k {\pi}_k$ is over all  
permutations between the qubits that lead to different terms.
A concrete example is the state 
$|{D_{3,1}}\rangle = (\ket{001} + \ket{010} + \ket{100})/\sqrt{3}$.  
Using the Dicke states, this projector can be rewritten as
\begin{align}
	P_{\text{sym}}^{(t)}
	= \sum_{m=0}^t |{D_{t,m}}\rangle \! \langle {D_{t,m}}|.
\end{align}

In order to explain the structure of $P_{\text{sym}}^{(t)}$ more generally, let us denote by $\text{Sym}(t)$ the symmetric group of a degree $t$ on 
the set $\{1,2,\ldots, t\}$ and  $W_\pi$ as a permutation operator on $\mathcal{H}_d^{\otimes t}$ representing a permutation $\pi = \pi(1) \ldots \pi(t) \in \text{Sym}(t)$ such that $W_\pi \ket{i_1, \ldots, i_t}
= \ket{i_{\pi(1)}, \ldots, i_{\pi(t)}}$.
Then one can write
$P_{\text{sym}}^{(t)}
= ({1}/{t!}) \sum_{\pi \in \text{Sym}(t)} W_\pi$.
Examples for $t=1$ and $t=2$ are
\begin{align}
	\int \mathrm{d}\mu({\psi}) \, \ket{\psi}\!\bra{\psi}
	&= \frac{\mathbbm{1}_d}{d},\\
	\int \mathrm{d}\mu({\psi}) \, \left(\ket{\psi}\!\bra{\psi}\right)^{\otimes 2}
	&= \frac{1}{d(d+1)} (\mathbbm{1}_d^{\otimes 2} + S),
\end{align}
where $S = \sum_{i,j}\ket{i}\!\bra{j} \otimes \ket{j}\!\bra{i}$ denotes the SWAP (or flip) operator with $S\ket{a}\otimes \ket{b} = \ket{b}\otimes \ket{a}$.
Eq.~(\ref{complex_projective_tdesign3}) implies uncertainty relations for any single-qudit state $\varrho$~\cite{ketterer_entropic_2020}.
Furthermore, another equivalent definition of complex projective $t$-designs is given by the condition
\begin{align}
	\frac{1}{K^2}\sum_{
		|\psi_i \rangle, |\psi_j \rangle \in D}
	\! \! \! \!
	|\langle \psi_i|\psi_j \rangle|^{2t}
	=\frac{1}{d_{\text{sym}}^{(t)}}.
	\label{complex_projective_tdesign4}
\end{align}
The left-hand side is called $t$-th frame potential.
According to the so-called Welch bound~\cite{welch_lower_1974,klappenecker_mutually_2005},
it is always greater than or equal to the right-hand side,
where the equality is saturated if and only if the set $D$ forms the complex projective $t$-designs.

Let us consider some examples of projective designs.
First, a trivial example of a complex projective $1$-design is a set of orthonormal basis vectors $\{\ket{i}\}_{i=1}^d$, which leads to
$(1/d)\sum_{i=1}^d\ket{i}\!\bra{i}= {\mathbbm{1}_d}/{d}$.

Second, a typical example of complex projective $2$-designs are so-called mutually unbiased bases (MUBs).
A collection $\{M_k\}$ of orthonormal bases $M_k = \{|{i_k}\rangle\}_{i=1}^d$ for a $d$-dimensional Hilbert space is called \textit{mutually unbiased} if
$|\langle i_k|j_l \rangle|^2 = {1}/{d}$,
for any $i,j$ with $k \neq l$, i.e. the overlap of any pair of vectors from different bases is equal~\cite{durt_mutually_2010}.
For the case of $d=2$, a set of MUBs is given by
$\{M_1, M_2, M_3\}$ with
$M_1 = \{\ket{0},\ket{1}\},
M_2 = \{\ket{+},\ket{-}\}$,
and
$M_3 = \{\ket{+i},\ket{-i}\}$.
Here, the bases $\{ \ket{0}, \ket{1} \}$, 
$\{ \ket{\pm}=(\ket{0} \pm \ket{1})/\sqrt{2} \}$, and
$\{ \ket{\pm i}=(\ket{0} \pm i\ket{1})/\sqrt{2} \}$
are the normalised eigenvectors of
$\sigma_z$, $\sigma_x$, and $\sigma_x \sigma_z$.
In general, the size of a maximal set of MUBs for a given dimension $d$ is an open problem and only partial answers are known. Moreover, this has been recognised as one of the five most important open problems in quantum information theory~\cite{horodecki_five_2020}.
It is known that for an arbitrary dimension $d$ the maximum number of MUBs cannot be more than $d+1$~\cite{weiner_gap_2013}. For prime-power dimensions $d=p^r$, sets of $d+1$ MUBs have been
constructed~\cite{ivonovic_geometrical_1981,noauthor_optimal_1989}.
For the dimensions $d=p^2$ and $d=2^r$ MUBs were experimentally implemented~\cite{wiesniak_entanglement_2011,seyfarth_construction_2011}.
The smallest dimension which is not a power of a prime and where the maximal number of MUBs is unknown is $d=6$~\cite{zauner_grundzuge_1999}.
Note that any collection of $(d+1)$ MUBs saturates the Welch bound and therefore forms a complex projective 2-design~\cite{klappenecker_mutually_2005}.

\subsubsection{Unitary $t$-designs}
\label{sec:unitary_tdesigns}

In the case of qubits, spherical designs are suited to evaluate integrals over random unitaries of measurement settings as those can be mapped to rotations on the Bloch sphere.
For higher dimensional systems, however, such a mapping no longer exists and the randomised scenario can be addressed by general unitary designs.
A set of unitaries $G = \{U_i: U_i \in \mathcal{U}(d)\}_{i=1}^K$ forms a \textit{unitary $t$-design} if
\begin{align}
	\frac{1}{K}\sum_{U_i \in G} P_t(U_i)
	=\int \mathrm{d}\mu(U)\, P_t(U),
	\label{unitary_tdesign1}
\end{align}
for any homogeneous polynomial function $P_t$ in $2d^2$ variables with degree $t$ (that is, on the elements of unitary matrices in $\mathcal{U}(d)$ with degree $t$ and on their complex conjugates with degree $t$), where $\mathrm{d}\mu(U)$ is the Haar unitary measure on $\mathcal{U}(d)$.
For details about unitary $t$-design, see Refs.~\cite{gross_evenly_2007,dankert_exact_2009,scott_optimizing_2008,low_pseudo_2010}.
Similarly to complex projective designs, there are several equivalent definitions of unitary $t$-designs.
One is given by
\begin{align}
	\frac{1}{K}\sum_{U_i \in G} U_i^{\otimes t} X (U_i^\dagger)^{\otimes t}
	&=\int \mathrm{d}\mu(U)\, U^{\otimes t} X (U^\dagger)^{\otimes t},
	\label{unitary_tdesign2}
\end{align}
for any operator $X \in \mathcal{H}_d^{\otimes t}$.
An important observation here is that if we set $\{\ket{\psi_i}\} = \{U_i \ket{0}\}$, then Eq.~(\ref{unitary_tdesign2}) leads to Eq.~(\ref{complex_projective_tdesign2}), i.e.~any  unitary $t$-design gives rise to a quantum state $t$-design.
The converse is not necessarily true, even if a set of unitaries creates a state design via $\{\ket{\psi_i}\} = \{U_i \ket{0}\}$, it does not constitute a unitary design. 
This simply follows from the fact that a relation like 
$\{\ket{\psi_i}\} = \{U_i \ket{0}\}$ does not determine
the $U_i$ in a unique way.

In order to find the analog of Eq.~(\ref{complex_projective_tdesign3}), note that the right-hand side in Eq.~(\ref{unitary_tdesign2}) commutes with all unitaries $V^{\otimes t}$ for $V \in \mathcal{U}(d)$, due to the left and right invariance 
of the Haar measure. According to the Schur-Weyl duality, if an operator $A \in \mathcal{H}_d^{\otimes t}$ obeys $\left[A, V^{\otimes t}\right] = 0$ for any $V \in \mathcal{U}(d)$, then $A$ can be written as a linear combination of subsystem permutation operators $W_\pi$ (while the converse statement is also true) \cite{roberts_chaos_2017}.
Thus, one has
\begin{align}
	\int \mathrm{d}\mu(U)\, U^{\otimes t} X (U^\dagger)^{\otimes t}
	= \sum_{\pi \in \text{Sym}(t)} x_\pi W_\pi,
	\label{eq:twirlingop}
\end{align}
where each of $x_\pi$ can be found with the help of the so-called Weingarten calculus~\cite{kostenberger_weingarten_2021, collins_weingarten_2022}.
As an example, we have
\begin{align}
	\int \mathrm{d}\mu(U)\, UXU^\dagger
	&= \frac{\tr(X)}{d}\mathbbm{1}_d,\\
	\int \mathrm{d}\mu(U)\, U^{\otimes 2} X (U^\dagger)^{\otimes 2}
	&= \frac{1}{d^2-1}
	\left\{
	\left[\tr(X)-\frac{\tr(XS)}{d}\right]\mathbbm{1}_d^{\otimes 2}
	- \left[\frac{\tr(X)}{d}-\tr(XS)\right]S
	\right\},
	\label{eq:twodesignformula}
\end{align}
where $S$ is the SWAP operator. We remark that the left-hand side in Eq.~(\ref{eq:twirlingop}) is called a twirling operation and it is a CPTP map. A quantum state obtained from the twirling operation is called a Werner state and is invariant under any $U \otimes U \otimes  \cdots \otimes U$ \cite{vollbrecht_entanglement_2001,eggeling_separability_2001}. For two particles, states of the form (\ref{eq:twodesignformula}) were the first states where it was shown that entanglement does not imply Bell nonlocality~\cite{werner_quantum_1989}.
For calculations with operators of the form (\ref{eq:twodesignformula}) and $X = X_1 \otimes X_2$ it is useful to note the so-called SWAP trick:
$\tr[(X_1 \otimes X_2)S] = \tr(X_1 X_2)$.
Moreover, the SWAP trick can be generalised using cyclic permutation operators,
e.g.~for a cyclic permutation operator $W_{\text{cyc}}$ with
$W_{\text{cyc}} \ket{x_1, x_2, \cdots, x_n} = \ket{x_2, \cdots, x_n, x_1}$, it holds that
$\tr[(X_1 \otimes X_2 \otimes \cdots \otimes X_n)W_{\text{cyc}}]
= \tr(X_1 X_2 \cdots X_n)$, see Refs.~\cite{horodecki_method_2002,ekert_direct_2002,harrow_random_2009,huber_positive_2021} for details and Refs.~\cite{huber_refuting_2023, rico_entanglement_2023} for the applications.
Cases with $t=3, 4$ are explicitly described in Example 3.27, and Example  3.28 in Ref.~\cite{zhang_matrix_2015}.
For more details, see~\cite{garcia_quantum_2021,brandao_models_2021,wyderka_complete_2022}. 

Moreover, yet another equivalent definition of unitary $t$-designs is given in Ref.~\cite{gross_evenly_2007}
\begin{align}
	\frac{1}{K^2}\sum_{U_i, U_j \in G}
	\left|\tr(U_i U_j^\dagger)\right|^{2t}
	=\begin{cases}
		t! \,
		&\text{for} \, d\geq t, \\
		\frac{(2t)!}{t! (t+1)!} \,
		&\text{for}\, d=2,
		\label{unitary_tdesign3}
	\end{cases}
\end{align}
where the left-hand side is called $t$-th frame potential and the right-hand side gives its minimal value similar to the Welch bound in complex spherical designs. The more general cases of these lower bounds were discussed in Ref.~\cite{rains_increasing_1998}. The frame potential is often employed as a useful measure to quantify the randomness of an ensemble of unitaries in terms of out-of-time-order correlation functions in quantum chaos~\cite{roberts_chaos_2017, hunter_chaos_2018}.

For the scenario of $n$-qubit systems, an example of a unitary $1$-design is the Pauli group $\mathcal{P}_n$, the group of all $n$-fold tensor products of single-qubit Pauli matrices $\{\mathbbm{1}_2, \sigma_x, \sigma_y, \sigma_z\}$.
This group does not form a unitary $2$-design~\cite{roy_unitary_2009},
however, note that we used Pauli measurements in Sec.~\ref{sec:designs} as a form of a spherical design.
In contrast, the Clifford group $\mathcal{C}_n$, a group of unitaries with the property $C\in \mathcal{C}_n$ if $CPC^\dagger \in \mathcal{P}_n$ for any $P \in \mathcal{P}_n$, is known to be a unitary $2$-design in this scenario.
Furthermore, it has been shown that the Clifford group also forms a unitary $3$-design, but not a unitary $4$-design~\cite{webb_clifford_2016,zhu_clifford_2016}.

Yet another approach to creating unitary designs was discussed in Ref.~\cite{toth_efficient_2007}. This algorithm approximates multipartite twirling operations by repeatedly applying random unitaries from a given set. In this way, $2^M$ unitaries are constructed from $M$ unitaries and are shown to efficiently realise a design, even if the original unitaries are not uniformly distributed according to the Haar measure or they are chosen randomly from a discrete set of unitaries. Finally approximate $1$- and $2$-designs can also be efficiently created with random circuits composed of two-qubit gates \cite{harrow_random_2009,brando_local_2016}.

\subsubsection{Applications to randomised measurements}
Finally, we show the usefulness of unitary designs in the scheme of randomised measurements.
For the sake of simplicity, we focus on a three-qudit state $\varrho_{ABC}$ and consider how to obtain its full-body sector length $S_3$ from the unitary 2-design.
Note that one can straightforwardly generalise this approach to the sector lengths $S_k$ of a $n$-qudit state for any $1\leq k \leq n$.

Let us consider the product observable
$\mathcal{M} = \lambda_a \otimes \lambda_b \otimes \lambda_c$
in the second-order moment, Eq.~(\ref{eq:generalmoment}),
for any choice of generalised Gell-Mann matrices with $a,b,c=1,\ldots,d^2-1$.
Substituting the generalised Bloch decomposition of $\varrho_{ABC}$ in Eq.~(\ref{eq-nqudits}) into the second-order moment, one finds
\begin{equation}
	\mathcal{R}^{(2)}_{\mathcal{M}}(\varrho_{ABC})
	= \frac{N_{3,d,2}}{d^6}
	\sum_{j_{A},j_{B},j_{C}=1}^{d^2-1}
	\sum_{k_{A},k_{B},k_{C}=1}^{d^2-1}
	T_{j_{A} j_{B} j_{C}}
	T_{k_{A} k_{B} k_{C}}
	\tr\left[(\Tilde{A}_{jk} \otimes \Tilde{B}_{jk}
	\otimes \Tilde{C}_{jk})
	(\lambda_{a}^{\otimes 2} \otimes
	\lambda_{b}^{\otimes 2}\otimes
	\lambda_{c}^{\otimes 2}
	)\right],
	\label{eq:threequditsectorevalu}
\end{equation}
where we used that
$[\tr(M)]^k = \tr(M^{\otimes k})$ for any matrix $M$ and integer $k$
and we denoted the twirling result as
$\Tilde{X}_{jk} = \int \mathrm{d}\mu(U_X)\, U_X^{\otimes 2}
(\lambda_{j_{X}} \otimes \lambda_{k_{X}})(U_X^\dagger)^{\otimes 2}$
for $X=A,B,C$.
Now, $\Tilde{X}_{jk}$ can be simply evaluated using the formula in Eq.~(\ref{eq:twodesignformula}) and reads:
$\Tilde{X}_{jk} = \delta_{j_X k_X}(dS-\mathbbm{1}_d^{\otimes 2})/(d^2-1)$,
where we employed the SWAP trick and the properties of the generalised Gell-Mann matrices
$\tr(\lambda_j) = 0$ and $\tr(\lambda_j \lambda_k)=d\delta_{jk}$.
As the last step, by inserting this form into the second moment in Eq.~(\ref{eq:threequditsectorevalu}) and choosing the normalisation constant as $N_{3,d,2} = (d^2-1)^3$, one obtains $\mathcal{R}^{(2)}_{\mathcal{M}} = S_3$, i.e. the tripartite sector length for qudits.

An important lesson from this result is that randomised measurements of the second moment are an indirect implication related to the SWAP operator.
For higher-order cases, the permutation operators $W_\pi$ will emerge according to the Schur-Weyl duality in Eq.~(\ref{eq:twirlingop}).
This will play an important role in estimating the purity of a state, the overlap between two states, and PT moments, for details see Secs.~\ref{sec:general_determination_uni_inv} and \ref{sec:pt_estimation}.

\subsection{Criteria for $n$-qubit entanglement}
\label{sec:n_qubit_entanglement}

In Sec.~\ref{sec:moments}, we discussed the entanglement detection for a two-qubit state based on the second moment from randomised measurements.
This can be generalised to the case of $n$ parties, where the correlation function of $\varrho$ is a straightforward generalisation of Eq.~(\ref{eq:2-qubit_correlation_function}), namely
\begin{equation}
	E(\mathbf{u}_1, \dots, \mathbf{u}_n) = \langle r_1 \dots r_n \rangle
	= \tr(\varrho \, \mathbf{u}_1 \cdot \sigma \otimes \dots \otimes \mathbf{u}_n \cdot \sigma).
\end{equation}
This is a special case of Eq.~(\ref{eq:correlation_function_general}) with $d=2$ and the product observable 
$\mathcal{M}_{\text{P}} = \sigma_z \otimes \cdots \otimes \sigma_z$,
where we denote the randomised Pauli matrix as $\mathbf{u} \cdot \sigma = U\sigma_z U^\dagger$.
Choosing the normalisation constant $N_{n,2,2}$ in Eq.~(\ref{eq:generalmoment}) as $3^n$ we can write the second moment as
\begin{equation}
	\mathcal{R}^{(2)} = \left( \frac{3}{4 \pi} \right)^n \int \mathrm{d} \mathbf{u}_1 \dots \int \mathrm{d} \mathbf{u}_n \, [E(\mathbf{u}_1, \dots, \mathbf{u}_n)]^2.
	\label{eq:secondmoment}
\end{equation}
Similarly to Sec.~\ref{sec:designs}, this integral can be simply evaluated using spherical $2$-designs
\begin{equation}
	\mathcal{R}^{(2)}= \sum_{j_1, \dots ,j_n=1,2,3} \tr[\varrho \  (\sigma_{j_1}\otimes \cdots \otimes \sigma_{j_n})]^2.
\end{equation}
Note that this quantity coincides with the full-body sector length $S_n$ introduced in Sec.~\ref{sec:sector_lengths}.
With this expression, one can analytically find an entanglement criterion.
Since the second moment $\mathcal{R}^{(2)}$ (that is, the sector length~$S_n$) is convex in a state and invariant under LU transformations, the maximal value over $n$-qubit fully separable states is, without the loss of generality, achieved by a pure product state $\ket{0}^{\otimes n}$.
This immediately yields the entanglement criterion~\cite{tran_quantum_2015}
\begin{align}
	\mathcal{R}^{(2)}
	> 1 \Rightarrow \varrho \textrm{ is entangled}.
	\label{EQ_D2_ENT}
\end{align}
Similar criteria have been presented Refs.~\cite{hassan_separability_2008,
	hassan_experimentally_2008,hassan_geometric_2009,tran_correlations_2016}.
In Sec.~\ref{sec:n-quditfullseparability}, this inequality will be extended to detect high-dimensional entanglement based on second moments.

The original result presented in (\ref{EQ_D2_ENT})  was derived without the notion of spherical $t$-designs~\cite{tran_quantum_2015}.
Moreover, this condition was shown to be necessary and sufficient for entanglement in pure states.
The sufficient criterion for mixed states in terms of $\mathcal{R}^{(2)}$ can still be formulated for any $d$, where the moment is invariant under not only local unitaries but also the choice of local operator basis, e.g.~Gell-Mann or Heisenberg-Weyl matrices. 
Given that $\mathcal{R}^{(2)}$ faithfully captures whether a state is entangled or not, it is natural to ask if it is an entanglement monotone~\cite{horodecki_quantum_2009}.
This question is relevant even for pure states where one asks whether state $\ket{\psi}$, endowed with $\mathcal{R}^{(2)}(\psi) \ge \mathcal{R}^{(2)}(\phi)$, can be converted via LOCC to another state $\ket{\phi}$.
It turns out that such conversions are possible for bipartite systems in any dimensions, where the proof utilises Nielsen's majorisation criterion~\cite{nielsen_conditions_1999}, but there exist multipartite quantum states which can be converted via LOCC to states with larger $\mathcal{R}^{(2)}$~\cite{tran_correlations_2016}.
Therefore, in general, the second moment $\mathcal{R}^{(2)}$ is not an entanglement monotone.
As a counterexample, consider the state $\ket{\psi} = (\ket{0} \ket{\psi^-} + \ket{1} \ket{\psi^+})/ \sqrt{2}$, where $\ket{\psi^{\pm}} = (\ket{01} \pm \ket{10}) / \sqrt{2}$ are the Bell states, that admits non-zero correlation functions $T_{zxx} = T_{zyy} = -1$, leading to $S_3(\psi) = 2$. 
Now measure the first qubit in the computational basis, do nothing if the outcome is ``0'' and apply $\sigma_z$ to the second qubit if the result was ``1''.
In both cases, one deterministically ends up in the pure state $\ket{\phi} = \ket{0} \ket{\psi^-}$ with the increased $S_3(\phi) = 3$.

\subsection{Specialised criteria for two-qubit entanglement}
\label{sec:mixed_states}

In the case of mixed states of two qubits, $\mathcal{R}^{(2)}$ no longer provides a necessary and sufficient criterion for entanglement and higher-order moments may be used for improved criteria. Here, we discuss two approaches specific to this scenario, where the first one is motivated by considering states in the Bell-diagonal form and the second one represents a refined method to access the PPT criterion using more complex LU-invariant quantities, partially based on non-product observables.
Additionally, also moments of the state itself are presented as a useful resource in this scenario.

\subsubsection{Bell-diagonal states}
\label{sec:Belldiagonalstate}

As the name suggests, Bell-diagonal states of two qubits, $\varrho_{BD}$, can be represented as a mixture of the four 
Bell states.
In terms of Pauli matrices, they have the form~\cite{horodecki_information-theoretic_1996}
\begin{align}
	\varrho_{\text{BD}} = \frac{1}{4} \left({\mathbbm{1}_2}^{\otimes 2} + \sum_{j=x,y,z} T_{jj} \sigma_j \otimes \sigma_j \right),
\end{align}
and any state with a diagonal correlation matrix and maximally mixed marginals is Bell-diagonal.
Since a Bell-diagonal state is parameterised only by the three values $(T_{xx}, T_{yy}, T_{zz})$, it allows for a much simpler analytical treatment than general states.
For instance, the PPT criterion as the necessary and sufficient condition for two-qubit separability can be rewritten as $\sum_{j=x,y,z} |T_{jj}| \leq 1$ ~\cite{horodecki_information-theoretic_1996}.

Additionally and crucially, any two-qubit state $\varrho_{AB}$ can be mapped to a Bell-diagonal state by local operations which conserve the values of the moments $\mathcal{R}^{(t)}$ and do not increase the amount of entanglement present in the state~\cite{wyderka_learning_2020}.
Thus, any criterion derived for a Bell-diagonal state that is based solely on these moments is also valid for arbitrary states.
The mapping from a general state to a Bell-diagonal state proceeds as follows.
The four Bell states are eigenstates of the two observables
$g_1 = \sigma_x \otimes \sigma_x$ and  $g_2 = \sigma_z \otimes \sigma_z$
with all the four possible combinations of eigenvalues $\pm 1$.
The map $\varrho \mapsto (\varrho + g_i \varrho g_i)/2$ amounts to applying
the local unitary transformation $\varrho \mapsto g_i \varrho g_i$ with
probability $1/2$.
If a general state is expressed in the basis of Bell states as
$\varrho = \sum_{ij} \alpha_{ij} |BS_i \rangle  \langle BS_j|$, then
applying the above map for $g_1$ and $g_2$ removes all the off-diagonal terms,
since for at least one $g_i$ the Bell states $\ket{BS_i}$ and $|BS_j \rangle$
have a different eigenvalue, so $\varrho$ is mapped to the Bell-diagonal state
$\varrho_{BD} = \sum_{i} \alpha_{ii} |BS_i \rangle  \langle BS_i|$.
This kind of depolarization is not specific to Bell states, it can be applied
to several other multipartite scenarios, like GHZ-symmetric or graph-diagonal states~\cite{dur_multiparticle_2001,guhne_multiparticle_2011,eltschka_entanglement_2012}.

For a Bell-diagonal state $\varrho_{\text{BD}}$, the second and fourth moments of
the product observable $\mathcal{M}_\text{P}=\sigma_z \otimes \sigma_z$ are given
by \cite{ketterer_characterizing_2019}
\begin{equation}
	\mathcal{R}^{(2)}
	=\frac{1}{9}
	\sum_{j=x,y,z} T_{jj}^2,
	\qquad
	\mathcal{R}^{(4)}
	=\frac{2}{75}
	\sum_{j=x,y,z} T_{jj}^4 +
	\frac{27}{25}[\mathcal{R}^{(2)}]^2,
\end{equation}
respectively.
Now, based on the separability constraint $\sum_{j=x,y,z} |T_{jj}| \leq 1$,
for a given value of $\mathcal{R}^{(2)}$ one can 
maximise and minimise
analytically the value of $\mathcal{R}^{(4)}$ for separable states. 
That is, the task is formulated as the following optimisation problem for a simple polynomial with three variables
\begin{align}
	\max_{T_{jj}}/\min_{T_{jj}} \ \
	&\mathcal{R}^{(4)}
	\quad
	\text{s.t.} \quad
	\mathcal{R}^{(2)} = \frac{1}{9}
	\sum_{j=x,y,z} T_{jj}^2,
	\quad
	\sum_{j=x,y,z} |T_{jj}| \leq 1,
	\quad
	0 \leq |T_{jj}| \leq 1.
\end{align}
In this way we obtain a separability region in the parameter space spanned by
$\mathcal{R}^{(2)}$ and $\mathcal{R}^{(4)}$. This approach allows to
detect entanglement that cannot be detected by the second moment itself,
which is illustrated in Fig.~\ref{imai21}. Moreover, using additionally
the sixth moment, a necessary and sufficient condition for entanglement
of two-qubit Bell diagonal states can be found, see the Appendix in
Ref.~\cite{ketterer_characterizing_2019}.

\subsubsection{Evaluating the PPT criterion for two qubits}
\label{sec:PPT_criterion}

For general two-qubit states, one can consider the randomised measurement scheme with non-product observables $\mathcal{M}_\text{NP}$. This scheme then allows for the complete characterisation of two-qubit entangled states. First, to detect two-qubit entanglement completely, we need to access 
the PPT criterion in the randomised measurement scheme.
The PPT condition $\varrho_{AB}^{\Gamma_B} \geq 0$ for a 
two-qubit state $\varrho_{AB}$, discussed in Sec.~\ref{subsec:pt_moments}, is equivalent to $\mathrm{det}(\varrho_{AB}^{\Gamma_B}) \geq 0$, since only one eigenvalue 
of $\varrho_{AB}^{\Gamma_B}$ becomes negative if the state is entangled. In Ref.~\cite{augusiak_universal_2008}, it has been found that $\mathrm{det}(\varrho_{AB}^{\Gamma_B})$ can be expressed as
\begin{align}
	\det(\rho^{\Gamma_B})
	= \frac{1}{24}
	(1-6p_4+8p_3+3p_2^2-6p_2),
\end{align}
where
$
p_2 = (1+x_1)/4,
p_3 = (1+3x_1+6x_2)/16,
p_4 = (1+6x_1+24x_2+x_1^2+2x_3+4x_4)/64,
$
with
$
x_1 = I_2+I_4+I_7,
x_2 = I_1+I_{12} ,
x_3 = I_2^2 -I_3,
x_4 = I_5+I_8+I_{14}+I_4 I_7.
$
Here $I_k$ are some of the Makhlin invariants~\cite{makhlin_nonlocal_2002}, which 
form a complete family of invariants under local 
unitaries.

In Ref.~\cite{wyderka_complete_2022}, it was shown 
that such Makhlin invariants can be accessed by the 
randomised measurement scheme. For instance,
in the case with the product observable 
$\mathcal{M}_{\text{P}} = \sigma_z \otimes \sigma_z$,
one has 
$\mathcal{R}_{\mathcal{M}_{\text{P}}}^{(2)} = I_2 = \sum_{j,k=x,y,z}T_{jk}^2$,
where $T = (T_{jk})$ is the correlation matrix given in Eq.~(\ref{bloch_two_qubit}) and $I_2$ corresponds to the sector length $S_2$.
It is important to note here that while the third 
moment for local observables on qubits vanishes, 
that is $\mathcal{R}_{\mathcal{M}_{\text{P}}}^{(3)} = 0$, 
this is not the case for the non-product observable $\mathcal{M}_{\text{NP}} = \sum_{j=1}^3 \sigma_j \otimes \sigma_j$. Indeed,  one can obtain
$\mathcal{R}_{\mathcal{M}_{\text{NP}}}^{(3)} = I_1 = \det{T}$.
In a similar way, other Makhlin invariants can be obtained via randomised measurements.
This result directly implies the possibility of detecting any two-qubit entanglement. Further details about the Makhlin invariants are discussed in Sec.~\ref{sec:makhlin_invariants}.

\subsubsection{State moments for entanglement detection}
\label{sec:moments_for_entanglement_detection}

The methods discussed so far were a direct implementation of the statistical moments $\mathcal{R}^{(t)}$, i.e.~of the moments of the distribution of correlation values. 
However, also the moments of the state itself can be used to derive certain separability bounds and detect entanglement. 
These quantities are LU invariant and hence a perfect candidate for the randomised measurement schemes. 

Lawson et al. have experimentally demonstrated the usefulness of the higher-order state moments for entanglement detection~\cite{lawson_reliable_2014} in that matter. 
For two-qubit density matrices, the Pauli decomposition of the $t$-th power of the density matrix, $\varrho^t$, can be used to define polynomials of correlation matrix elements, denoted as $Q_t$.
In particular, for $t=2$, the expression directly resembles what we call the sector length, i.e.~$Q_2 = S_2 =\mathcal{R}^{(2)}$.
For $Q_3$, the authors consider  $Q_3=\langle\sigma_x\sigma_z\rangle\langle\sigma_y\sigma_y\rangle\langle\sigma_z\sigma_x\rangle
-\langle\sigma_x\sigma_y\rangle\langle\sigma_y\sigma_z\rangle\langle\sigma_z\sigma_x\rangle
-\langle\sigma_x\sigma_z\rangle\langle\sigma_y\sigma_x\rangle\langle\sigma_z\sigma_y\rangle
+\langle\sigma_x\sigma_x\rangle\langle\sigma_y\sigma_z\rangle\langle\sigma_z\sigma_y\rangle
+\langle\sigma_x\sigma_y\rangle\langle\sigma_y\sigma_x\rangle\langle\sigma_z\sigma_z\rangle
-\langle\sigma_x\sigma_x\rangle\langle\sigma_y\sigma_y\rangle\langle\sigma_z\sigma_z\rangle$,
which depends on third-order terms of two-body correlations, only.
Notice that $Q_3$ is indeed equal to $-\det{T} = -I_1$, which is one of the Makhlin invariants discussed in the previous section.
Also, the higher-order expressions $Q_4$ and $Q_5$ were defined in Ref.~\cite{lawson_reliable_2014}, where one can simply write them as
$Q_4 = I_2^2 - I_3$
and
$Q_5 = - I_1 I_2$
for the Makhlin invariants $I_2, I_3$
that will be defined in Eq.~(\ref{eq:threeinvariants}) in Sec.~\ref{sec:makhlin_invariants}. 
The authors perform numerical simulations to first establish a lower bound on concurrence, which quantifies the entanglement of two-qubit states~\cite{hill_entanglement_1997}, based on the second and higher-order correlation matrix polynomials. 
\begin{figure}[htb!]
	\includegraphics[width=\textwidth]{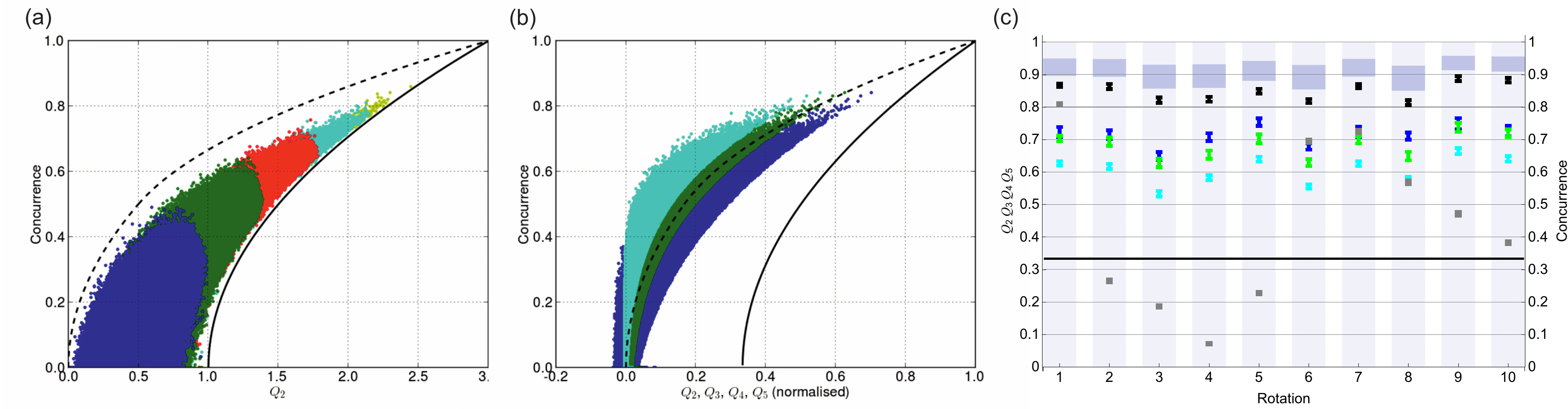}
	\caption{\label{lawson14}$(a)$ Numerical simulations of two-qubit states indicate that the concurrence is bounded from above (dashed) and below (solid) using the reference-frame independent quantity $Q_2$. 
		Colours visualise the purity of the state, which grows with growing values of $Q_2$. 
		$(b)$ Normalised higher-order quantities depending on two-body correlations ($Q_3$ in dark blue, $Q_4$ in green, $Q_5$ in light blue) may allow tighter bounds on the concurrence. $(c)$ Experimental measurements of $Q_2$, \dots, $Q_5$ (color coding as in panel b) of the maximally entangled state $\left|\phi^-\right \rangle$ with about $91\%$ fidelity (rotation $1$) and nine local unitary rotations of it. The thereby deduced blue-shaded measurements of the moments allow to bound the concurrence (gray-blue shaded intervals around $0.9$) which is well above the separability threshold of $1/3$.
		Figure taken from Lawson et al. \cite{lawson_reliable_2014}.}
\end{figure}
As shown in Fig.~\ref{lawson14}, for a given value of $Q_2$, the concurrence is bounded from both sides.
Furthermore, only states with higher purities can achieve higher values of $Q_2$ as visible from the colour encoding (increasing purities from dark blue with $\mathcal{P}\leq0.5$ over green, red and light blue to yellow with $\mathcal{P}\in[0.8,0.9]$).
In a similar fashion, the concurrence of simulated two-qubit states is shown against the normalised $Q_3$, $Q_4$ and $Q_5$ in Fig.~\ref{lawson14}. This strongly suggests that large respective values of the latter lead to tighter bounds on the concurrence.
For the experimental demonstration, Lawson \emph{et al.} use a commercial spontaneous down-conversion source and perform local-unitary rotations on one of the two qubits using waveplates.
As the state is assumed to be a maximally entangled $\left|\phi^-\right \rangle=\left(|00 \rangle - |11\rangle \right)/\sqrt{2} $ state, this is formally equivalent to a rotation of another qubit and demonstrates the direct experimental accessibility of the $Q_t$ polynomials, see Fig.~\ref{lawson14}.
Finally we remark that the upper and lower bounds on the concurrence of multiparticle quantum systems have been quantified in the randomised measurement framework~\cite{ohnemus_quantifying_2023}.

\subsection{Bipartite systems of higher dimensions}
\label{sec:bipartite_higher_dimensions}
Although it is not trivial to generalise the methods presented so far to higher dimensional systems, at least for two-qudit states, i.e. the case of $n=2$ with local dimension $d$ and product observables $\mathcal{M} = M_A \otimes M_B$, several entanglement criteria were found.
As in the qubit case, a basic approach aims to derive conditions based on the second moment of the distributions of correlations and a more refined approach takes higher orders into account, allowing even to certify weakly entangled states such as bound entanglement.

\subsubsection{Second moments}
\label{sec:second_moments}
In the qubit case with $d=2$, the moments can be easily evaluated by virtue of the concept of spherical designs discussed in Sec.~\ref{sec:quantum_tdesigns}.
This is based on the fact that unitary transformations on qubits can be regarded as orthogonal rotations on the Bloch sphere, due to the connection between $\text{SU}(2)$ and $\text{SO}(3)$ groups.
In the qudit (higher dimensional) case,
on the other hand, we lack a similar connection because the notion of a
Bloch sphere is not available.
Accordingly, not all possible observables are equivalent under random unitaries. However, as long as they are traceless the second moments do not depend on the choice of observables~\cite{tran_correlations_2016,imai_bound_2021}.
In fact, one can evaluate the second moments and turn them to the sector lengths.
In the following, we denote that
$\mathcal{R}_{A}^{(2)} = S_1^A$,
$\mathcal{R}_{B}^{(2)} = S_1^B$,
$\mathcal{R}_{A}^{(2)}
+ \mathcal{R}_{B}^{(2)} = S_1$, and
$\mathcal{R}_{AB}^{(2)} = S_2$,
where $S_1$ and $S_2$ are the sector lengths in Eq.~(\ref{eq:sectorlengths}).

In Ref.~\cite{tran_correlations_2016}, it has been shown that
any two-qudit separable state obeys
\begin{align}
	\mathcal{R}_{AB}^{(2)}
	\leq (d-1)^2,
	\label{eq:twoquditsecondmo}
\end{align}
with the normalisation constant $N_{2,d,2} = (d^2 -1)^2$.
This was further improved by including the marginal moments  $\mathcal{R}_A^{(2)}$ and $\mathcal{R}_B^{(2)}$~\cite{imai_bound_2021}.
The modified bound valid for all separable states is given as
\begin{align}
	\mathcal{R}_{AB}^{(2)}
	\leq 
	d-1 + (d-1) \mathcal{R}_A^{(2)}
	-\mathcal{R}_B^{(2)}.
	\label{eq:secondoptimal}
\end{align}
Again, any violation implies that the state is entangled.
This detection method was shown to be strictly stronger than the criterion in Eq.~(\ref{eq:twoquditsecondmo}).
The criterion in Eq.~(\ref{eq:secondoptimal}) is equivalent to the second-order Rényi entropy criterion \cite{elben_renyi_2018} stating that any separable state obeys
$H_2(\varrho_A) \leq H_2(\varrho_{AB})$,
$H_2(\varrho_B) \leq H_2(\varrho_{AB})$,
with $H_2(\varrho)$ defined in Sec.~\ref{subsection:LUinva}.
The $H_2(\varrho)$ has been estimated by local randomised measurements in Refs.~\cite{elben_renyi_2018, brydges_probing_2019},
where an ion-trap quantum simulator was used to perform measurements of the Rényi entropy.
We note that the criterion in Eq.~(\ref{eq:secondoptimal}) was extended to detect the Schmidt number~\cite{imai_work_2023} and that higher-order Rényi entropy criteria are also present in the literature~\cite{nielsen_conditions_1999,vollbrecht_conditional_2002,guhne_entropic_2004,bengtsson_geometry_2017}.
As a final remark, the violation of Eq.~(\ref{eq:secondoptimal}) does not detect a weak form of entanglement known as bound entanglement, which cannot be distilled into pure maximally entangled states~\cite{horodecki_mixedstate_1998} and cannot be verified by the PPT criterion~\cite{hiroshima_majorization_2003}.

\subsubsection{Fourth and higher-order moments}
\label{sec:higher_order_moments}
Several entanglement criteria, the ones based on second moments and, of course, all the ones based on PT moments from randomised measurements fail to detect bound entangled states.
In this section, we explain that higher-order moments are able to detect such weakly entangled states.

Let us begin by noting again that higher-order moments $\mathcal{R}^{(t)}$ for $t>2$ from randomised measurements in high dimensions ($d>2$) depend on the choice of the observable, unlike the case of qubits or second moments in high dimensions.
Ref.~\cite{imai_bound_2021} has offered a systematic method to address this problem.
The key result is that one can find observables $\mathcal{M} = M_A \otimes M_B$ such that the moments $\mathcal{R}^{(t)}$ coincide with alternative moments obtained as uniform averages over a high-dimensional sphere, the so-called pseudo-Bloch sphere, with
\begin{align}
	\mathcal{S}^{(t)} = N_{d,t}
	\int \mathrm{d}\mathbf{u}_a
	\int \mathrm{d}\mathbf{u}_b
	\left\{\tr[\varrho_{AB}
	(\mathbf{u}_a\cdot \mathbf{\lambda}) \otimes
	(\mathbf{u}_b \cdot \mathbf{\lambda})]\right\}^t.
	\label{eq:orthogonalmoment}
\end{align}
Here, $\mathbf{u}_a, \mathbf{u}_b$ denote $(d^2-1)$-dimensional unit real vectors uniformly distributed over the pseudo-Bloch sphere, $\mathbf{\lambda} = (\lambda_1, \cdots, \lambda_{d^2-1})$ is the vector of generalised Gell-Mann matrices, and $N_{d,t}$ is a suitable normalisation constant.
Also, the observables $M_A$ and $M_B$ are defined by a suitable choice of the eigenvalues for the coincidence between $\mathcal{R}^{(t)}$ and $\mathcal{S}^{(t)}$ to hold~\cite{imai_bound_2021,wyderka_probing_2023}.
For example, in the case of $d = 3$, the observable can be simply chosen by $M_A = M_B = \text{diag}(\sqrt{3/2},0,-\sqrt{3/2})$~\cite{wyderka_probing_2023}.
It is essential that the moments $\mathcal{S}^{(t)}$ are invariant not only over all local unitaries but also over all changes of local operator basis $\mathbf{\lambda}$, meaning the independence of the specific choice of observable.

\begin{figure*}[htb!]
	\includegraphics[width=\textwidth]{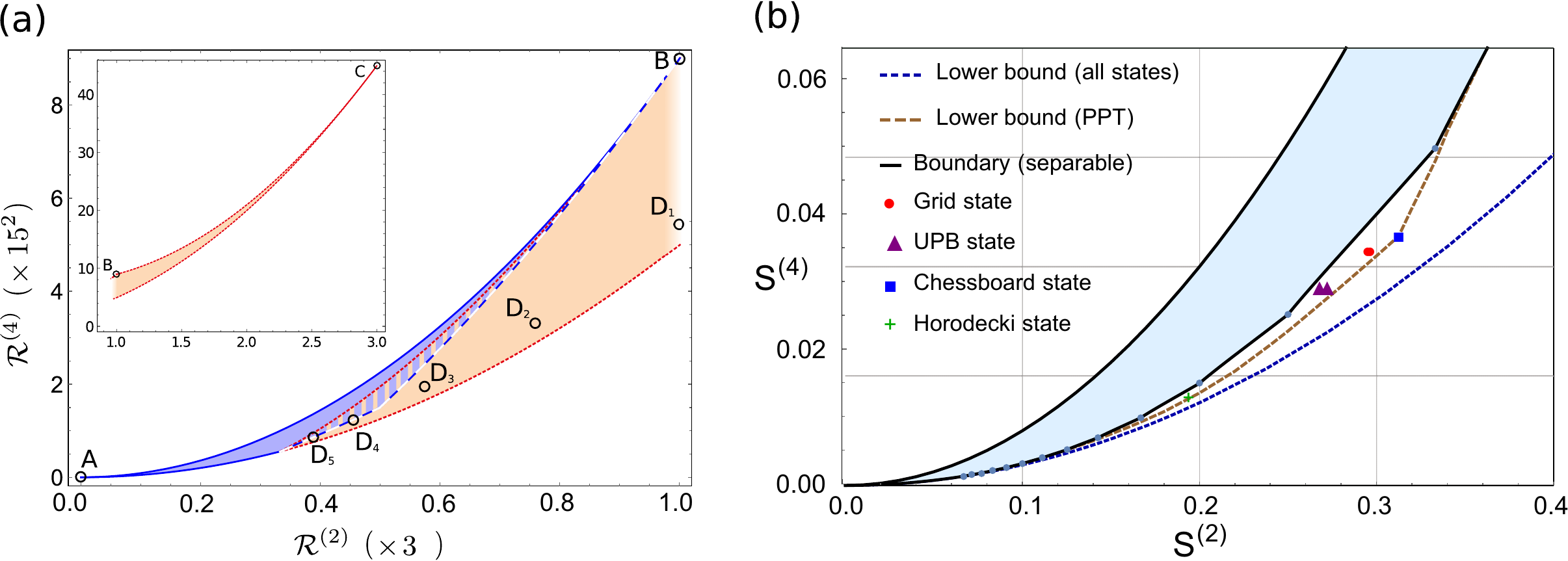}
	\caption{ 
		Left panel presents the set of Bell-diagonal states in the space spanned by the second and fourth moments $(\mathcal{R}^{(2)}, \mathcal{R}^{(4)})$ with suitable normalisation of randomised measurements ~\cite{ketterer_characterizing_2019}.
		Separable states are contained in the area surrounded by blue solid lines, while entangled states are outside and are contained in the area surrounded by red dotted lines.
		The striped region contains both separable and entangled states.
		Labelled circles represent the maximally mixed state (A), the pure product states (B), the Bell states (C), and the reduced states of the $n$-qubit Dicke state for $n=3,\ldots,7$ with two excitations ($\text{D}_1$) to ($\text{D}_5$).
		Right panel shows the set of two-qutrit states in the space spanned by the second and fourth moments,
		denoted as $(\mathcal{S}^{(2)}, \mathcal{S}^{(4)})$ in Eq.~(\ref{eq:momentsimai}), with suitable normalisation of randomised measurements~\cite{imai_bound_2021}.
		Separable states are contained in the light-blue area, while several bound entangled states (denoted by coloured symbols) are outside, implying that they can be detected with the method based on randomised measurements discussed in Sec.~\ref{sec:higher_order_moments}.
		Figures taken from Ketterer at al.~\cite{ketterer_characterizing_2019} and Imai et al.~\cite{imai_bound_2021} respectively.}
	\label{imai21}
\end{figure*}

In fact, the moments $\mathcal{S}^{(t)}$ for $t=2,4$ for any dimension can be evaluated analytically and are simply expressed as
\begin{equation}
	\mathcal{S}^{(2)}
	= \sum_{i=1}^{d^2-1} \tau_i^2,
	\qquad
	\mathcal{S}^{(4)}
	= 2 \sum_{i=1}^{d^2-1} \tau_i^4 + (\mathcal{S}^{(2)})^2,
	\label{eq:momentsimai}
\end{equation}
where the normalisation constant $N_{d,t}$ in Eq.~(\ref{eq:orthogonalmoment}) is chosen as $N_{d,2} = (d^2-1)^2/V^2$ and $N_{d,4} = (d^4-1)^2/(3V^2)$ with the surface area $V = 2\sqrt{\pi}^{d^2-1}/\Gamma[(d^2-1)/2]$ and Euler’s gamma function $\Gamma[z]$ and $\tau_i$ are singular values of the two-body correlation matrix $T = (T_{ij})$ with $T_{ij} = \tr[\varrho_{AB} \lambda_i \otimes \lambda_j]$ for $i,j=1,\ldots, d^2-1$.
This results from the fact that the moments $\mathcal{S}^{(t)}$ are invariant under local orthogonal rotations of the matrix $T$.
Accordingly, in a similar manner to Sec.~\ref{sec:Belldiagonalstate}, one can consider the space spanned by the moments $(\mathcal{S}^{(2)}, \mathcal{S}^{(4)})$ and formulate separability criteria in this space.
As a suitable constraint for this purpose, the so-called de Vicente criterion proposed in Ref.~\cite{de_vicente_separability_2007} was used.
This criterion states that any two-qudit separable state obeys $\|T\|_{\text{tr}} = \sum_{i=1}^{d^2-1} \tau_i \leq d-1$,
where $\|\cdots\|_{\text{tr}}$ denotes the trace norm,  invariant under orthogonal transformations.
The task is then to perform the optimisation
\begin{align}
	\max_{\tau_{i}}/\min_{\tau_{i}} \ \
	&\mathcal{S}^{(4)}
	\quad
	\text{s.t.} \quad
	\mathcal{S}^{(2)}
	= \sum_{i=1}^{d^2-1} \tau_i^2,
	\quad
	\sum_{i=1}^{d^2-1} \tau_i \leq d-1,
	\quad
	0 \leq \tau_i \leq d-1.
\end{align}
This results in the set of admissible values $(\mathcal{S}^{(2)}, \mathcal{S}^{(4)})$ for separable states in any dimension $d$, which allows for the detection of various bound entangled states as illustrated in Fig.~\ref{imai21}.
As further generalisations, the characterisation of the Schmidt number as dimensional entanglement has been discussed using this method in Refs.~\cite{wyderka_probing_2023,liu_characterizing_2022}.

The above criterion to detect bound entanglement has been implemented experimentally for two-qutrit chessboard states
$\varrho_{\text{ch}} =(1/4) \sum_{i=1}^4\ket{V_i}\!\bra{V_i}$,
which are written as a mixture of four states $\ket{V_i}$~\cite{bruss_construction_2000}.
The chessboard state was created by first generating two-qubit polarisation-entangled photon pairs through a spontaneous parametric down-conversion process and subsequently transforming them to two-qutrits $\ket{V_i}$ via dimension-expanding local operations implemented with motorised rotating half-wave plates and quarter-wave plates.
The experimentally prepared chessboard state $\varrho_{\text{ch}}^{\text{exp}}$ has a fidelity beyond $98\%$ with $\varrho_{\text{ch}}$, and the white noise level $p=0.129$.
For this state, the second and fourth moments $(\mathcal{S}^{(2)}, \mathcal{S}^{(4)})$ were computed, and its entanglement was verified in Ref.~\cite{bound_experi_fourth_dete}.

\subsection{Multipartite entanglement structure}
\label{sec:multipartite_entanglement}

The previous discussion was focused on verifying the presence of entanglement in bipartite quantum systems.
In multipartite systems, the structure of entanglement can vary significantly between states culminating in genuine multipartite entanglement (GME).
In this section, we present a series of criteria for the analysis of multipartite entanglement which are all based on functions of both the full as well as the marginal second moments.
We also discuss the results of a four-qubit experiment in which one of these criteria is applied to detect several types of entanglement.

\subsubsection{Full separability}
\label{sec:n-quditfullseparability}

The detection of high-dimensional multipartite entanglement was discussed using the $k$-body sector length $S_k$.
In Refs.~\cite{aschauer_local_2003,markiewicz_detecting_2013,klockl_characterizing_2015,tran_quantum_2015,huber_some_2018}, it has been shown that any $n$-qudit fully separable state obeys
\begin{align}
	S_k \leq \binom{n}{k}(d-1)^k,
	\label{eq:skalone}
\end{align}
where $S_k$ is the $k$-body sector length.
Violation of this bound implies that the state is entangled as can be easily demonstrated, for instance, in graph states.
Note that this criterion can be seen as a generalisation of Eq.~(\ref{eq:twoquditsecondmo}) to sector lengths between a number of observers smaller than $n$ and any dimension.
One can also consider linear combinations of various sector lengths as it was shown in Ref.~\cite{miller_sector_2022} that
\begin{align}
	\sum_{k=0}^n[(d-1)n - dk]S_k \geq 0
\end{align}
holds for any $n$-qudit fully separable state.
This criterion is strictly stronger (detects more entangled states) than the one in Eq.~(\ref{eq:skalone}) and can be understood as the $n$-qudit generalisation of Eq.~(\ref{eq:secondoptimal}).

\subsubsection{$k$-separability}
\label{sec:k_separability_drom_SL}

In order to recall the notion of $k$-separability~\cite{dur_separability_1999,dur_classification_2000}, let us first consider pure states.
A $n$-particle pure state is called $k$-separable if it can be written as
\begin{align}
	|{\psi_{k\text{-sep}}}\rangle =
	\ket{\phi_1} \otimes \ket{\phi_2} \otimes
	\cdots \otimes \ket{\phi_k}.
\end{align}
A mixed state is $k$-separable if it is a convex mixture of pure $k$-separable states, with different elements in the mixture possibly admitting different partitions into $k$ subsystems.
For $k=n$, this notion is equivalent to the full separability.
For example, the following state is $100$-qubit $12$-separable:
\begin{align}
	|{\psi_{100, 12}}\rangle
	=\ket{\text{GHZ}_{20}}^{\otimes 3}
	\otimes
	\ket{\text{GHZ}_{10}}^{\otimes 2}
	\otimes
	\ket{\text{GHZ}_{5}}^{\otimes 2}
	\otimes
	\ket{\text{Bell}}^{\otimes 5}.
\end{align}

In Ref.~\cite{ketterer_statistically_2022}, the hierarchical criteria for $k$-separability have been proposed using the full-body sector lengths $S_n$:
any $n$-qubit $k$-separable state obeys
\begin{align} \label{msepbound}
	S_n  \leq  
	\begin{cases}
		3^{k-1} \; 2^{n-(2k-1)}, \ &\text{if \ odd}\ n,\\
		3^{k-1} \; \Big(2^{n-(2k-1)}+1\Big), \ &\text{if \ even}\ n,
	\end{cases}
\end{align}
for $k=2,3,\ldots,
\lfloor (n-1)/2 \rfloor$.
A violation of the inequality for some $k$ implies that the state is at most $(k-1)$-separable.
In particular, if the state violates the inequality with $k=2$, then it is verified to be genuinely $n$-partite entangled for $n>4$.

\subsubsection{Tripartite entanglement}
\label{sec:three_partite_entanglement}

One idea to detect entanglement using second moments more efficiently is to consider linear combinations of full and marginal moments.
Note that the sector lengths themselves are convex functions, but their combinations do not necessarily have to be convex.
Recall that the relations between the sector lengths and the second moments are
$\mathcal{R}_{A}^{(2)}
+ \mathcal{R}_{B}^{(2)} + \mathcal{R}_{C}^{(2)}= S_1$ and
$\mathcal{R}_{AB}^{(2)}
+ \mathcal{R}_{BC}^{(2)} + \mathcal{R}_{CA}^{(2)}= S_2$,
and
$\mathcal{R}_{ABC}^{(2)} = S_3$, see Eq.~(\ref{eq:sectorlengths}).
In Ref.~\cite{imai_bound_2021}, it has been shown that any fully separable three-qudit state obeys
\begin{align}
	S_3 \leq d-1 + \frac{2d-3}{3}S_1 + \frac{d-3}{3}S_2,
	\label{eq:sectorthreefullsep}
\end{align}
whereas any three-qudit state which is separable for any fixed bipartition obeys
\begin{align}
	S_2 + S_3 \leq \frac{d^3-2}{2}(1+S_1).
	\label{eq:sectorthreebisep}
\end{align}
In the case of $d=2$, strong numerical evidence suggests that the above inequality also holds for mixtures of biseparable states with respect to different partitions~\cite{imai_bound_2021}, discussed in Eq.~(\ref{eq:biseparablethreepartite}).
Violation of (\ref{eq:sectorthreebisep}) would therefore imply genuinely three-qubit entanglement, but its analytical proof has not yet been provided.

It is essential to note that the criteria in Eqs.~(\ref{eq:sectorthreefullsep}, \ref{eq:sectorthreebisep}) for $d=2$ can be interpreted as the geometry of the three-qubit state space in terms of sector lengths $(S_1, S_2, S_3)$~\cite{wyderka_characterizing_2020, imai_bound_2021}.
In this space, both criteria were shown to be much more effective in certifying entanglement than criteria based only on the full sector length $S_3$.
In particular, Eq.~(\ref{eq:sectorthreebisep}) can detect multipartite entanglement for mixtures of GHZ states and W states, even if the three-tangle~\cite{coffman_distributed_2000} and the bipartite entanglement in the reduced subsystems vanish simultaneously~\cite{lohmayer_entangled_2006}.

The geometry of two-body correlations in three-qubit states, using the three coordinates $(\mathcal{R}_{AB}^{(2)}, \mathcal{R}_{BC}^{(2)}, \mathcal{R}_{CA}^{(2)})$, has also been studied in Ref.~\cite{shravan_2023}. Technically, these coordinates are elements of a decomposition of the sector length $S_2$, i.e. $S_2 = \mathcal{R}_{AB}^{(2)} + \mathcal{R}_{BC}^{(2)} + \mathcal{R}_{CA}^{(2)}$. It has been shown that any fully separable three-qubit state obeys
\begin{equation}
	\mathcal{R}_{XY}^{(2)} + \mathcal{R}_{YZ}^{(2)} - \mathcal{R}_{ZX}^{(2)} \leq 1,
\end{equation}
where $X,Y,Z = A,B,C$. Moreover, any three-qubit state which is separable
in a $X|YZ$ bipartition obeys not only the above inequality but also
\begin{align}
	3\mathcal{R}_{XY}^{(2)} + \mathcal{R}_{YZ}^{(2)} - \mathcal{R}_{ZX}^{(2)} &\leq 3,
	\\
	-\mathcal{R}_{XY}^{(2)} + \mathcal{R}_{YZ}^{(2)} + 3 \mathcal{R}_{ZX}^{(2)} &\leq 3.
\end{align}
A violation of these inequalities implies entanglement across the $X|YZ$ bipartition but does not imply genuine multipartite entanglement.

\subsubsection{Nonlinear functions of second moments}
\label{sec:nonlinear_finctions_of_moments}

In Ref.~\cite{knips_multipartite_2020}, another criterion based on products of marginal moments was formulated.
Instead of considering the factorisability of the correlation functions themselves, the factorisability of the second moments is considered with a purity-dependent bound. 
Namely, for two-qubit states, the inequality obeyed by all separable states reads 
\begin{equation}
	{\mathcal M}_{2} \equiv \mathcal{R}_{AB}^{(2)} - \mathcal{R}_A^{(2)}\mathcal{R}_B^{(2)} \le
	\begin{cases}
		(4\mathcal{P}-1)/9 & \mathrm{\;for\;} \mathcal{P}<\frac{1}{2}, \\
		4(1-\mathcal{P})\mathcal{P} / 9 & \mathrm{\;for\;} \mathcal{P}\geq\frac{1}{2},
	\end{cases}
\end{equation}
where
$\mathcal{P} = \tr(\varrho_{AB}^2)$ is the state's purity,
the normalisation constant for moments was chosen as $N_{n,d,t} = 1$ in Eq.~(\ref{eq:generalmoment}),
and a product observable $\mathcal{M} = \sigma_z \otimes \sigma_z$ was used.
A violation of this inequality implies the presence of entanglement between the parties.
Unlike the previously presented results, this criterion is expressed as the nonlinear combination of the second moments.
This nonlinearity enhances the detection power compared to the criterion discussed in Eq.~(\ref{EQ_SIMPLE_R2}).
Additionally to the purity-dependent inequality for two-qubit states, inequalities for three- and four-qubit states have been obtained using numerical simulations.
To indicate genuine tripartite (four-partite) entanglement, the value of ${\mathcal M}_{3}$ (${\mathcal M}_{4}$) has to overcome the bounds given by
\begin{eqnarray}
	{\mathcal M}_{3} & = & \mathcal{R}_{ABC}^{(2)} - \mathcal{R}_{A}^{(2)}\mathcal{R}_{BC}^{(2)}- \mathcal{R}_{B}^{(2)} \mathcal{R}_{AC}^{(2)} - \mathcal{R}_{C}^{(2)}\mathcal{R}_{AB}^{(2)} \le \tfrac{8}{27}(1-\mathcal{P})\mathcal{P},
	\label{EQ_3GEM} \\
	{\mathcal M}_{4} &= &
	\mathcal{R}^{(2)}_{ABCD}-\frac{1}{2} \sum_M \mathcal{R}^{(2)}_M \mathcal{R}^{(2)}_{\overline{M}} 
	\le \tfrac{8}{81}(1-\mathcal{P}^2),  \label{EQ_4GEM}
\end{eqnarray}
where the summation is performed over all subsets of $\lbrace ABCD \rbrace$ except for the full and empty set, and where $\overline{M}$ denotes the set complementary of $M$.
Although the simple form of the latter two expressions raises hope for a generalisation to $\mathcal{M}_n$, i.e.~for an expression comparing the second-order moment of the distribution of the correlations of an $n$-qubit state with the products of all second-order moments of the marginals, such an expression remained missing. 

This concept has been experimentally demonstrated using a pair of polarisation-entangled photons created by means of spontaneous parametric down-conversion~\cite{knips_multipartite_2020}.
The two entangled photons are sent into two separate interferometers, making both path and polarisation degree accessible~\cite{knips_multipartite_2016}.
In this experiment, no knowledge about or direct control of the measurement directions is required.
However, the angles of the wave plates which were required to set the measurement directions cannot be selected from a uniform distribution because this would lead to a non-uniform sampling of the local unitaries.
Hence, to ensure a uniform distribution according to the Haar measure, i.e.~to distribute the measurement directions following a uniform sampling of the local Bloch spheres, appropriate unitary transformations have been randomly picked and implemented by bringing the wave plates to the angles corresponding to the drawn unitary transformation.

\begin{figure}[htb!]
	\includegraphics[width=\textwidth]{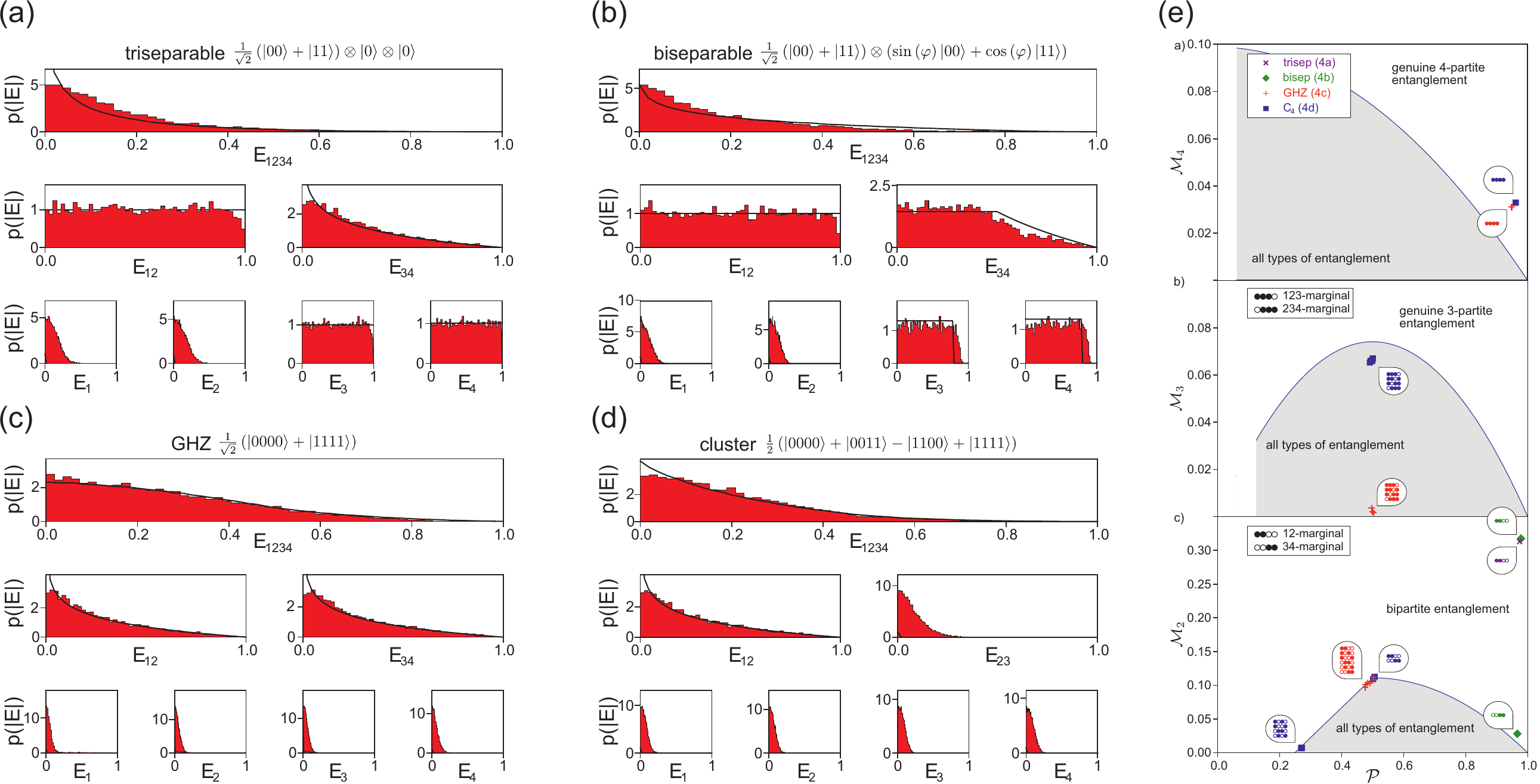}
	\caption{
		Panels $(a)$ to $(d)$ show histograms of experimental distributions of correlations averaged over random local measurements for a four-qubit triseparable state (a), biseparable state (b), GHZ state (c) and linear cluster state (d). Full correlations are shown alongside distributions of several of their marginal states. Here $p\left(|E|\right)$ denotes the probability distribution of modulus of the correlation value of the four-partite state (for $E_{1234}$) and some of its marginal states (for $E_{12}$, $E_1$, and so on). The solid lines indicate the distribution for the ideal states with infinite statistics. The rightmost panel presents the comparison of the second-order moment of the correlation distribution of an $n$-qubit state with all products of moments of marginals allowing the detection of genuine multipartite entanglement and revealing the entanglement structure of states.
		Figures taken from Knips et al.~\cite{knips_multipartite_2020}.}
	\label{fig:Knips20}
\end{figure}

Four different types of four-qubit states were experimentally studied: a triseparable state, a biseparable state, a GHZ state, and a linear cluster state. 
After applying random local transformations to each of the four qubits, projective measurements allowed to retrieve the statistics of correlation values as shown in Fig.~\ref{fig:Knips20}.

The shape as well as the factorisability of the particular distributions directly visualise the entanglement structure of the state.
For example, for the triseparable state $\ket{\textrm{Bell}}\ket{0} \ket{0}$ the distribution of the modulus of correlation values for the marginal of the first two qubits, i.e. $E_{12} \equiv \tr[(U_1^{\dagger}\sigma_1 U_1 \otimes U_2^{\dagger}\sigma_2 U_2 \otimes \mathbbm{1}\otimes \mathbbm{1}) \varrho ]$, is almost uniform and not explainable by a product of the single-qubit marginals (panels $E_1$ and $E_2$).
At the same time, the other two qubits individually already show large values for $E_3$ and $E_4$, respectively, sufficient to explain the distribution $E_{34}$.

This graphical analysis illustrates how a criterion for GME allows to probe the entire entanglement structure of a state when applied to different partitions and combinations of subsystems. 
The rightmost panel in Fig.~\ref{fig:Knips20} shows the results of this structural analysis for the four experimentally prepared states, summarizing which states and marginals are violating the respective bounds.
The top-most subplot ($\mathcal{M}_4$) indicates that both the GHZ and the linear cluster state are detected to be genuinely four-partite entangled, which is not detected for the biseparable or the triseparable state (as it should be).
$\mathcal{M}_2$, on the other hand, shows that for the biseparable state $\ket{\textrm{Bell}} \ket{\psi_{\textrm{ent}}}$ the two-qubit marginals are still entangled, whereas the triseparable state carries entanglement solely between the first two qubits.

\subsubsection{Discriminating $W$-class entanglement}
\label{sec:W_class}
In multipartite systems, the structure of entanglement becomes much richer and more complicated than in bipartite systems. In the bipartite state space, there exists an ordering in terms of quantum resource theories~\cite{chitambar_quantum_2019}.
In this sense, the maximally entangled state can be defined as the entangled state that enables creation of any bipartite entanglement by LOCC operations. On the other hand, in multipartite systems, such an ordering does not exist anymore and the notion of maximally entangled states cannot be defined uniquely. In fact, already three-qubit pure states are divided into two classes: the GHZ class and the W class. The GHZ state cannot be transformed to the W state and vice versa with LOCC operations, even if they
are not required to reach the state with probability one (so-called stochastic LOCC (SLOCC) operations)~\cite{dur_three_2000, acin_classification_2001}.
This distinction leads to different roles the GME states play in information processing tasks~\cite{cleve_how_1999,raussendorf_oneway_2001,lvovsky_optical_2009}.
For four-qubit states, there is already an infinite collection of such classes, which can be grouped into nine families~\cite{verstraete_four_2002}.
Therefore, in addition to entanglement detection, another interesting and important issue is to determine which class a given multipartite state belongs to.

The discrimination of $W$-class entanglement has been studied with two criteria based on the second and fourth moments of randomised measurements~\cite{ketterer_entanglement_2020}.
The first is an analytical upper bound for the second moment of a $n$-qubit $W$-class state:
\begin{align}
	S_n \leq
	5-\frac{4}{n},
\end{align}
where we set $S_n = \mathcal{R}^{(2)}$ and the inequality is saturated by a pure $W$ state: $\ket{W_n} = (\ket{10\cdots 0} + \ket{01\cdots 0} + \cdots +\ket{0\cdots 01})/\sqrt{n}$.
The violation of this condition implies that a multiqubit state is detected to be outside of the W class,
i.e.~it cannot be obtained from the W state by SLOCC.
The second criterion uses a linear combination of the second and fourth moments, with weights optimised based on the $n$-qubit state $\ket{W_n}$ and the biseparable state $\ket{W_{n-1}}\ket{\psi}$.
Furthermore, Ref.~\cite{ketterer_entanglement_2020} has provided the characterisation of three-qubit and four-qubit states using the second and fourth moments from an extensive numerical approach.

\subsubsection{Spin-squeezing entanglement from collective randomised measurements}
\label{sec:spinsqueezing}
So far, we have discussed the randomised measurement scheme based on the moments $\mathcal{R}^{(t)}_{\mathcal{M}}$ defined in Eq.~(\ref{eq:generalmoment}). In general, they can be obtained by measuring each of the local subsystems individually.
However, in some physical systems like ultra-cold atoms, Bose-Einstein condensates, or trapped ions, it may be impossible to measure and manipulate individual particle in an ensemble of quantum particles.
This is also a typical restriction in macroscopic systems.
For this reason, it is worth considering criteria based on the measurement of collective angular momentum denoted as $J_l = \tfrac{1}{2} \sum_{j=1}^n \sigma_l^{(j)}$, where $\sigma_l^{(j)}$ represent Pauli spin matrices for $l = x, y, z$ on the $j$-th subsystem.
This type of measurement allows certification of so-called spin-squeezing entanglement.
This notion is originally related to spin-squeezing parameters in quantum metrology~\cite{wineland_1992}, but more generally any system with entanglement which can be detected with $J_l$ and $J_l^2$ only is called ``spin squeezed''~\cite{wang_2003, korbicz_2005, toth_optimal_2007}.

Motivated by this experimental issue, we explain the concept of \textit{collective} randomised measurement, used to characterise spin-squeezing entanglement, by following the description in Ref.~\cite{imai_2023}.
Consider random rotations on a multiparticle system, implemented through a collective unitary $U^{\otimes n}$. Then one can obtain the expectation value and the variance with respect to such a random unitary given by
\begin{equation}
	\langle J_z \rangle_U = \tr[\varrho U^{\otimes n} J_z (U^\dagger)^{\otimes n}],
	\qquad
	(\Delta J_z)_U^2 = \langle J_z^2 \rangle_U - \langle J_z \rangle_U^2.
\end{equation}
Sampling over the collective random unitaries, the resulting distribution can be characterised using the moments
\begin{eqnarray}
	\mathcal{J}^{(t)}(\varrho)
	= \int \mathrm{d}\mu(U) \, \left[
	\alpha (\Delta J_z)_U^2 + \beta \langle J_z \rangle_U^2 + \gamma \right]^t,
\end{eqnarray}
where $\alpha, \beta, \gamma$ is to be optimised for best entanglement detection capabilities.
The Haar integral corresponds to the uniform randomisation over the \textit{collective Bloch sphere} in the three coordinates $(J_x, J_y, J_z)$.
Here, it is essential that moments are invariant under any collective unitary, that is, $\mathcal{J}^{(t)}(\varrho) = \mathcal{J}^{(t)}(V^{\otimes n} \varrho (V^\dagger)^{\otimes n})$ for all $V \in \text{SU}(2)$.
In Ref.~\cite{imai_2023} several entanglement criteria are presented which are collective-reference-frame-independent.
It has been shown that if a state is permutationally symmetric (bosonic), such as the W state or more generally the Dicke state, then its spin-squeezing entanglement is completely characterised by $\mathcal{J}^{(t)}$ for $t=1,2,3$.
This means that they yield necessary and sufficient spin-squeezing entanglement conditions.
Furthermore, even in the case of the non-symmetric states, the moment $\mathcal{J}^{(1)}$ evaluated as $\mathcal{J}^{(1)} = \sum_{l=x,y,z} (\Delta J_l)^2$ for $\alpha =3$ and $\beta=\gamma=0$ can be used to detect even the multipartite bound entanglement.
This follows from violation of the bound $\mathcal{J}^{(1)} \geq n/2$~\cite{toth_entanglement_2004} that any $n$-qubit separable state obeys.

\subsection{Finite datasets and single-setting entanglement detection}
\label{sec:finite_datasets}

Since in experimental practice always only finite datasets are available, statistical noise additionally complicates tasks such as the certification of entanglement, which is of course also true for the scenario of randomised measurements.
In this section, we present tools to analyse and quantify these statistical effects and discuss how they affect the moments of probability distributions making it possible to refine the various entanglement criteria accordingly.
For more details about the field of statistical analysis on quantum systems,
see Refs.~\cite{flammia_direct_2011,pallister_optimal_2018,elben_statistical_2019,elben_cross-platform_2020,yu_statistical_2022}.

\subsubsection{Estimation of correlation functions}
\label{sec:estimation_of_correlation_function}

Statistical noise in the scenario of randomised measurements has two sources.
The first is a result of the quantum nature of the system, for which the correlation tensor element $T_i$ (for fixed measurement settings collectively indexed by $i$) cannot be measured directly but has to be determined in a statistical manner via the repetition of a probabilistic measurement which can yield a set of discrete outcomes.
These outcomes are distributed around the mean value $\mu_i = T_i$ with a variance of $\sigma_i^2$, which in general is a function of $T_i$ as well.
Formally, this series of $m$ measurements can be expressed as a set of independent and identically distributed (i.i.d.) random variables $\{X_1,X_2,\ldots, X_m\}$ with the same mean and variance.
In this case the sample mean $\widetilde{X} = \sum X_i / m$ is an \textit{unbiased} and \textit{consistent estimator} for $\mu_i$, since its expectation value is $\mathbbm{E}(\widetilde{X}) = \mu_i$ and $\widetilde{X}$ approaches $\mu_i$ for increasing $m$.

The second type of statistical noise is the propagation of the fluctuation of the measured values of $T_i$ to any quantity calculated from them, such as for example the second moment of their distribution $\mathcal{R}^{(2)}$.
Note that in this case the fluctuation of each particular $T_i$ additionally combines with noise due to sampling of only a finite amount of different $T_i$'s (different settings).
For a set of $m$ measured correlation values $\{\widetilde{T}_1,\widetilde{T}_2,\ldots \widetilde{T}_m\}$ the estimator $\widetilde{\mathcal{R}}^{(2)} = \sum \widetilde{T}^2_i / m$ albeit consistent is, however, biased.
Even though each particular $\widetilde{T}_i$ is an unbiased estimator for the corresponding $T_i$, the value of $\widetilde{\mathcal{R}}^{(2)}$ is systematically increased due to taking the square.
Both the statistical fluctuation due to finite sample size as well as any systematic bias has to be taken into account when applying the entanglement criteria based on a bound violation, such as for example in Eq.~(\ref{EQ_D2_ENT}).
When this is properly considered, it turns out that even very limited data sets can lead to a valid conclusion about the entanglement in the system~\cite{dimic_single-copy_2018,saggio_experimental_2019,cieslinski_high_2022}.

A common way to quantify how much the values of a general estimator $\widetilde{\theta}$ deviate from the actual parameter $\theta$ is based on the likelihood. This is 
generally defined as the probability of the data given
some assumptions or models (e.g., the probability of tomographic data given some quantum state $\varrho$).
More specifically, one can consider the probability
that an estimator takes an observed value for
a given value of the parameter, 
$\mathrm{Prob}(\widetilde{\theta} | \theta)$.
It allows us to calculate the $p$-\textit{value}, i.e.~the probability that for a certain $\theta$, the estimated value of $\widetilde{\theta}$ will be sufficiently close.
The usual definition is
\begin{align}
	\text{Prob}( |\widetilde{\theta} - \theta| \geq \delta) \leq \alpha,
	\label{eq:pvalue}
\end{align}
where $\delta$ is called the \textit{error or accuracy},
$\delta/\theta$ the \textit{relative error},
$\alpha$ the \textit{statistical significance level},
and $\gamma = 1-\alpha$ the \textit{confidence level}.
A practical tool to estimate $p$-values are so-called \textit{concentration inequalities} such as the Chebyshev inequality which, for example, allows to estimate the probability that the sample mean $\widetilde{X}$, from a sample of $m$ observations, will be close to the mean value as
\begin{align}
	\text{Prob}(|\widetilde{X}-\mu| \geq \delta) \leq \frac{\sigma^2}{m \delta^2}.
\end{align}
This relation allows us to estimate the minimal number of measurements $m$ to achieve a certain significance level.

\subsubsection{Statistical significance and randomised measurements}
\label{sec:statistically_significant_test}
\begin{figure}[htb]
	\begin{center}
		\includegraphics[width=0.6\textwidth]{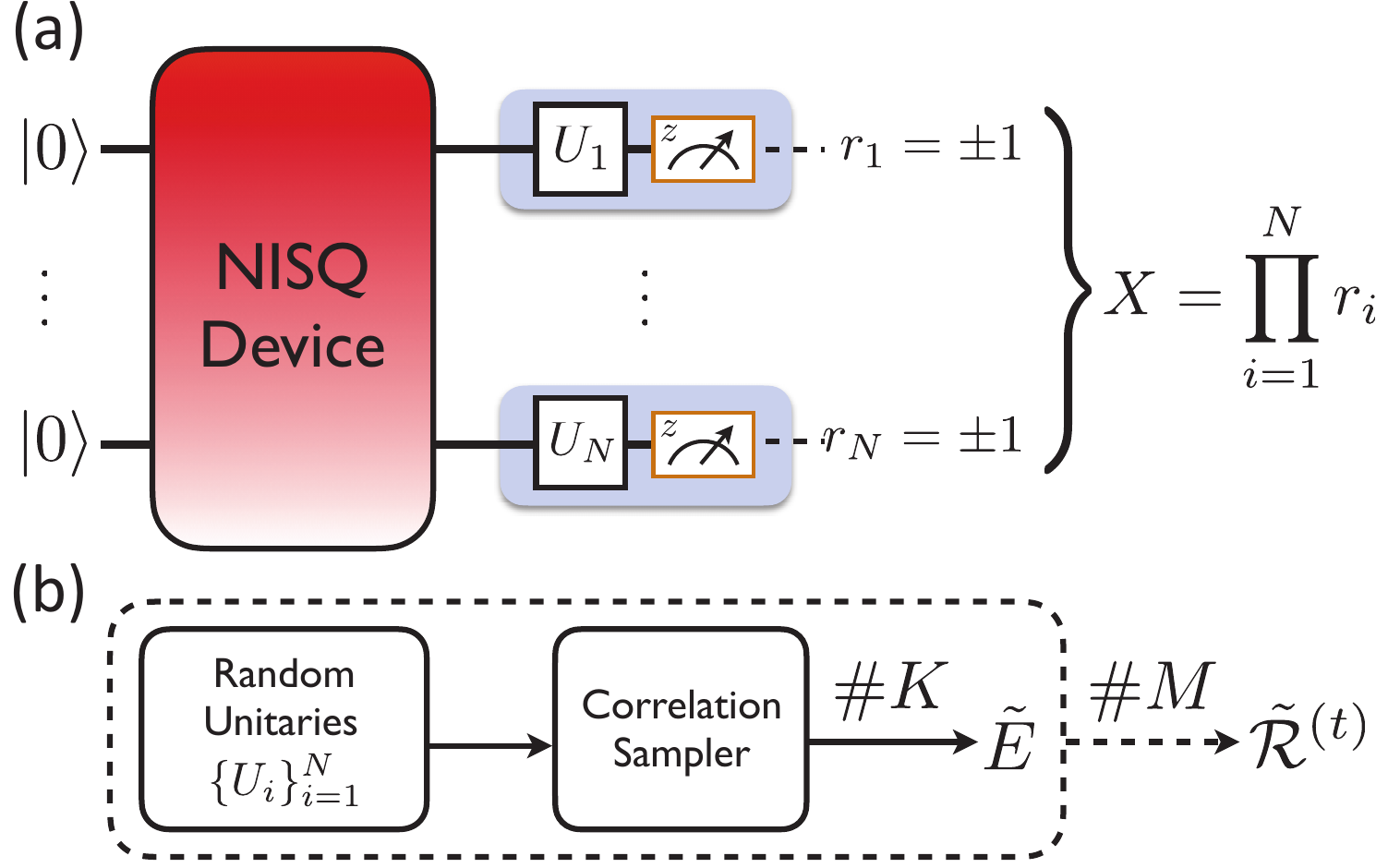}
		\caption{
			Characterisation of a Noisy Intermediate-Scale Quantum (NISQ) device using locally randomised measurements:
			(a) A measurement on $N$ qubits using random local bases defined by a set of local unitaries
			$\{U_i\}_{i=1}^N$, resulting in a correlation sample $X$.
			(b) $M$ repetitions of the measurement protocol described in (a), with each repetition using a different set of randomly sampled measurement bases.
			Every individual projective measurement is performed $K$ times and the overall scheme provides us with the moment estimator.
			Figure taken from Ketterer et al.~\cite{ketterer_statistically_2022}.
		}
		\label{Ketterer_fig1}
	\end{center}
\end{figure}

To evaluate the statistical effect when applying criteria based on quantities like $\mathcal{R}^{(t)}$, it is crucial to quantify how much the estimated value can deviate from the parameter of interest, as expressed in Eq.~(\ref{eq:pvalue}).
As before, we consider an experiment in which a sample of $M$ different measurement settings is chosen randomly and for each setting measurements are performed on an ensemble of $K$ state copies, see Fig.~\ref{Ketterer_fig1}.
An unbiased estimator for the moment can be given by $\Tilde{\mathcal{R}}^{(t)} = \frac{1}{M}\sum_{i=1}^M [\Tilde{E_t}]_i$.
In this expression, $\Tilde{E_t}$ denotes the unbiased estimator of $E^t$ (the $t$-th power of the correlation function $E$) that is obtained from the $K$ measurements and the subscript $i$ refers to the setting.

Even in the case where $[\Tilde{E_t}]_i$ cannot be assumed to be i.i.d. random variables, we can find deviation bounds such as Eq.~(\ref{eq:pvalue}) based on the variance of $\Tilde{\mathcal{R}}^{(t)}$ using the Chebyshev-Cantelli inequality
\begin{align}
	\text{Prob}(|\Tilde{\mathcal{R}}^{(t)}
	-\mathcal{R}^{(t)}| \geq \delta)
	\leq
	\frac{2 \text{Var}(\Tilde{\mathcal{R}}^{(t)})}
	{\text{Var}(\Tilde{\mathcal{R}}^{(t)}) + \delta^2}.
\end{align}
This leads to a minimal two-sided error bar
$
\delta_{\text{error}}
=\sqrt{({1+\gamma})/({1-\gamma})}
\sqrt{\text{Var}(\Tilde{\mathcal{R}}^{(t)})},
$
which guarantees the confidence level~$\gamma = 1-\alpha$.
For instance, in the estimation of the second moment $\mathcal{R}^{(2)}$ for an $n$-qubit state with a product observable, the expression
of the variance is
\begin{align}
	\text{Var}(\Tilde{\mathcal{R}}^{(2)})
	= \frac{1}{M}\left[
	A(K) \mathcal{R}^{(4)}
	+ B(K) \mathcal{R}^{(2)}
	+ C(K) - (\mathcal{R}^{(2)})^2
	\right],
	\label{eq:second_moment_variance}
\end{align}
where $A(K), B(K), C(K)$ are determined through the properties of the binomial distribution, and has been derived in Ref.~\cite{ketterer_statistically_2022}.
From this result, the total number of measurements for the precise estimation of $M_{\text{total}} = M \times K$ can be determined depending on the state under consideration and required accuracy $\delta$ and confidence level $\gamma = 1-\alpha$.

The dependence of the second moment on the squared correlations 
gives rise to a systematic error that must be taken into account. 
Reference~\cite{knips_multipartite_2020} proposes to mitigate this with the use of Bayesian methods. This requires establishing the probability $P(\Tilde{T})$ with which a given value of correlations $T$, estimated after $K$ trials, occurs in the experiment. The second moment written in this language is
\begin{equation}
	\int_{-1}^{1} \mathrm{d} T \, T^2 \, P(T) = \int_{-1}^1 \int_{-1}^1  \mathrm{d} T \, \mathrm{d}\Tilde{T} \, T^2 \, P(T|\Tilde{T}) \, P(\Tilde{T}). 
\end{equation}
The Bayes theorem then gives $P(T|\Tilde{T}) = P(\Tilde{T} | T) \tilde P(T) / P(\Tilde{T})$, where $\tilde P(T)$ represents the prior assumption about the unknown ideal distribution $P(T)$.
In practice this prior can be chosen as estimated $P(\Tilde{T})$ and the conditional probability $P(\Tilde{T} | T)$ can be assumed to be a normal distribution centred at $T$ and with variance $(1-T^2) / K$, leading to an updated estimation of the second moment $\tilde{\mathcal{R}}^{(2)}$.

\subsubsection{Certification of entanglement}
\label{sec:n_qubit_entanglement_finite_data}

\begin{table}[b]
	\centering
	\caption{Probability of detecting entanglement for $n$-qubit GHZ state with a single random measurement per party and $p$-value of $4.6\%$.
		Number $K$ denotes the number of trials based on which the correlation function is estimated.
		The table is taken from Ref.~\cite{tran_correlations_2016}.}
	\label{TAB_GHZ_K1000}
	\begin{tabular}{r|| c c c c c c c c}
		$n$                     & 3   & 4    & 5    & 6    & 7    & 8    & 9    & 10  \\ \hline 
		$K=1000$                & \, 26\% & \, 44\% & \, 47\% & \, 57\% & \, 52\% & \, 48\% & \, 41\% & \, 34\%\\
		$K \rightarrow \infty $ & \, 26\% & \, 44\% & \,  48\% & \, 63\% & \,  67\% & \, 77\% & \, 80\% & \, 86\%
	\end{tabular}
\end{table}

In the case of entanglement detection with finite statistics the estimated second moment $\tilde{\mathcal{R}}^{(2)}$ may happen to be larger than $1$ also for a product state, as discussed in Ref.~\cite{tran_quantum_2015,tran_correlations_2016}. 
In such a scenario, an entanglement indicator can be formulated as 
\begin{equation}
	\tilde{\mathcal{R}}^{(2)} > 1 + \delta \implies \textrm{likely } \ket{\psi} \textrm{ is entangled},
	\label{EQ_CRIT_MK}
\end{equation}
where $\delta$ determines the confidence of entanglement detection.
Given sufficiently big $M$ and $K$, a normal distribution with a standard deviation $\Delta_{M,K}$ approximates the distribution of values of $\tilde{\mathcal{R}}^{(2)}$ for any state.
A significance level of $5\%$ can be satisfied by
setting $\delta=2 \, \Delta_{M,K}$ in which case the probability that a separable state will yield a value within the specified bound is $95.4\%$.
Hence, a measurement of a value higher than $1 + 2 \Delta_{M,K}$ detects entanglement with a $p$-value of $4.6\%$ and confidence level of at least $95\%$.

The statistical analysis gives rise to the possibility of entanglement detection with a single randomly chosen measurement setting.
Table~\ref{TAB_GHZ_K1000} presents the probability that $n$-qubit GHZ state violates the bound $\mathcal{R}^{(2)}_{M,K}>1 + \delta$, with a single randomly chosen setting ($M=1$) and $p$-value of $4.6\%$. 
The case of ideal quantum predictions ($K \to \infty$) is compared with $K = 1000$ repetitions. 
For the ideal predictions, the probability of violation grows with the number of qubits whereas for finite statistics the detection probability first increases but then decays. 
See~\cite{tran_correlations_2016} for the explanation of $n$ above which the probability decays.

\section{Functions of states}
\label{SEC_FUNCTIONS}

Addressing the task of entanglement detection is vital for many quantum information applications.
Nevertheless, it is not the only relevant question for quantum state analysis.
One may also be interested in the functions of states, which not only can be used to verify entanglement but also to provide a description of other state features.
Many quantum mechanical notions, such as the purity or PT moments explained above, are non-linear functions of the state and require in their standard definition the knowledge of the complete density matrix for their calculation, see Eq.~(\ref{sectorandpurity}) and Eq.~(\ref{eq:pt_moments}) respectively. 
This, however, poses a challenge in experimental settings due to the exponential growth in the number of measurements required for state tomography as the number of particles increases. 

Randomised measurements have emerged as a valuable approach in this domain, enabling the extraction of these quantities from experimental data with considerably fewer resources.
They can be accessed through properly designed functions of probabilities over outcomes of measurements in product bases, statistical moments and other post-processing methods.
The following section provides an overview of approaches for the estimation of state functions through the use of these techniques.
In addition, we will explain the notion of shadow tomography in Sec.~\ref{sec:shadow_tomography}, where randomised measurements also play a pivotal role.

\subsection{Determination of invariant properties of states}
\label{sec:general_determination_uni_inv}

A range of unitarily invariant quantities has been shown to be accessible by randomised measurements. In contrast to the methods discussed in Sec.~\ref{SEC_ENTANGLEMENT}, which rely mainly on the determination of statistical moments of the distribution of correlations, the approaches in this and the following subsections are based on the direct analysis of the occurring frequencies or probabilities.

\subsubsection{Theoretical results}

In the most general scenario, one starts with a set of (potentially nonlocal) unitary transformations $\lbrace U \rbrace$ and a product basis $\lbrace |s \rangle \rbrace =\lbrace |s_1 \dots s_n \rangle \rbrace $ in which the measurements are performed.
Without loosing
generality, one can assume this to be the computational basis, as all quantities of interest in this section are invariant under a local basis change.
Then, the probabilities $P_U(s)=\tr(U \varrho U^{\dagger} \ket{s}\bra{s})$ of a string of results can be inferred from the experimental data.
Consequently, properly designed functions of $P_U(s)$ averaged, e.g.~over a Haar-distributed set of unitaries $\lbrace U \rbrace$, allow the extraction of multiple state functions.

Since the purity is invariant under any unitary transformation, it can even be accessed if the distribution is averaged under random \textit{global unitary} transformations. Indeed, it was already shown early that the purity of a state can be expressed as a linear function of the ensemble average of $P_U(s)^2$ \cite{van_enk_measuring_2012}. 
Experimentally, however, the implementation of {\it global} random unitary transformations in systems with local interactions requires significant resources \cite{ohliger_efficient_2012} and a simpler protocol is desirable.

Indeed, if {\it local} unitary rotations are considered it was shown in Ref.~\cite{brydges_probing_2019, elben_renyi_2018,elben_statistical_2019} 
that the purity of $n$-qubit states can be expressed as
\begin{equation}
	\mathcal{P}(\varrho)=2^n \sum_{s,s'}(-2)^{-D[s,s']}\overline{P_{U}(s)P_{U'}(s')}.
	\label{local_purity}
\end{equation}
Here, $U$ is a random local unitary according to the 
Haar measure and $\overline{\cdots}$ denotes the 
ensemble average over all pairs $U, U'$ of these 
unitaries. Furthermore, $D[s,s']$ is the Hamming 
distance between two-bit strings, denoting the number 
of elements in which two $n$-tuples of measurement 
results differ. Note that for the qubit case, the 
measurements in $x, y,$ and $z$ direction form a 
spherical 3-design, so one can replace the 
random unitaries by three measurements in these 
directions, see Sec.~\ref{sec:designs}.
Also, note that the purity $
\mathcal{P}(\varrho) = \tr{(\varrho^2)} = 
\tr(\varrho \otimes \varrho S)$ can be written as an 
expectation value of two copies of a state, which connects 
Eq.~(\ref{local_purity}) to the SWAP trick in Sec.~\ref{sec:quantum_tdesigns}.

For two qubits, this expression can be understood in terms of the sector length. As mentioned above, we can 
consider local Pauli measurements, resulting in
nine possible measurement combinations. Let us focus on 
a single term where, for definiteness, $\sigma_z$ is measured on both particles. In this case, the Hamming distance can take the values 0, 1 
or 2. Consequently, the terms in the sum have prefactors 
$1, -1/2$ or $1/4$. The occurring frequencies can be derived from products of the probabilities $P(00), P(01), P(10),$ 
and $P(11)$ of the results of a $\sigma_z \otimes \sigma_z$ measurement. These products can also be expanded in terms 
of the squared expectation values  $\langle\sigma_z \otimes \sigma_z\rangle^2$, $\langle\sigma_z \otimes \mathbbm{1} \rangle^2$, $\langle \mathbbm{1} \otimes \sigma_z\rangle^2$, 
and  $\langle \mathbbm{1} \otimes \mathbbm{1} \rangle^2$. 
Indeed, one finds after a short calculation
\begin{align}
	\sum_{s,s'} (-2)^{-D[s, s']} P(s) P(s') & =
	1\times \left(P(00)^2 + P(01)^2 + P(10)^2 + P(11)^2 \right) 
	\nonumber 
	\\
	&-\frac{1}{2} \left(P(00)P(01) + P(10)P(11) 
	+ \dots +
	P(10)P(11) \right)
	\nonumber 
	\\
	&+\frac{1}{4} (P(00)P(11)) + P(11)P(00)
	+ P(01)P(10) + P(10)P(01))
	\nonumber 
	\\
	&
	= \frac{9}{8} \langle\sigma_z \otimes \sigma_z\rangle^2
	+ \frac{3}{8} (\langle\sigma_z \otimes \mathbbm{1} \rangle^2+\langle \mathbbm{1} \otimes \sigma_z\rangle^2)
	+
	\frac{1}{8}\langle \mathbbm{1} \otimes \mathbbm{1} \rangle^2.
	\label{elben_to_sectors}
\end{align}
Averaging these terms over all nine possible 
measurement combinations would introduce additional 
prefactors occurring due to LU invariance of sector 
lengths, see Sec.~\ref{sec:n_qubit_entanglement}.
Then Eq.~(\ref{elben_to_sectors}) contains a sum over 
9 different tensor products of Pauli measurements.
Each of those will recover one term of the type 
$\langle\sigma_i \otimes \sigma_j\rangle^2$, 
certain triples of them will give the marginal 
terms like $\langle\sigma_i \otimes \mathbbm{1} \rangle^2$ giving the same weight as the two-body correlation 
$\langle\sigma_i \otimes \sigma_j\rangle^2$ and all nine  contribute to the term $\langle \mathbbm{1} \otimes \mathbbm{1} \rangle^2$. Knowing this, it is clear that the formula (\ref{elben_to_sectors}) recovers the purity, which 
is proportional to the total sector length in Eq.~(\ref{sectorandpurity}).

This reasoning can be further generalised to $n$-qudit systems~\cite{elben_renyi_2018,brydges_probing_2019}.
While Eq.~(\ref{local_purity}) was formulated for global systems, it can be easily applied to access the information about the purity of a subsystem, by considering random unitary operations and projective measurements on the subsystem $i$ 
only. Therefore, the probabilities $P_{U_i}(s_i)$, explicitly given as $\tr(U_i \varrho_i U_i^{\dagger} \ket{s_i}\bra{s_i})$, allow to recover the reduced state's purity.
In the case of global unitary operations, the discussed 
approach recovers the original results presented by van 
Enk and Beenakker, but note that in this case the Hamming distance needs to be redefined \cite{van_enk_measuring_2012}.

Other interesting quantities are state overlaps and 
state fidelities. As shown in Ref.~\cite{elben_cross-platform_2020}, randomised measurements allow the 
cross-platform estimation of 
\begin{equation}
	\mathcal{F}(\varrho_1,\varrho_2)
	=\frac{\tr(\varrho_1 \varrho_2)}{\text{max}\lbrace \mathcal{P}(\varrho_1),\mathcal{P}(\varrho_2) \rbrace},
	\label{eq:2_fidelity}
\end{equation}
with subscripts $1$ and $2$ referring to the states of two 
different quantum devices, labelled by $1$ and $2$.
The above fidelity between mixed states was first proposed 
in Ref.~\cite{liang_quantum_2019} . It fulfils all of Jozsa's axioms~\cite{jozsa_fidelity_1994} and can be interpreted 
as a Hilbert-Schmidt product of the two states normalised 
by their maximal purity. Contrary to the widely used Uhlmann fidelity~\cite{uhlmann_transition_1976}, it is easier to 
compute. Evaluation of Eq.~(\ref{eq:2_fidelity}) 
is done via the randomised measurement of cross-correlations. That means that in Eq.~(\ref{local_purity}) 
different randomly chosen unitaries $U$ and 
$U'$ are applied to the output state of the two platforms
independently.
Again, this can be understood as a slight generalisation of  Eq.~(\ref{elben_to_sectors}); it can also be understood
as a version of the SWAP trick in the form of 
$ \tr{(\varrho_1 \varrho_2)} = \tr(\varrho_1 \otimes \varrho_2 S)$.

In this spirit, the presented approach to randomised measurements is not limited to informational concepts only.
It can also provide protocols for the measurement of many-body topological invariants (MBTIs) of symmetry-protected-topological phases.
Likewise the R\'enyi entropy, MTBIs are non-linear functions of the reduced density matrices.
Ref.~\cite{elben_many-body_2020} proposed a measurement scheme for MTBIs associated with the partial reflection, time reversal and internal symmetry \cite{pollmann_detection_2012} of one-dimensional interacting bosonic systems.
Both of these quantities can be estimated using ensemble averages in a similar manner as in Eq.~(\ref{local_purity}).

\subsubsection{Experimental implementations}

\begin{figure*}[t]
	\centering
	\includegraphics[width=0.8\textwidth]{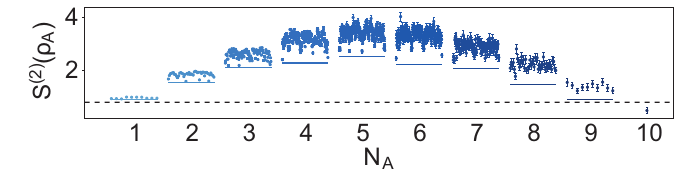}
	\caption{Experimental data for the second-order Rényi entropy
		measurement in all partitions of a $10$-qubit state realised with
		$^{40}\mathrm{Ca}^+$ ions. The number of ions in partition $A$
		has been denoted as $N_A$ and  $N_A=10$ corresponds to the global
		state. The dashed black  horizontal line corresponds to three
		standard deviations  above the full system’s entropy ($N_A=10$)
		and solid blue lines to three standard deviations below the
		minimal subsystem’s entropy for a given $N_A<10$.  The data
		demonstrate entanglement across all the $2^9-1=511$ bipartitions.
		The randomised protocol involved $M=500$ random unitary
		operations~\cite{mezzadri_how_2007}, each followed by $K=150$
		repetitions of measurement. Figure taken from Ref.~\cite{brydges_probing_2019}.
	}
	\label{fig:Brydges19}
\end{figure*}

A significant amount of experimental work was invested into demonstrating the feasibility of the above discussed techniques. Starting from the result presented in Eq.~(\ref{local_purity}), the second-order Rényi entropy defined in  Eq.~(\ref{eq:Rényientropy}) was used to measure the entanglement entropy in an ion system. The measurements were performed for all partitions of a $10$ qubit state
in a 20 qubit trapped-ion quantum simulator using $^{40}\mathrm{Ca}^+$~\cite{brydges_probing_2019}, see Fig.~\ref{fig:Brydges19}
for a part of the results.
This experiment involved an application of $M=500$ unitary operations, generated numerically according to an algorithm given in Ref.~\cite{mezzadri_how_2007}, each followed by $K=150$ repetitions of measurement.

The measurement of the entropy over the time evolution of a state allows to study the dynamical properties of quantum many-body systems.
A ballistic (linear) entropy growth is present for an interacting quantum system without disorder which will reach thermalisation, whereas a logarithmic entropy growth is expected for a system with strong disorder and sufficiently short-ranged interactions. 
The latter system will exhibit many-body localisation~\cite{nandkishore_many_2015}. 
Measurements of the entropy growth in Ref.~\cite{brydges_probing_2019} showed the diminishing effect of local, random disorder on the growth rate at early times compared to a system without disorder indicating localisation.
Furthermore, the evolution of the second-order Rényi mutual information
was detected providing another indicator of localisation due to the presence of
disorder in the system.

Another quantity which is related to multipartite entanglement of many-body systems is the quantum Fisher information (QFI)~\cite{helstrom_quantum_1969,braunstein_statistical_1994,pezze_quantum_2018,liu_quantum_2019}.
The inverse of the QFI sets a fundamental limit on the accuracy of parameter-estimation measurements and thus quantifies the potential of a quantum state in metrological applications~\cite{toth_quantum_2014}.
In Ref.~\cite{yu_experimental_2021} the QFI was estimated via randomised measurements using two different platforms: a nitrogen-vacancy (NV) center spin in diamond and a superconducting four-qubit state provided by IBM Quantum Experience.
For the one-qubit NV system, the dynamical evolution of the QFI was measured showing the applicability of the scheme to both pure and mixed quantum states. In the latter case, for multi-qubit states a lower bound on the QFI is set \cite{rath_quantum_2021}.

Randomised measurements also provide a powerful and experimentally feasible method to probe quantum dynamics.
One example is quantum information scrambling which describes how initially localised quantum information becomes, during time evolution, increasingly nonlocal and distributed over the entire system under consideration.
Scrambling can be measured via the decay of an out-of-time-order (OTO) correlation function.
Randomised measurements allowed for the measurement of these OTO correlation functions in a four-qubit nuclear magnetic resonance based quantum simulator~\cite{nie_detecting_2019} and in a system of 10 trapped ions with local interactions of tunable range~\cite{joshi_quantum_2020}.

\subsection{Estimation of higher PT moments}
\label{sec:pt_estimation}

As explained in Sec.~\ref{subsec:pt_moments}, bipartite entanglement of $\varrho_{AB}$ can be detected using the so-called PT moments, denoted as $p_k = \tr[(\varrho_{AB}^{\Gamma_B})^k]$.
The second PT moment $p_2$ is nothing but the state's purity and its estimation from randomised measurements was already discussed in the previous section.
Further analysis of a state, of course, involves higher-order PT moments that provide essential information about the state.
For example, it was shown that they become especially useful for extracting the logarithmic negativity~\cite{gray_machine-learning-assisted_2018}, and in analysing entanglement properties in many-body systems~\cite{vitale_symmetry-resolved_2022,carrasco_2024}.
So the question arises how to obtain higher PT moments from randomised measurements.
To address this, one needs to develop strategies of randomised measurements, such as generalisations of the SWAP trick mentioned in Sec.~\ref{sec:quantum_tdesigns}.
In the following, we will focus particularly on the third $p_3$ moment, reviewing the results given in Refs.~\cite{zhou_single-copies_2020, elben_mixed-state_2020}.

Let us begin by noticing that $p_3$ can be rewritten using a Hermitian operator $M_{\text{neg}}$ acting on three copies of $\varrho_{AB}$,
\begin{equation}
	p_3 = \tr\left[\varrho_{AB}^{\otimes 
		3} M_{\text{neg}} \right],
	\qquad
	M_{\text{neg}} = \frac{1}{2}
	\left(W_{\text{cyc}}^A \otimes W_{\text{inv}}^B + W_{\text{inv}}^A \otimes W_{\text{cyc}}^B
	\right),
	\qquad
	W_{\text{inv}}^X
	= (W_{\text{cyc}}^X)^{-1}
	=(W_{\text{cyc}}^X)^{T},
	\label{eq-p3representation}
\end{equation}
where
$W_{\text{cyc}}^X$ is the cyclic operator with
$W_{\text{cyc}}^X
\ket{s_1, s_2, s_3} = \ket{s_2, s_3, s_1}$ acting on the subsystem $X=A,B$ of the three copies,
for details see Sec.~\ref{sec:quantum_tdesigns}.
With this expression, one can understand the operator $M_{\text{neg}}$ in the scheme of randomised measurements.
First, note that the third moment $\mathcal{R}_{\mathcal{M}}^{(3)}$ defined in Eq.~(\ref{eq:generalmoment}) with an observable $\mathcal{M}$ acting on the whole system $AB$ can be reformulated as
\begin{equation}
	\mathcal{R}^{(3)}_{\mathcal{M}}(\varrho_{AB})
	=N_{2,d,3}
	\tr\left[
	\varrho_{AB}^{\otimes 3}
	\Phi_A^{(3)} \otimes \Phi_B^{(3)}(\mathcal{M})
	\right].
	\label{eq:thirdmomentpt}
\end{equation}
Here we denoted
$\Phi_X^{(t)}(X) = \int \mathrm{d}\mu(U_X)\,
U^{\otimes t} X^{\otimes t} (U^\dagger)^{\otimes t}$
as the $t$-fold twirling operation on system $X=A,B$, see Sec.~\ref{sec:unitary_tdesigns}.
This representation gives an alternative interpretation of the $t$-th moment $\mathcal{R}_{\mathcal{M}}^{(t)}(\varrho)$ as an expectation of a locally-twirled observable on $t$ copies of the state $\varrho$.
So, the question arises how to choose the observable $\mathcal{M}$ such that the desired $M_{\text{neg}}$ 
is achieved by twirling, 
$\Phi_A^{(3)} \otimes \Phi_B^{(3)}(\mathcal{M})$.

To proceed, notice that the operator $M_{\text{neg}}$ 
can be also decomposed into
$
M_{\text{neg}} = \frac{1}{4}
(M_+^A \otimes M_+^B
- M_-^A \otimes M_-^B),
$
where
$M_{\pm}^X = 
W_{\text{cyc}}^X \pm  W_{\text{inv}}^X$
for $X=A,B$.
In Ref.~\cite{zhou_single-copies_2020}, each of these terms was shown to be accessible with randomised measurements.
More precisely, there exist a product observable
$\mathcal{M}_\text{P} = \mathcal{M}_A \otimes \mathcal{M}_B$
and a non-product (Bell-basis) observable
$\mathcal{M}_\text{Bell}$
such that
\begin{equation}
	\Phi_A^{(3)} \otimes \Phi_B^{(3)}(\mathcal{M}_\text{P})
	= M_+^A \otimes M_+^B,
	\qquad
	\Phi_A^{(3)} \otimes \Phi_B^{(3)}(\mathcal{M}_\text{Bell})
	= M_-^A \otimes M_-^B.
\end{equation}
This immediately leads to the result
$M_{\text{neg}} = (1/4) \Phi_A^{(3)} \otimes \Phi_B^{(3)}(\mathcal{M}_\text{P} - \mathcal{M}_\text{Bell})$, showing that the operator $M_{\text{neg}}$ can be realised by combining the product and non-product observables from randomised measurement schemes.

In a similar manner, higher-order PT moments can be evaluated as an expectation value of a permutation operator acting on $k$ copies of $\varrho_{AB}$ \cite{elben_mixed-state_2020}
\begin{equation}
	p_k=\tr \left[
	\varrho_{AB}^{\otimes k}
	W_{\text{cyc}}^A
	\otimes
	W_{\text{inv}}^B
	\right],
\end{equation}
where $W_{\text{cyc}}^A$ and $W_{\text{inv}}^B$ are cyclic operators
$W_{\text{cyc}}^X
\ket{s_1, \cdots, s_k}
=\ket{s_2, \cdots, s_{k}, s_1}$ and $W_{\text{inv}}^X
\ket{s_1, \cdots, s_k}
=\ket{s_k, s_1, \cdots, s_{k-1}}$ acting on the subsystem $X = A, B$ of the $k$ copies.

In the following, we shortly explain another way to estimate the higher moments $p_k$ based on tomographic methods shown in Refs.~\cite{elben_statistical_2019, huang_predicting_2020}, for more details see also Sec.~\ref{sec:shadow_tomography} below.
A key idea is to create the unbiased estimator
$\hat{\varrho}_{AB}^{(r)}$ for $\varrho_{AB}$
from a data set of projective measurement results
with different random unitaries $r=1,2,\ldots, M$.
Based on this, one can define
the unbiased estimators for the PT moments as
\begin{equation}
	\hat{p}_k=\frac{1}{k!}
	{M \choose{k}}^{-1}
	\sum_{r_1 \neq \cdots \neq r_k}
	\tr \left[
	W_{\text{cyc}}^A
	\otimes
	W_{\text{inv}}^B \hat{\varrho}_{AB}^{(r_1)}
	\otimes \cdots \otimes  \hat{\varrho}_{AB}^{(r_k)}
	\right],
\end{equation}
where $r_i$ labels the data acquired from the measurements performed after the action of $r \,$th unitary. Implementing these techniques estimates the PT moments, therefore allowing for the certification of entanglement through $p_k$-PPT and the optimal $p_k$-OPPT criteria discussed in Sec.~\ref{subsec:pt_moments}.

\subsection{Moment-based permutation criterion}
\label{sec:permutation_moments}

In the previous section, we explained how the randomised measurement scheme can access the PT moments using the trick involving cyclic operations.
As a further development, Ref.~\cite{liu_detecting_2022} has generalised the concept of PT moments to the so-called permutation moments to enhance the power of entanglement detection.
Let us begin by recalling that any bipartite quantum state $\varrho_{AB}$ can be expressed in the computational basis as
$\varrho_{AB} = \sum_{i,j,k,l} \varrho_{ij,kl} \ket{i}\! \bra{j}_A \otimes \ket{k}\! \bra{l}_B$,
for the row indices $i,k$ and the column indices $j,l$. The partially transposed state is represented by
$\varrho_{AB}^{\Gamma_B} = \sum_{i,j,k,l} \varrho_{ij,lk} \ket{i}\! \bra{j}_A \otimes \ket{k}\! \bra{l}_B$, which corresponds to mapping the indices from $(ijkl)$ to $(ijlk)$ by exchanging $k$ and $l$.
More generally, one can define a different matrix $\Gamma_{\pi}(\varrho_{AB})$, by changing the indices $(ijkl)$ with an arbitrary permutation $\pi$. Ref.~\cite{horodecki_separability_2006} has shown that any separable state obeys
\begin{equation}
	\|\Gamma_{\pi}(\varrho_{AB})\|_{\text{tr}}
	= \sum_{i} \lambda_i \leq 1,
\end{equation}
where $\lambda_i$ denote the singular values of $\Gamma_{\pi}(\varrho_{AB})$.
A violation of this inequality implies the presence of entanglement.
The cases of $\pi = (1243)$ and $(1324)$ correspond to the PPT criterion~\cite{peres_separability_1996, horodecki_separability_1996} and the computable cross norm or realignment (CCNR) criterion~\cite{rudolph_further_2005, chen_matrix_2002}, respectively.

This criterion can also be implemented through randomised measurements. Below we describe systematic methods to obtain the lower bound on the norm $\|\Gamma_{\pi}(\varrho_{AB})\|_{\text{tr}}$, following Ref.~\cite{liu_detecting_2022}. The main idea is to introduce the permutation moments defined as
\begin{equation}
	M_{2k}^{\pi}(\varrho_{AB}) =
	\tr\left[
	(\Gamma_{\pi} \Gamma_{\pi}^\dagger)^k
	\right]
	= \sum_{i} \lambda_i^{2k}.
\end{equation}
In the case of $k=1$, the second moment $M_{2}^{\pi}$ is equivalent to a purity $\tr(\varrho_{AB}^2)$ for any $\pi$. Due to that, the first nontrivial permutation moment is the $M_{4}^{\pi}$. Interestingly, it can be used to detect entanglement through the following criterion
\begin{equation}
	E_{4}^{\pi}(\varrho_{AB})
	= \sqrt{\frac{q(qM_{2}^{\pi} + r)}{q+1}}
	+ \sqrt{\frac{M_{2}^{\pi} - r}{q+1}} >1
	\implies
	\mbox{$\varrho_{AB}$ is entangled,}
\end{equation}
where
$q = \lfloor (M_{2}^{\pi})^2/M_{4}^{\pi} \rfloor$
and $r = \sqrt{q(q+1)M_{4}^{\pi}-q(M_{2}^{\pi})^2}$.
To show that the fourth moment can be accessed through Haar unitary integrals, we consider a specific example. 
For $\pi=(1324)$ one can express the $M^{(1324)}_4(\varrho_{AB})$ on four state copies as
\begin{align}
	M_{4}^{(1324)}(\varrho_{AB}) = \tr\left[
	S_A^{(1,2)} \otimes S_A^{(3,4)} \otimes S_B^{(1,4)} \otimes S_B^{(2,3)}
	\varrho_{AB}^{\otimes 4}
	\right],
\end{align}
where $S_X^{(i,j)}$ denotes the SWAP operator acting on the subsystem $X=A,B$ among the $i$-th and $j$-th state copy with $i,j=1,2,3,4$.
This implies a connection with randomised measurements since as discussed in Refs.~\cite{brydges_probing_2019, elben_renyi_2018,elben_statistical_2019}, there exists a postprocessing operator $\mathcal{O}$ that implements the SWAP operator via Haar integrals such that
$\int \mathrm{d}\mu(U)\, U^{\otimes 2} \mathcal{O} (U^\dagger)^{\otimes 2} = S$ and $\mathcal{O} = \sum_{s, s^\prime} O_{s, s^\prime} \ket{s}\! \bra{s} \otimes \ket{s^\prime}\! \bra{s^\prime}$.
This leads to the
\begin{align}
	M_{4}^{(1324)}(\varrho_{AB}) =
	\int \mathrm{d}\mu(U_A)
	\int \mathrm{d}\mu(V_A)
	\int \mathrm{d}\mu(U_B)
	\int \mathrm{d}\mu(V_B) \,
	E_{U_A \otimes U_B}
	E_{U_A \otimes V_B}
	E_{V_A \otimes U_B}
	E_{V_A \otimes V_B},
\end{align}
where
$E_{X_A \otimes Y_B} =
\tr[\varrho_{AB} (X_A \otimes Y_B)^\dagger \mathcal{O} (X_A \otimes Y_B)]$
for unitaries $X,Y = U, V$. Notice that $M_{4}^{(1324)}$ has a structure that is  different from the moment $\mathcal{R}_{\mathcal{M}}^{(4)}$ defined in Eq.~(\ref{eq:generalmoment}). It should be noted that Ref.~\cite{liu_detecting_2022} has extended the above result to higher-order moments and multipartite systems. Moreover, it was also used to estimate the enhanced CCNR criterion by defining $M_{4}^{(1324)}$ on $(\varrho_{AB} - \varrho_{A} \otimes \varrho_{B})$, which can help to improve the detection power, and especially, to detect bound entanglement.

\subsection{Makhlin invariants}
\label{sec:makhlin_invariants}

The previous sections described the determination of the purity of states or PT moments using randomised measurements, while Sec.~\ref{SEC_ENTANGLEMENT} discussed the statistical moments $\mathcal{R}^{(t)}$ as
correlation-type quantities such as sector lengths.
All the resulting quantities were LU invariant and 
from arguments as in Eqs.~(\ref{elben_to_sectors})
and (\ref{eq-p3representation}) one can infer that arbitrary LU invariants could possibly be measured with randomized measurements.
The question now arises how randomised measurements
can be used to completely characterise the LU orbit 
of a quantum state. For two qubits, there is indeed
a complete set of LU invariants, the so-called Makhlin invariants and here we explain how they can be determined with randomised measurements \cite{wyderka_complete_2022}.

Let us begin by recalling that two quantum states $\varrho$ and $\sigma$ are called LU equivalent if and only if one can be transformed into the other by local unitary operation $U_A \otimes U_B$,
that is,
$\varrho = (U_A \otimes U_B)\sigma (U_A^\dagger \otimes U_B^\dagger)$.
Clearly, two LU equivalent states have the same values of quantities invariant under LU operations.
Conversely, one may ask whether there is a (finite)
set of invariants, such that two states are LU 
equivalent if they have the same values for these
invariants. In two-qubit systems, this question 
was answered by Makhlin \cite{makhlin_nonlocal_2002}.
It has been shown that two two-qubit states $\varrho$ 
and $\sigma$ are LU equivalent if and only if they 
have equal values of $18$ LU invariants $I_1, \ldots ,I_{18}$,
nowadays called the Makhlin invariants.

Using the notation in Eq.~(\ref{bloch_two_qubit}) from Sec.~\ref{SEC_NUTSHELL},
the first three invariants are given by:
\begin{align}
	I_1 = \det(T),
	\ \ \ 
	I_2 = \tr(TT^\top),
	\ \ \ 
	I_3 = \tr(TT^\top TT^\top).
	\label{eq:threeinvariants}
\end{align}
Only these invariants are already sufficient to compute 
a potential violation of the CHSH quantity, given by $S(\varrho) = 2\sqrt{\lambda_1 + \lambda_2}$, as 
discussed in Sec.~\ref{sec.nutshell.bell}.
This is because the two largest eigenvalues $\lambda_1$ 
and $\lambda_2$ of the matrix $TT^\top$ can be obtained from its characteristic polynomial and therefore can be computed from $I_1$, $I_2$, and $I_3$.

Reference~\cite{wyderka_complete_2022} has demonstrated that these invariants can be accessed via the moments $\mathcal{R}^{(t)}$ from randomised measurements, providing a tool to certify Bell nonlocality in a reference-frame-independent manner. Using a similar approach, one can derive a lower bound on the teleportation fidelity of 
the state~\cite{horodecki_general_1999, wyderka_complete_2022, guhne_geometry_2021}.

It is worth noting that $I_2$ and $I_3$ are also 
invariant under the partial transposition of a state, 
while $I_1$ flips the sign. This distinction corresponds
to the fact that $I_2$ and $I_3$ can be obtained using randomised measurements with the product observable $\mathcal{M}_{\text{P}} = \sigma_z \otimes \sigma_z$, whereas $I_1$ comes from the non-product observable $\mathcal{M}_{\text{NP}} = \sum_{i=x,y,z} \sigma_i \otimes \sigma_i$.
In addition, the invariant $I_{14} = \tr(H_a T H_b^\top T^\top)$ also flips the sign under partial transposition, where $(H_x)_{ij} = \sum_{k=x,y,z} \epsilon_{ijk} x_k$ represents the elements of a skew-symmetric matrix constructed from the  Bloch vectors of the reduced states $x_k = a_k, b_k$ and the Levi-Civita symbol $\epsilon_{ijk}$.
Also, the invariant $I_{14}$ was shown to be obtained from combinations of different non-product observables:
$\mathcal{M}_{\text{NP}}^{\pm} = \mathbbm{1} \otimes \sigma_x + \sigma_x \otimes \mathbbm{1} + \sigma_y \otimes \sigma_z \pm \sigma_z \otimes \sigma_y$.
The LU invariants $I_1$ and $I_{14}$ are sensitive to partial transposition and play a crucial role in implementing the PPT criterion via randomised measurements, as explained in Sec.~\ref{sec:PPT_criterion}.

Let us shortly describe the experimental scheme to 
obtain the LU invariants $I_1$, $I_2$, and $I_3$ from 
the moments $\mathcal{R}^{(t)}$ of randomised 
measurements  \cite{wyderka_complete_2022}. The creation 
of two-qubit states was implemented by polarisation-entangled photon pairs from an entangled photon source.
This generates signal and idler photon pairs via four-wave 
mixing in a dispersion-shifted fiber. In the detector 
station, each photon is detected with an efficiency of 
about $20\,\%$ and dark count probabilities about $4 \times 10^{-5}$ per gate.

\begin{figure*}[h!]
	\includegraphics[width=\textwidth]{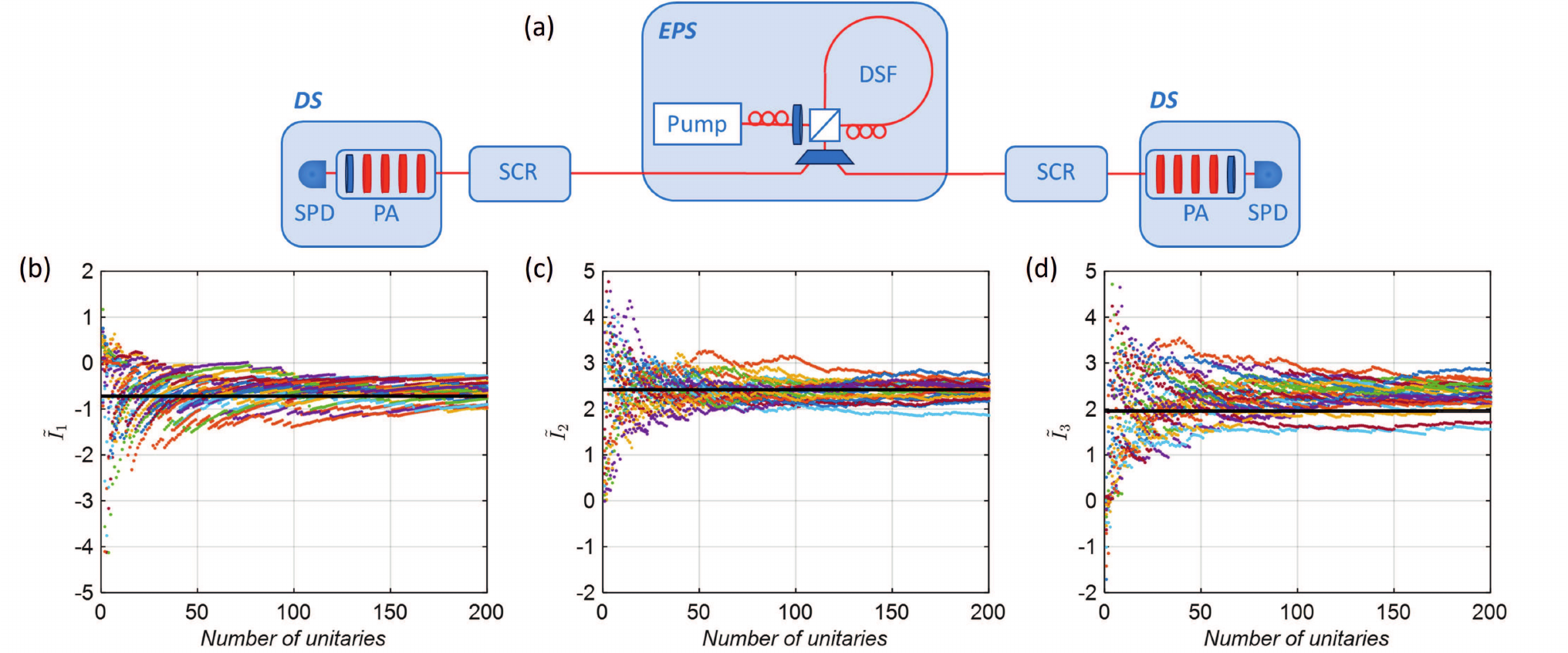}
	\caption{Determination of Makhlin invariants via randomised measurements. $(a)$ A schematic experimental setup.  The source of two-qubit entangled states (EPS) sends a signal and idler photon pairs via four-wave mixing in a dispersion-shifted fibre (DFS) to the detection station (DS). Before the measurement, random unitaries using polarisation scramblers (SCR) involving random polarisation state rotations were applied to the state. 
		Panels $(b), (c)$ and $(d)$ present the experimental data for unbiased estimators of Mahklin $I_1,I_2$ and $I_3$ LU invariants, respectively, plotted as a function of the number of applied unitaries. Figure taken from Ref.~\cite{wyderka_complete_2022}.}
	\label{Wyderka22b} 
\end{figure*}

The experimental generation of random unitaries has been accomplished using polarisation scramblers as depicted in Fig.~\ref{Wyderka22b}. These can rapidly change the vector in the Bloch sphere and create random polarisation rotations.
Clearly, it is necessary to check that a set of unitaries 
generated in this way is really Haar random. Here, it is sufficient to show that the used random unitaries form a unitary $t$-design with an appropriate order $t$, depending on the degrees of the polynomial used in the evaluation. One can confirm the degree of the Haar randomness for a set of unitaries by computing the frame potential, discussed in Sec.~\ref{sec:quantum_tdesigns}. Recall that only unitary $t$-designs achieve the minimal value of the frame potential and correspond to the $t$-th moments $\mathcal{R}^{(t)}$ from randomised measurements.
Then, to determine the LU invariants $I_1$, $I_2$, and $I_3$, one has to evaluate the frame potential for a finite set of drawn unitaries and compare it with the minimal value.

Finally, the LU invariants $I_1, I_2, I_3$ were measured using their unbiased estimators $\Tilde{I}_1, \Tilde{I}_2, \Tilde{I}_3$ from the experimental data of the set of unitaries generated, and their statistical behaviour is illustrated in Fig.~\ref{Wyderka22b}.
This analysis aimed to certify Bell's nonlocality and assess the usefulness of quantum teleportation.

\subsection{Randomised measurements and shadow tomography}
\label{sec:shadow_tomography}

Let us finally explain how randomised measurements can be 
used to obtain information about a quantum state in the 
framework of shadow tomography. Generally, quantum state tomography refers to the task of determining the entire 
density matrix from measurement data~\cite{james_measurement_2001,dariano_quantum_2003,paris_quantum_2004}. As the number of free parameters in the density 
matrix scales exponentially with the number of particles, 
this becomes in practice unfeasible already for a modest 
number of qubits. 

Several methods have been suggested to circumvent the
scaling problem. This includes matrix-product-state tomography~\cite{cramer_efficient_2010} and compressed sensing~\cite{gross_quantum_2010}, which work well if 
the state under scrutiny obeys certain constraints, e.g.
it has a high purity. Another example is permutationally invariant tomography~\cite{toth_permutationally_2010, moroder_permutationally_2012}, where relevant subspaces 
of the entire Hilbert space are probed. Shadow tomography, proposed theoretically in Ref.~\cite{aaronson_shadow_2020} 
and suggested as a practical procedure in Ref.~\cite{huang_predicting_2020},
is also a scheme to reduce the experimental effort, but here 
the idea is to redefine the task of tomography in a meaningful way.

In standard tomography, one determines a density matrix $\varrho$, 
and this is typically used for predicting the expectation 
values of observables that were not measured. Shadow tomography can be understood as a method to predict future measurements 
with high probability as good as possible~\cite{aaronson_shadow_2020, huang_predicting_2020}.
In a first data collection phase, 
measurements are carried out and for each result an (unbiased)
estimator of the quantum state is recorded. The collection of 
all these estimators is called the classical shadow of $\varrho$ 
and, importantly, storing these data can be achieved with a moderate effort. As the name suggests, this task does not aim to recover the full density matrix $\varrho$, but only the shadow that $\varrho$ casts on the measurements. In the second phase, called the prediction phase, the classical shadow is used to predict other observables that were not measured in the data collection phase. Two questions are relevant: First, how does one obtain estimators for the quantum state from a single measurement result? Second, how to choose the measurements in the data collection phase in order to predict many other measurements later with high accuracy?

Let us elaborate on these points explicitly, following the description of Ref.~\cite{nguyen_optimizing_2022}. Assume that some generalised measurement (or positive operator-valued measure, POVM) is carried out on a quantum state defined by the density matrix $\varrho$. This measurement consists of a collection of effects $E=\{E_i\}$ which are positive ($E_i \geq 0$) and normalised ($\sum_i E_i = \mathbbm{1}$) and the probabilities for the outcomes are given by $p_i = \tr(\varrho E_i).$ This POVM defines a map from the space of density matrices to the probability distributions via
\begin{equation}
	\Phi_E(\varrho) =\{\tr(\varrho E_i)\},  
\end{equation}
and the adjoint $\Phi_E^\dagger$ maps probability 
distributions to  operators. In general, one can ask 
how to infer from an observed probability distribution 
the state $\varrho.$ Performing a single measurement 
of the POVM leads to a single outcome $i$, so that the distribution over the outcome probabilities $q_i$ corresponds 
to a $\delta$-type probability distribution with 
$\{q_i\}=\delta_{ik}$. The problem of how to assign an estimator 
to such a single-point distribution is indeed known and solved in the fields of data science and machine learning~\cite{bishop_pattern_2006}. 
The least square estimator is the so-called shadow estimator $\chi$ and is as a map given by: 
\begin{equation}
	\chi = (\Phi_E^\dagger \Phi_E)^{-1} \Phi_E^\dagger.
\end{equation}
To assure the existence of the inverse of $C_E=\Phi_E^\dagger \Phi_E$,
it is assumed that the POVM is informationally complete, that is, the state $\varrho$ can be reconstructed from the full set of probabilities $p_i$. With the given definition
of $C_E$ one also directly finds that $C_E(\varrho)=\sum_{k} \tr(\varrho E_k) E_k$.
Then, for the $\delta$-type distribution $\{q_i\}=\delta_{ik}$ we have that 
$\Phi_E^\dagger(\{q_i\}) = E_k$, leading to the estimate of $\varrho$ from
a single data point, 
\begin{equation}
	\hat{\varrho}_k = C_E^{-1}(E_k).    
\end{equation}
This estimator, however, does not need to be a proper quantum state; especially, $\hat{\varrho}_k$ may have negative eigenvalues.

So far, the discussion was general and the POVM $E$ was not specified. 
The key point is now that randomised measurements can be viewed 
as such
a POVM. If one measures randomly one of the three Pauli matrices 
$\sigma_x$, $\sigma_y$, or $\sigma_z$ on a single qubit, this can be
seen as a six-outcome POVM with the 
effects $\{\ket{x^\pm}\bra{x^\pm}/3, 
\ket{y^\pm}\bra{y^\pm}/3, \ket{z^\pm}\bra{z^\pm}/3 \}$. For these effects, one can directly calculate that the shadow estimator for outcome $k\pm$ is
given by
\begin{equation}
	\label{eq:singlequbitshadow}
	\hat{\varrho}_k = 3 |{k^\pm}\rangle \langle{k^\pm}| - \mathbbm{1}.
\end{equation}
This is not a positive semidefinite density matrix, but this 
does not affect it usefulness for predicting future measurements. 
If random Pauli measurements are carried out on $n$ qubits, the 
global POVM has $6^n$ outcomes. Most importantly, the shadow estimator is the tensor product of the single-qubit estimators 
as in Eq.~(\ref{eq:singlequbitshadow}). This product structure allows to store the multi-qubit shadow efficiently.

Let us now discuss the error of this procedure, if one wishes to estimate an arbitrary observable $X$ from the collection of shadows $\{ \hat{\varrho}_{k_\ell} \}$, 
where $\ell= 1, \dots , M$ denotes the number of measurements.
First, the mean value $(1/M) \sum_{\ell} \tr(X \hat{\varrho}_{k_\ell})$
converges to $\langle X \rangle = \tr(X \varrho)$. The precision of the estimation is quantified by the variance in a single experiment, 
$\mathrm{Var}[\tr(X \hat{\varrho})] = 
\sum_k \tr(X \hat{\varrho}_{k})^2 \tr(\varrho E_k) - \langle X \rangle^2$, which is upper bounded by the shadow norm
\begin{equation}
	\Vert X \Vert^2_E := \Vert
	\sum_{k} \tr(X \hat{\varrho}_{k})^2 E_k \Vert_{\rm op},
\end{equation}
where $\Vert \dots \Vert_{\rm op}$ denotes the largest eigenvalue of an operator. Given a set $\mathcal{X}$ of observables that one wishes to predict, the quality of a shadow tomography scheme can be
quantified by $\kappa^2_E = \max\{\Vert X \Vert^2_E, X \in \mathcal{X}\}$.
Clearly, this depends on the set $\mathcal{X}$ and the POVM $E$ chosen.
Here, it turns out that the random choice of three Paulis per qubit is often
the optimal choice of the POVM~\cite{nguyen_optimizing_2022}.

Finally, it is also worth noting that shadow tomography with local randomised
Pauli measurements is a well suited tool for estimating observables $X$ which
act on few qubits only. If $X$ acts on $L$ qubits, then it is determined
by the mean values of $3^L$ tensor products of Paulis. In a shadow tomography 
scheme  with $M$ randomised settings, each of these settings has, on average,
been measured $M/3^L$ times. This provides sufficient information if $M$ is
large enough, but note that the required $M$ does not depend on the number 
$n$ of particles. This idea can be translated to rigorous statistical statements for local observables, see Ref.~\cite{huang_predicting_2020} for details.
The optimisation of purity measurement and other multi-copy observables using classical shadows have also been discussed in Refs.~\cite{rath_importance_2021, vermersch_enhanced_2023}.
Extension to shadow process tomography is elaborated on in Ref.~\cite{kunjummen_shadow_2023}.

The scheme proposed in Ref.~\cite{huang_predicting_2020} was implemented experimentally using four-qubit GHZ states encoded with polarisation-entangled photons~\cite{zhang_experimental_2021}.
Three schemes with uniform, biased and derandomised classical shadows were compared to conventional ones that sequentially measure each state function using importance sampling or observable grouping, where the derandomised classical shadow method was shown to outperform other advanced measurement schemes.
For other experimental implementations see~Refs.~\cite{struchalin_experimental_2021,stricker_experimental_2022,rath_entanglement_2023}.


\section{Non-local correlations}
\label{SEC_BELL}

\label{sec:non_classicasl_correlations}

In Sec.~\ref{sec.nutshell.bell} we introduced the notion of non-local correlations and the framework of Bell-type inequalities to test them.
In general, correlations not only depend on the quantum state under consideration, but also on the choice of measurements.
As already discussed in the simple example of the CHSH inequality (\ref{CHSH}), in order to achieve maximal violation, the measurement settings must be chosen carefully.
This requires great care in the preparation of the experiment.
In the case of multi-observer Bell tests, where the efficiency of the experimental setup is much more challenging to maintain, the situation becomes even more complicated.

In this chapter, we outline how nevertheless it is possible to study quantum correlations also with randomly selected measurements.
In general, to verify the presence of correlations that violate a Bell inequality (and its extensions to multipartite scenarios), it is necessary to measure several combinations of local measurement settings across the relevant parties for an entangled state.
For example, let us consider the case of the CHSH inequality in a randomised measurement scenario. To measure the first expectation value $E(\mathbf{u}_1,\mathbf{u}_2)$, the first observer measures in the randomly chosen setting $\mathbf{u}_1$, while the second observer uses the randomly chosen setting $\mathbf{u}_2$.
For the second expectation value, the second observer can switch to the randomly selected setting $\mathbf{u}^\prime_2$, while the first observer remains at setting $\mathbf{u}_1$.
It becomes clear that for the third expectation value the second observer now has to return to the previously employed setting $\mathbf{u}_2$ in order to measure it jointly with a new setting of the first observer.
Therefore, in all such experiments it is necessary to be able to revisit previously employed settings.
Note that in typical implementations of so-called ``loophole-free'' Bell tests which exclude Bell-local models as possible explanations of quantum phenomena, it is additionally necessary to switch between the local settings on a shot-to-shot basis without predetermining the subsequent setting.

The type of randomness which can be present when trying to violate a Bell-type inequality, thus, cannot consist of a fluctuating noise or a total lack of control over the settings.
Rather, we distinguish random but fixed rotations between the reference frames of different observers (often denoted in the literature as ``misaligned devices'') or random fixed rotations when choosing different settings within each local frame (denoted as ``uncalibrated devices''). 
These random rotations need to stay fixed on the timescale of data acquisition for at least one run of the respective experiment.
The only exception to this is average correlation discussed in section~\ref{sec.corr.avgCorr}, which is based on the first moment of the distribution of correlation functions and is thus compatible with full randomness, as in the criteria of sections~\ref{SEC_ENTANGLEMENT} and \ref{SEC_FUNCTIONS}.

\subsection{Probability of violation}

In general, when measurement settings are chosen randomly, the violation of Bell-type inequalities such as the CHSH inequality is no longer assured, even for a highly entangled state.
Thus, a natural way to quantify the strength of quantum correlations in a particular system is to ask what is the probability of violation of any CHSH inequality if observers choose observables at random without caring about the precise determination of the optimal measurement settings.
Such a probability, also called the ``volume of nonlocality'' or ``nonlocal fraction''~\cite{barasinski_genuine_2020,barasinski_experimentally_2021}, can be expressed by
\begin{equation}
	\mathcal{P}_V^{\rm CHSH} = \int_{\Omega}  \mathrm{d} \Omega 
	\,
	f_{\rm CHSH}(\Omega) , 
	\label{PV}
\end{equation}
where the integration is over all parameters that define the observable (local measurement settings) and $f_{\rm CHSH} = 1$ for settings that violate any of the CHSH inequalities and $f_{\rm CHSH} = 0$ otherwise.
Apart from CHSH inequalities, the probability of violation $\mathcal{P}_V^I$ can of course also be formulated for any other family of Bell-type inequalities $I$.
While in general the probability of violation depends on the choice of measure $\mathrm{d}\Omega$, the Haar-measure emerges as a natural choice, since in this case the probability of violation becomes invariant under local unitary operations applied to the state \cite{lipinska_towards_2018}.

In Ref.~\cite{liang_nonclassical_2010} the probability of violation of the CHSH inequality by the Bell state $|{\phi^+} \rangle = (|00 \rangle + |11 \rangle)/\sqrt{2}$ was derived analytically as $\mathcal{P}_V^{\rm CHSH} = 2(\pi-3) \approx 28.32\%$.
Furthermore, also in Ref.~\cite{liang_nonclassical_2010}, the probabilities of violation for complete sets of MABK \cite{mermin_extreme_1990,ardehali_bell_1992,belinskii_interference_1993} and WWW\.{Z}B \cite{weinfurter_four-photon_2001,werner_all-multipartite_2001,zukowski_bells_2002} inequalities by the $n$-particle GHZ state have been determined numerically.
The violation increases with the number of particles but seems to converge asymptotically (for large $n$) to a value strictly below unity. 

\subsection{Strength of non-locality}
\label{sec.corr.strength}

The non-locality of a state $\varrho$ is quantified not only via the probability of violation but also through the \textit{strength of non-locality} $\mathcal{S}$~\cite{de_rosier_strength_2020}.
This quantity is based on testing how robust the correlations are under the addition of white noise.
The strength is obtained by considering the noisy state $\Tilde{\varrho}$ parameterized by the noise parameter $v$, so-called visibility, with
\begin{equation}
	\Tilde{\varrho}(v) = v \varrho + (1-v)\frac{1}{d^n} \mathbbm{1}
\end{equation}
and finding the critical visibility $v=v_\mathrm{crit}$ below which no inequality is violated for a given set of measurement settings.
The value $v_\mathrm{crit}$ provides the strength of non-locality via $\mathcal{S} = 1 - v_\mathrm{crit}$.
The critical visibility minimized over measurements is denoted as $v_\mathrm{crit}^\mathrm{min}$ and corresponds to the maximal strength of non-locality $\mathcal{S}^\mathrm{max}$.
For values below $v_\mathrm{crit}^\mathrm{min}$ the probability of violation reduces to $0$.

The strength of non-locality $\mathcal{S}$ does not provide complete information about the non-local properties of the state, primarily as it only quantifies for which $v$ the probability of violation $\mathcal{P}_V$ reduces to $0$ and not how $\mathcal{P}_V$ depends on variations in $v$.
This information is captured, at least partially, via the \textit{average strength of non-locality} $\bar{{\mathcal{S}}}$, given as the expectation value
\begin{equation}
	\bar{{\mathcal{S}}} = \int_{0}^{\mathcal{S}^{max}} \mathcal{S} g(\mathcal{S}) \mathrm{d}\mathcal{S}, 
\end{equation}
where $g(\mathcal{S})$ is the probability density over values of $\mathcal{S}$ when choosing measurement directions according to $\mathrm{d}\Omega$, normalized such that $\int_{0}^{\mathcal{S}^{max}} g(S)d\mathcal{S} = 1$.
Examples of $g(\mathcal{S})$ are presented in Ref.~\cite{de_rosier_strength_2020}.
As the number of settings per party tends to infinity,  $\mathcal{S}$ becomes more and more independent of the choice of measurement directions $\Omega$ and $g(\mathcal{S}) \to \delta(\mathcal{S}^{max}-\mathcal{S})$~\cite{de_rosier_strength_2020}.
In this limit $\bar{\mathcal{S}}$ and $\mathcal{S}_\mathrm{max}$ become equivalent.

It is also possible to define a similar quantifier for states, the so-called ``trace-weighted nonlocality strength''~\cite{patrick_nonlocality_2022}, which is based on the trace distance of Bell correlations~\cite{brito_quantifying_2018} instead of the strength of non-locality of Bell correlations.
Notably, just as the probability of violation, both strengths are invariant under LU.
Furthermore, it can be shown that both are strictly positive for all pure entangled states in a setup with at least two binary-outcome measurements per party.
The proof is based on the fact that any pure multipartite entangled state violates some two-setting two-outcome Bell inequality~\cite{yu_all_2012,laskowski_highly_2015}.

\subsection{Generalized Bell-type inequalities}
\label{ns}

The Bell scenario can be generalised to a larger number of observers $n$, measurement settings $m$, and a larger dimension of the local Hilbert space $d$.
There are many examples of Bell-type inequalities for many different experimental situations \cite{brunner_bell_2014}.
However, full sets of tight Bell inequalities (Bell-Pitovsky polytopes) are known only in a few cases~\cite{scarani_bell_2019}.
Moreover, the analytical determination of probabilities of violation for other states and other Bell inequalities is challenging due to the nature of the integral in Eq.~(\ref{PV}).

A numerical method based on linear programming can, however, successfully address this problem.
It is known (see, e.g. Ref.~\cite{fine_hidden_1982}) that  there exists a Bell-local explanation of correlations if 
and only if there exists a joint probability distribution $p_\mathrm{BL}$ over the results of all settings for all observers, from which the experimental probabilities can be predicted via marginalisation.
Thus, for a given set of measurement settings the numerical search for this probability distribution, under the constraint that it generates the experimental correlations, is equivalent to testing all possible Bell-type inequalities for these particular observables.
If such a distribution can be found, it implies that the correlations are Bell-local and thus no inequality can be violated~\cite{kaszlikowski_violations_2000}.
Conversely, if the joint distribution cannot be found, the correlations are shown to be Bell non-local. In the numerical computations, this is of course true up to the numerical precision.
Note that this method does neither yield nor require knowledge of the exact form of Bell-type inequalities for a given experimental situation (which are often unknown), but nevertheless effectively allows to test all of the conceivable inequalities.

By sampling the measurement settings with sufficient statistics according to the measure $\mathrm{d}\Omega$, the numerical method allows approximating the unconditioned $\mathcal{P}_V$, which contrary to $\mathcal{P}^I_V$ is independent of the choice of a specific family of Bell-type inequalities $I$.
At the same time, the method is also extremely well suited to determine and maximise the strength of non-locality $\mathcal{S}^\mathrm{max}$.
This numerical approach was used for the first time to the GHZ states and two measurement settings in Ref.~\cite{liang_nonclassical_2010} and comprehensively for various qubit and qutrit states and multiple measurement settings in Ref.~\cite{de_rosier_multipartite_2017}.
It can be straightforwardly generalised to a larger number of particles, measurement settings, and subsystem dimensions, see, e.g.~Ref.~\cite{gruca_nonclassicality_2010}.

While the computational approach is equivalent to testing all possible Bell inequalities, it should be noted that already a single family of inequalities is sufficient to detect non-locality in most cases~\cite{de_rosier_strength_2020}.
This optimal family with $\mathcal{P}^\mathrm{opt}_V$ is obtained by extending the CHSH inequality to more observers and settings, e.g.~via a lifting procedure~\cite{pironio_lifting_2005}.
It turns out that for sufficiently many parties and settings $\mathcal{P}_V \approx \mathcal{P}^\mathrm{opt}_V$, as for example already for the three-qubit W state, $\mathcal{P}_V = 54.893\%$, while $\mathcal{P}^\mathrm{opt}_V = 50.858\%$.

\subsection{Properties of the probability of violation}
In the following, we briefly present the most important properties of the probability of violation and the notion of typicality of non-locality.\\

\noindent\textit{Dependence on number of measurement settings.---}It was shown in Ref.~\cite{de_rosier_multipartite_2017} that the probability of violation increases rapidly with the number of measurement settings per party.
For the GHZ state, already for two parties and five settings, $\mathcal{P}_V$ is close to $100\%$.
This fact can be explained in two ways.
Firstly, from a statistical point of view, as the number of settings increases, so does the chance of finding suitable pairs of settings that lead to violation. 
Secondly, new inequalities emerge involving all the additional measurement settings.
Additionally, in Ref.~\cite{lipinska_towards_2018} it is shown in general that for any pure bipartite entangled state, the probability of violation tends to unity when the number of measurement settings tends to infinity.\\

\noindent\textit{Multiplicativity of probability of violation.---}The probability of violation is multiplicative~\cite{de_rosier_multipartite_2017},
in the sense that the probability $\mathcal{P}_\mathrm{BL} = 1 - \mathcal{P}_V$ to choose settings allowing a Bell-local explanation is multiplicative over subsystems with
\begin{equation}
	\mathcal{P}_\mathrm{BL} (\varrho_1 \otimes \varrho_2) = \mathcal{P}_\mathrm{BL} (\varrho_1)\, \mathcal{P}_\mathrm{BL} (\varrho_2).
\end{equation}
Here, $\mathcal{P}_{\mathrm{BL}} (\varrho_1 \otimes \varrho_2)$ refers to the probability of Bell locality on a $(k+l)$-particle system, if $\varrho_1$ and $\varrho_2$ are states of $k$ and $l$ particles, respectively. Note that due to the phenomenon of superactivation of nonlocality this relation does not hold, if the subsystems are combined, e.g., if $\varrho_1 = \varrho_2$ is a two-particle system and $\varrho_1 \otimes \varrho_2$ is considered to be a two-particle system of higher dimension.

This multiplicativity relation immediately hints at the increase of $\mathcal{P}_V$ with system size.
Consider for example a product of $n$ two-qubit GHZ states ${\mathrm{GHZ}_2}^{\otimes n}$.
Using the result $\mathcal{P}_V\left(\mathrm{GHZ}_2\right) = 2(\pi-3)$ obtained above, the probability of violation becomes
\begin{equation}
	\mathcal{P}_V\left({{\rm GHZ_2}^{\otimes n}}\right) 
	= 1 - \mathcal{P}_\mathrm{BL}\left({{\rm GHZ_2}^{\otimes n}}\right)
	= 1 - \left(\mathcal{P}_\mathrm{BL}\left({\rm GHZ_2}\right)\right)^n
	= 1 - (7-2\pi)^n,
\end{equation}
which converges to unity for large $n$.\\

\noindent\textit{Maximal probability of violation.---}It was shown that the $n$-particle GHZ state maximises the probability of violation for a special set of two-outcome inequalities containing only full $n$-particle correlation functions when $n$ is even~\cite{lipinska_towards_2018}.
More generally, if we are not restricted to any particular type of inequality and consider the full set of possible two-setting Bell inequalities (in practice only realisable via the numerical method discussed in Sec.~\ref{ns}) it turns out that the GHZ states do not exhibit the highest $\mathcal{P}_V$.
Note that due to the numerical nature of the considerations, it is difficult to explicitly find the state which maximises $\mathcal{P}_V$. \\ 

\noindent\textit{Maximal non-locality and maximal entanglement.---}
One commonly used measure of non-locality, namely the robustness to white noise, was already introduced in Section~(\ref{sec.corr.strength}). However, it has many disadvantages.
One flagship example is the discrepancy between the maximal violation of the two-setting $d$-outcome CGLMP inequality~\cite{collins_bell_2002} and its violation by the $d \times d$ maximally entangled state.
It turns out that some asymmetric states tolerate a greater admixture of white noise while remaining non-local than the maximally entangled states.
For the probability of violation, however, the value is maximised by the maximally entangled states and the anomaly disappears at least for $d \leq 10$ \cite{fonseca_measure_2015,fonseca_survey_2018}.
In addition, it is proven in Ref.~\cite{lipinska_towards_2018} that for two qubits in a pure state the probability of violation for bipartite full-correlation Bell inequalities is an entanglement monotone.\\

\noindent \textit{Witness for genuine multipartite entanglement.---}The probability of violation can also serve as a witness of genuine multipartite entanglement~\cite{bancal_device-independent_2011}.
For example, for $n=3$ and two settings per party, $\mathcal{P}_V > 2(\pi - 3)$ certifies that the state is truly multipartite entangled~\cite{de_rosier_multipartite_2017}.
For a larger number of particles and a larger number of settings, similar criteria can also be formulated. 
However, so far they are only based on numerics, see, e.g.~Ref.~\cite{yang_device-independent_2020}. \\

\noindent \textit{Typicality of non-locality.---}The notion of probability of violation also allows to address the question how typical non-locality is not only under variation of observables but also of states.
In Ref.~\cite{de_rosier_multipartite_2017}, it was shown that the typical $n$-qubit states present in many quantum information problems for $n \geq 5$ exhibit Bell non-local correlations for almost any choice of observables ($\mathcal{P}_V >99.99\%$).
In Ref.~\cite{de_rosier_strength_2020}, through a sampling of the whole state space of pure states it was demonstrated that for a random pure state
the probability of violation strongly increases with the number of qubits and already for $N \geq 6$ it is greater than $99.99\%$.  

\subsection{Average correlation}
\label{sec.corr.avgCorr}

An alternative to considering the probability of violation $\mathcal{P}_V$ has been proposed in Ref.~\cite{tschaffon_average_2023}.
It builds on the intuition that for example the CHSH inequality is violated if the correlations $E(\mathbf{u}_a,\mathbf{u}_b)$ are sufficiently high for at least two pairs of different settings.
Thus, instead of testing any particular Bell-type inequality, \textit{average correlation} $\Sigma$ is calculated with
\begin{align}
	\Sigma = \int_{\Omega}  \mathrm{d}\Omega \; |E(\mathbf{u}_a,\mathbf{u}_b)|,
\end{align}
i.e.~it corresponds to the first moment of the distribution of the modulus of the correlation function.
In Ref.~\cite{tschaffon_average_2023}, it is shown analytically that for bipartite systems $\Sigma > 1/(2\sqrt{2})$ implies Bell non-locality and $\Sigma < 1/4$ excludes it, i.e.~a state with $\Sigma < 1/4$ cannot violate any CHSH inequality.
Compared to testing of Bell inequalities, this method has the advantage that it can be applied even in scenarios with fluctuating randomness where previous settings cannot be revisited.
Moreover, as shown in Ref.~\cite{tschaffon2024average}, the concept of average correlation proves useful not only in the analysis of Bell non-locality but also in the case of the two-qubit entanglement. In particular, if $\Sigma > 1/4$, the state is entangled. Conversely, it is separable if $\Sigma < 1/6$.

\subsection{Genuine multipartite non-locality}
\label{GMNL}

The notion of Bell-locality readily extends from two to an arbitrary number of parties.
In a Bell-local scenario, there are only classical correlations between any of the parties and it is not possible to violate any Bell-type inequality for any partition of the system.
The set of such multipartite Bell-local correlations is commonly denoted by $\mathcal{L}$.
Just as with the concept of genuine multipartite entanglement, however, the notion of Bell-nonlocality can be refined to \textit{genuine multipartite non-locality} (GMNL).
It allows to distinguish cases where there are only some Bell non-local correlations in the system, from cases where all parties share suitable correlations.

Starting from the bipartite definition, there are several options how to define GMNL, the first of which was proposed by Svetlichny~\cite{svetlichny_distinguishing_1987}.
In a straightforward formal analogy to the definition of Bell locality, it defines non-GMNL tripartite correlations as those which admit the decomposition
\begin{equation}
	P(a,b,c|x,y,z) = \sum_{\lambda} q_{\lambda} P(a,b|x,y,\lambda)P(c|z,\lambda) + \sum_{\mu} q_{\mu} P(a,c|x,z,\mu)P(b|y,\mu) + \sum_{\nu} q_{\nu} P(b,c|y,z,\nu)P(a|x,\nu),
	\label{EQ_SVET}
\end{equation}
where $\lambda, \mu, \nu$ are shared parameters that correlate measurement outcomes, the outcome of Alice when she chooses setting $x$ is denoted by $a$ and similarly pairs $y$, $b$ and $z$, $c$ denote the settings and outcomes of Bob and Charlie. 
Note that in each term the statistics of measurement results of one party conditionally factors out from the statistics of the other parties. 
In general, for $n$ observers, the set of these correlations is denoted by $\mathcal{S}_2$, where the subscript indicates factorisability into at least two parts.
As pointed out in~\cite{almeida_multipartite_2010,bancal_definitions_2013} this definition is rather strict since no additional assumptions are made about the joint probability distributions such as $P(a,b|x,y,\lambda)$.
In particular they can correspond to signalling correlations, which allow for superluminal communication between parties.
Note that the signalling is only a formal consequence of the decomposition since $P(a,b,c|x,y,z)$ of course does not contain any signalling.
This potentially unphysical nature of the decomposition motivates a different definition of non-GMNL correlations~\cite{almeida_multipartite_2010,bancal_definitions_2013}, namely those that can be decomposed as Eq.~(\ref{EQ_SVET}), but with the additional requirement that all elements of the decomposition need to remain non-signalling, i.e.~physical. The corresponding set of non-GMNL correlations is denoted as $\mathcal{NS}_2$.
For any $n \ge 3$, this condition is strictly stronger than just requiring non Bell-local correlations, and strictly weaker than Svetlichny's definition and therefore gives rise to the following strict inclusions $\mathcal{L}\subsetneq\mathcal{NS}_2\subsetneq\mathcal{S}_2$.

Based on these definitions, it is now straightforward to define the probability of violation $\mathcal{P}_{V}$, just as in the case of standard multipartite non-locality, as
\begin{equation}
	\mathcal{P}_{V}(\varrho,S) = \int f(\varrho, \Omega)~\mathrm{d}\Omega,
\end{equation}
where
\begin{equation}
	f(\varrho, \Omega) =
	\begin{cases}
		1, & \text{if settings lead to correlations outside the set $S$, }\\
		0,& \text{otherwise.}
	\end{cases}       
\end{equation}
The case of $S = \mathcal{L}$ corresponds to the discussion above and $S = \mathcal{NS}_2$ as well as $S = \mathcal{S}_2$ correspond to the probability of violation given the alternative definitions of GMNL, respectively.
Note that just as in the standard case, for each definition of GMNL either specific families $I$ of inequalities can be distinguished or a general probability for any type of inequality can be considered.
A numerical study of these probabilities has been performed for specific inequalities $I$ in~\cite{barasinski_genuine_2020,senel_demonstrating_2015}.
For each setup (defined by the multipartite state $\varrho$, the number of settings per party $m$, and the set $\mathcal{S}$), we can single out a dominant inequality $I$ that gives the best lower bound value of $\mathcal{P}_V$.

\begin{table*}[t]
	\centering
	\caption{\label{tab_PV} Numerically obtained probability of violation $\mathcal{P}_V$ for any Bell inequality  and $\mathcal{P}_V^I$ for a given inequality $I$  (in $\%$) for different multipartite non-locality scenarios (denoted by $\mathcal{L}$, $\mathcal{NS}_2$ and $\mathcal{S}_2$) for two settings per party ($m=2$) and the two emblematic states $\text{GHZ}_3$ and $\text{W}_3$.
		The numerical values are obtained in Ref.~\cite{pandit_optimal_2022}, except for those where a reference is given after the value.}
	\begin{tabular}{cllll} \hline \hline
		$ $  & $\mathcal{P}_V(\text{GHZ}_3)$ & $\mathcal{P}_V^I(\text{GHZ}_3)$ & $\mathcal{P}_V(\text{W}_3)$ & $\mathcal{P}_V^I(\text{W}_3)$ \\ \hline
		$\mathcal{L}$  & 74.69~\cite{de_rosier_multipartite_2017} &  70.00~\cite{barasinski_experimentally_2021} &   54.89~\cite{de_rosier_multipartite_2017}  &  50.86 \\
		$\mathcal{NS}_2$ &    11.57   &  10.63  &  3.730  &    3.231  \\
		$\mathcal{S}_2$  &    0.5413  &  0.5353 &  0.0085  &   0.0030  \\ 
		\hline \hline
	\end{tabular}
\end{table*}

In Ref.~\cite{pandit_optimal_2022}, the probability of violation $\mathcal{P}_V$ was investigated for three-party GHZ and W states with increasing number of measurement settings per party for the different tripartite scenarios (for results, see Tab.~\ref{tab_PV} for $m=2$, and see Ref.~\cite{pandit_optimal_2022} for $m>2$).
The following conclusions about $\mathcal{P}_V$ can be drawn based on Tab.~\ref{tab_PV} and on additional numerical data in the tables of Refs.~\cite{barasinski_experimentally_2021,pandit_optimal_2022}.
Importantly, $\mathcal{P}_V$ steadily increases with the number of measurement settings $m$~\cite{pandit_optimal_2022}.
In particular, the probability of violating the $\mathcal{L}$ and $\mathcal{NS}_2$ conditions is greater than $99.9\%$ in the respective cases $m>3$ and $m>5$, for both $\text{W}_3$ and $\text{GHZ}_3$ states.
For $\mathcal{S}_2$, however, this percentage is much smaller, especially for the $\text{W}_3$ state.
This trend is observed even up to $m=6$. 

The numerical computations in Refs.~\cite{barasinski_experimentally_2021,pandit_optimal_2022} also demonstrate that in each of the three discussed tripartite scenarios, one can clearly distinguish a so-called dominant inequality. This is a facet of the respective convex sets $\mathcal{L}$, $\mathcal{NS}_2$, $\mathcal{S}_2$
that defines a family of equivalent inequalities which are most often violated for a given Bell-type experiment.
As seen in Tab.~\ref{tab_PV}, the estimated $\mathcal{P}_V^I$ is surprisingly close to the $\mathcal{P}_V$ value for almost all two-setting ($m=2$) scenarios.
In fact, the same tendency is observed for larger $m$ as well~\cite{barasinski_experimentally_2021,pandit_optimal_2022}.
In particular, a very small difference between the $\mathcal{P}_V^I$ and $\mathcal{P}_V$ can be observed for $\mathcal{L}$ and $\mathcal{NS}_2$ scenarios up to $m=6$.
The worst match between the two probabilites is found in the $\mathcal{S}_2$ scenario for the $\text{W}_3$ state.

\subsection{Guaranteed violation for partial randomness}

It was observed that partial local alignment can lead to a significant increase in the probability of violation, culminating in the \textit{guaranteed violation}, i.e.~cases for which $\mathcal{P}_V = 100\%$ (apart from trivial cases when, e.g. all measurement directions are the same).
Under suitable conditions, the guaranteed violation is robust against noise and experimental deficiencies~\cite{shadbolt_guaranteed_2012}. 

\subsubsection{Locally orthogonal settings}

As mentioned above, for the case of two-qubit maximally entangled state and the CHSH inequality, when observers cannot align their measurement settings (neither between them or locally) and thus choose them randomly, the probability of violation of the CHSH inequality is approximately $28.3\%$.
If, however, local calibration is possible such that each party can choose two orthogonal settings the probability increases to $41.3\%$~\cite{liang_nonclassical_2010}.
For an $n$-qubit GHZ state and two measurement settings per site, the $\mathcal{P}_V$ increases and reaches a value greater than $99.9\%$ for $n \geq 4$.

\subsubsection{Local measurement triads}

If the number of orthogonal settings is further increased to three, i.e.~we allow orthogonal \textit{triads} of measurement directions,
the probability of violation effectively reaches $100\%$ already for two parties.  
The only case when a violation would not occur is if the triads would happen to be perfectly aligned~\cite{shadbolt_guaranteed_2012,wallman_observers_2012}.
For multipartite Bell scenarios, guaranteed violation was shown numerically in Refs.~\cite{wallman_observers_2012,yang_device-independent_2020} for up to 8 qubits.
In addition, the numerical study in Ref.~\cite{yang_device-independent_2020} suggests that the multipartite correlations arising from the randomly generated triads certify with almost certainty for $n=3$ and $n=4$ parties the existence of genuine multipartite entanglement possessed by the $\text{GHZ}_n$ state. That is, even in the absence of an aligned reference frame, a device-independent way of certifying genuine multipartite entanglement~\cite{bancal_device-independent_2011} is possible. 
In particular, for the specific cases of three and four parties, results, which were obtained from semidefinite programming, suggest that these randomly generated correlations always reveal, even in the presence of a non-negligible amount of white noise, the genuine multipartite entanglement possessed by these states~\cite{pandit_optimal_2022}. 
In other words, provided local calibration can be carried out to good precision, a device-independent certification of the genuine multipartite entanglement contained in these states can, in principle, also be carried out in an experimental situation without sharing a global reference frame.

\subsubsection{MUBs in higher dimensional systems}

The pairs of orthogonal measurements as well as the triads are instances of mutually unbiased bases (MUB) for qubits.
This is generalised in Ref.~\cite{tabia_bell_2022}, where the probability of violation is tested on maximally entangled bipartite $d\times d$ systems using local MUBs.
Note that MUBs are known to provide maximal quantum violation of certain Bell inequalities, including the CHSH inequality (see, e.g.~Ref.~\cite{tavakoli_mutually_2021}).
Using higher dimensional random MUB measurements in the case of $d=3$ and $d=4$, near guaranteed Bell violation was obtained~\cite{tabia_bell_2022}.

\subsubsection{Planar measurements with single aligned direction}
Apart from local calibration there is even less randomness when the observers are allowed to share one measurement direction while leaving the other still randomly unaligned.
If in this case the observers choose locally orthogonal measurement settings in the shared plane, any of the CHSH inequalities is always violated~\cite{wallman_generating_2011} (except for a set of zero measure).
In the case of $n$ qubits and the set of MABK inequalities, it has been shown that the probability of violating any of the MABK inequalities by a factor of $\epsilon \sqrt{2}$ is equal to $\mathcal{P}_V^{MABK} = (4/\pi) \arccos \epsilon $.
This leads to the conclusion that there is a guaranteed (albeit sometimes only infinitesimal) violation of the MABK inequality.

\subsection{Experimental demonstrations}
\label{sec.corr.exp}

Experiments with entangled photons serve as the prototype for studies in the context of non-local correlations with randomised measurements~\cite{palsson_experimentally_2012,shadbolt_guaranteed_2012,wang_experimental_2016,andreoli_experimental_2017,barasinski_experimentally_2021}.
Note that the scenario with partial alignment where two parties share only a single measurement direction is of immediate practical relevance in photonic systems.
If, for example, the photons are transmitted via polarisation-maintaining fibers the parties all share a linear polarisation basis, with random phase rotations between the two basis states.
Similarly, for photons distributed along free-space links between rotating objects such as satellites the linear polarisation is maintained, but the relative orientation of the linear polarisation bases is unknown.
The two-qubit experiments of Refs.~\cite{palsson_experimentally_2012,shadbolt_guaranteed_2012} both use CHSH inequalities to show the presence of Bell correlations.
Three-qubit experiments have more varying approaches, where in Ref.~\cite{wang_experimental_2016} the Svetlichny inequality is employed to show GMNL and in Ref.~\cite{barasinski_experimentally_2021} a representative type of Bell-inequality and a representative so called ``hybrid''-inequality are considered, where the latter is less strict allowing for some Bell-type correlations between subsets of parties.
Finally, Ref.~\cite{andreoli_experimental_2017} tackles the scenario of entanglement swapping and tests the corresponding bilocality inequality.
All these inequalities are based on observing two measurement settings per party and although some experimental schemes involve more settings this does not lead to different inequalities, but rather allows to consider several versions of each inequality in post processing by differently combining the results for different choices of pairs of settings at each observer.
The inequality with the strongest violation is then chosen as the result.

Since the observation of generalised Bell correlations requires to revisit the same settings, randomness of settings has to be introduced in a controlled or at least reproducible manner.
Most of the experiments use polarisation entanglement, and therefore measurements are realised by polarisation-dependent beam splitting~\cite{wang_experimental_2016,andreoli_experimental_2017} or polarisation filtering~\cite{palsson_experimentally_2012,barasinski_experimentally_2021} preceded by controllable waveplates, which can apply all possible unitary transformations allowing to access any desired direction on the Bloch sphere.
In the somewhat different scenario of Ref.~\cite{andreoli_experimental_2017}, additionally a liquid crystal device is employed, which adds unknown rotations between the two parts of the experiment.
Even in Ref.~\cite{barasinski_experimentally_2021}, where also the path degree is employed, the final measurement is done as a polarisation measurement after that path state has been transferred to polarisation.
In Ref.~\cite{shadbolt_guaranteed_2012}, a significantly different scheme is applied, where the path entanglement is measured by sending each photon into an interferometer realised using integrated optics.
In this platform, a phase shift can be induced both before and inside of the interferometer, which again allows to access all possible measurement directions for the Bloch sphere of the path qubit.
These path-state rotations in integrated optics are much less controllable than in the case of polarisation in free space and as such the technical approach from \cite{shadbolt_guaranteed_2012} itself motivates an approach with random measurements.

While in principle all theoretical results are based on Haar-randomness, not all experiments fully realise this requirement.
For experiments based on polarisation measurements, full Haar-randomness can in principle be easily achieved by appropriately choosing the distribution of settings for the waveplates.
However, Refs.~\cite{wang_experimental_2016} and \cite{barasinski_experimentally_2021} remain unclear whether a true Haar-random sampling was actually implemented.
In Ref.~\cite{palsson_experimentally_2012}, while the settings do not correspond to a Haar-random sampling, appropriate statistical weights are introduced when calculating the probability to violate the CHSH inequality.
Also, it is acknowledged in Ref.~\cite{andreoli_experimental_2017} that the evenly distributed choice of settings for the waveplates does not translate into a Haar-random sampling, the consequences of which, however, are not discussed further by the authors.
In contrast, the authors of Ref.~\cite{shadbolt_guaranteed_2012} admit explicitly that their measurement scheme is somewhat biased, but conclude that even without any correction their results correspond to the theoretical predictions sufficiently well.

Finally, several of the experiments also address the question how typical experimental imperfections affect their results.
The main contributions affecting the recorded correlations are reduced fidelity of the entangled states and statistical noise due to the limited number of state copies often encountered in experiments with multiple photons.

\subsubsection{One axis aligned}
The least random scenario, where each local frame is calibrated and additionally the parties share a common reference direction is addressed in Refs.~\cite{palsson_experimentally_2012} and \cite{wang_experimental_2016}, for two and three qubits, respectively.
In Ref.~\cite{palsson_experimentally_2012} the choice of two locally orthogonal settings achieves the theoretical prediction of guaranteed violation in almost all trials, where statistical noise and an imperfect state can fully explain the small fraction of cases where it was not achieved.
For the three-qubit GHZ state in Ref.~\cite{wang_experimental_2016} triples of equally distributed measurements on the plane of the Bloch sphere are performed (denoted as ``Y-Shaped triads''), after which all combinations of possible Svetlichny inequalities are considered.
The higher number of chosen settings increases the probability to measure in directions sufficiently suited to observe GMNL.
Also here the theoretical prediction of a certain violation is confirmed by the experimental results.

\begin{figure}[h!]
	\includegraphics[width=\textwidth]{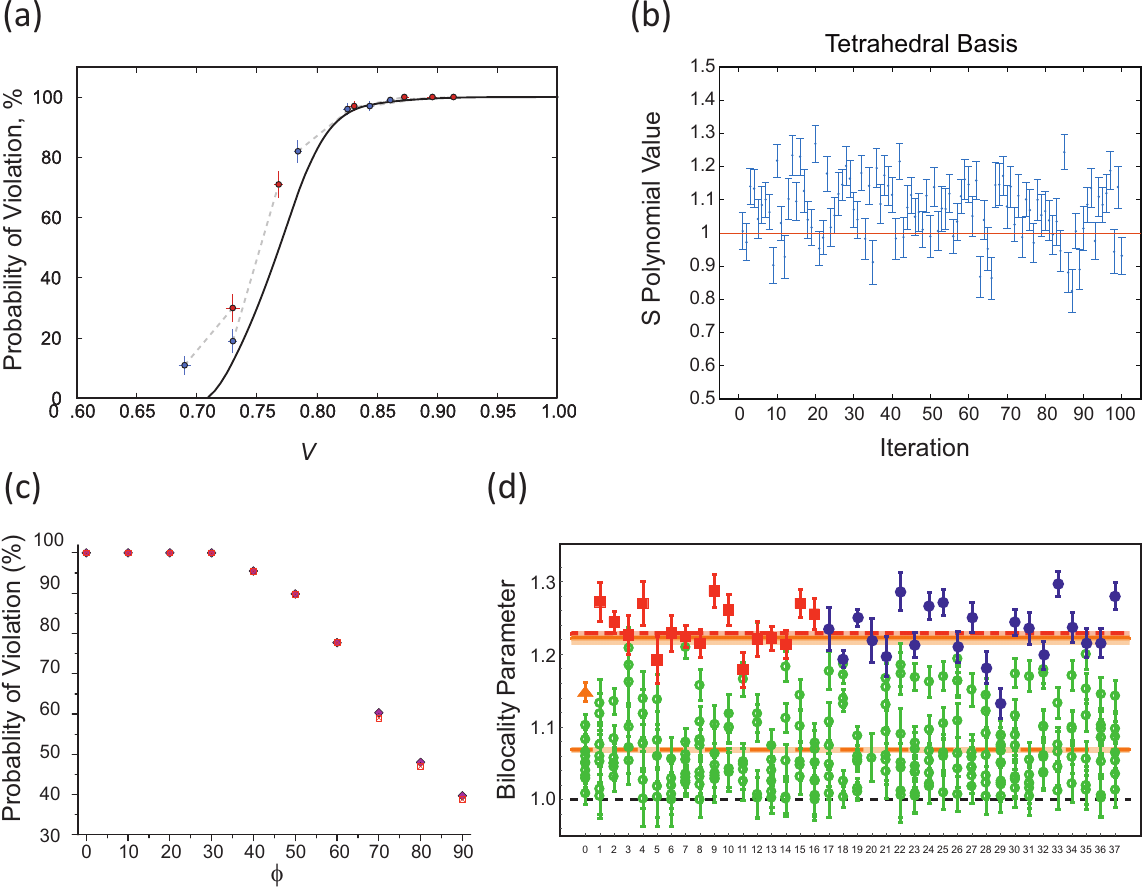}
	\caption{Experimental investigations of Bell violation with random settings.
		a) Increase of probability of violation (to guaranteed violation) if the fidelity of the state (quantified via visibility $V$) is increased~\cite{shadbolt_guaranteed_2012}.
		b) Violation of Svetlichny inequality with high probability for three qubits with four tetrahedral settings each~\cite{wang_experimental_2016}.
		Although only $58$ of the $100$ datapoints lie at least one standard deviation above the violation threshold of `1', about $76$ of the values themselves are above the threshold.
		Considering the reduced fidelity of $97\%$ the probability of violation is expected to decrease to about $69\%$, which means that the results fit the theoretical prediction very well given the experimental imperfections and uncertainty.
		c) Decrease of probability of violation (from guaranteed violation) if alignment of single axis is gradually lost~\cite{palsson_experimentally_2012}.
		The angle $\phi$ characterises the maximally allowed angle of misalignment of the shared axis with the maximum misalignment for $\phi = 90^\circ$.
		d) Certain violation of bilocality inequality in the entanglement swapping scenario~\cite{andreoli_experimental_2017}.
		The bold datapoints correspond to different ways of introducing unitary rotations between the reference frames, with the orange triangle corresponding to just the identity operation, the red square to Haar randomly sampled directions and the blue dots to a potentially biased sampling based on random experimental parameters.
		For each set of unitaries the green empty circles show other non-maximal violations for different versions of the bilocality inequality.}
	\label{fig.bell.exp}
\end{figure}

\subsubsection{Only local calibration - no alignment between parties}
For the next level of randomness, where any relative alignment between the parties is absent,  Ref.~\cite{palsson_experimentally_2012} shows that the prediction of a guaranteed violation is well confirmed up to experimental imperfections.
Additionally, it is explored how the scheme of using two mutually unbiased settings fails to achieve a violation with certainty as the two parties gradually loose the alignment of one common direction, see Fig.~\ref{fig.bell.exp}a.
However, even in the case of total misalignment, still a violation with a probability of $42\%$ is obtained, fitting closely with the theoretical prediction.
Conversely, in Ref.~\cite{shadbolt_guaranteed_2012} the scheme with triplets of mutually unbiased settings at each location (``orthogonal measurement triads'') yields a certain violation when accounting for experimental imperfections, even without any alignment.
The authors additionally investigate how this probability is affected by state deterioration by artificially introducing a temporal delay in the entangling CNOT process, which reduces the coherence of the state.
As shown in Fig.~\ref{fig.bell.exp}b, there exists a relatively large range of reduced coherence where a violation with almost certainty is still achieved.
In the three-qubit experiment of Ref.~\cite{wang_experimental_2016}, the usage of tetrahedral bases at each location achieves a violation with a probability of roughly $58\%$, as shown in Fig.~\ref{fig.bell.exp}c, which seems somewhat far away from the theoretical prediction of $88\%$.
However, also this deviation is explained very well by reduced state fidelity and high statistical noise not unusual for a three-photon experiment.
Finally, the entanglement swapping scenario from Ref.~\cite{andreoli_experimental_2017} considers three parties, $A$, $B$, and $C$, where $C$ establishes entanglement between $A$ and $B$ via a projective measurement on two qubits, each of which is entangled with a qubit at $A$ and $B$ respectively.
While there is full alignment between the measurements at $A$ and the $A$-side of $C$ ($C^A$), both the other measurement at $C$ ($C^B$) and the measurement at $B$ itself have randomly rotated reference frames.
As in Ref.~\cite{shadbolt_guaranteed_2012}, the experimental scheme of Ref.~\cite{andreoli_experimental_2017} involves a measurement of orthogonal triads for the two parties with randomly rotated frames $C^B$ and $B$.
As shown in Fig.~\ref{fig.bell.exp}d, the theoretical prediction of a certain violation of the corresponding bilocality inequality is confirmed very well.

\subsubsection{Complete lack of calibration}
The most extreme case of randomness, when also no information about the relative orientation of local settings is present, is considered in Ref.~\cite{shadbolt_guaranteed_2012} for two qubits and in Ref.~\cite{barasinski_experimentally_2021} for the three-qubit GHZ state.
The approach of Ref.~\cite{shadbolt_guaranteed_2012} is to simply measure a certain number $m$ of different settings per party, where a higher number of trials increases the probability that a combination of settings allowing a violation will occur.
Indeed, the authors show that with as little as four or five completely randomly chosen settings per party, violations are observed with close-to-unit probability.
In Ref.~\cite{barasinski_experimentally_2021}, a representative Bell-inequality and a less strict hybrid inequality, which allows for some nonlocality in the model, are experimentally tested with trials that always involve two settings per party.
The authors verify that the theoretical predictions of $61.1\%$ probability of violation for the Bell inequality and $5.7\%$ for the hybrid inequality are consistent with the experimental results within the margins of error.


\section{Conclusions}
\label{SEC_CONCLUSIONS}

We reviewed aspects of randomised measurements and their applications in quantum information.
It was discussed in detail how quantum entanglement is characterised and witnessed via statistical properties of correlations averaged over random measurement settings. 
This includes higher-dimensional entanglement, bound entanglement, spin-squeezing entanglement and different classes of multipartite entanglement.
The methods described here are systematic and capable of revealing subtle entanglement properties of complex multipartite and higher-dimensional systems.
Whenever possible, theoretical criteria were discussed in parallel with their experimental implementations taking into account the effects of finite statistics. Quantum designs were also reviewed in detail as means to simplify the implementation of averaging over the random measurements.

These methods were also shown to provide many other characteristics of quantum states as illustrated with estimations of non-linear functions of density matrices that include purity, fidelity, many-body topological and local unitary invariants, among others.
Our last focus was on the role that randomised measurement settings play in violation of Bell inequalities. In particular, we covered the typicality of violation, related quantifiers of non-classicality, and the impact of restricted randomness in both bipartite and multipartite scenarios.

All these results clearly demonstrate that randomised measurements are powerful theoretical and experimental tools that provide a practical advantage as well as a fundamental understanding of certain state properties. At the same time, they are not yet fully explored and exploited. 
To conclude we would like to mention a few related open problems.

The field of entanglement detection via randomised measurements has evolved towards using higher moments of the correlation distribution showing steady progress in the number and subtleties of detected entangled states. It is therefore natural to ask if given all the moments or a joint probability distribution of averaged correlations for all marginals, entanglement of any mixed state could be witnessed in a reference-frame-independent way. A similar problem is whether general LU invariance can be decided on the basis of randomised measurements only. For two qubits this is indeed possible, as we reviewed such reconstructions of Makhlin invariants, but the problem is open for higher dimensions
and multipartite systems. 
In both situations, the complete sets of invariants are unknown and identified invariants have not been phrased explicitly in terms of randomised measurements.
Moving on from the informational properties of quantum states, randomised measurements should be applied to the characterisation of quantum processes and are likely useful in other domains. A natural field is thermodynamics where information-theoretical tools have already been applied very successfully.

On the practical side, Haar randomness is the requirement of all discussed schemes. It therefore becomes important to develop methods confirming the degree of randomness realised in experiments and to construct efficient ways of generating high-dimensional Haar random unitaries. Theoretical protocols should be studied for other than Haar-random choices of settings. Finally, it is intriguing to see random measurements applied to study $k$-body marginal properties and to measure them on macroscopic quantum samples. Altogether, we hope to stimulate further progress in this domain that will lead to new fundamental discoveries and practical protocols with randomised measurements.

\section*{Declaration of Competing Interest}

The authors declare that they have no known competing financial interests or personal relationships that could have appeared to influence the work reported in this paper.

\section*{Acknowledgements}
We thank
Jan L. Bönsel,
Borivoje Daki{\'c},
Qiongyi He,
Marcus Huber,
Daniel E. Jones,
Andreas Ketterer,
Waldemar K{\l}obus,
Brian T. Kirby,
Shuheng Liu,
Zhenhuan Liu,
Xiongfeng Ma,
Simon Morelli,
Stefan Nimmrichter,
Aniket Rath,
Peter J. Shadbolt,
Shravan Shravan,
Jens Siewert,
G\'eza T\'oth,
Minh Cong Tran,
Michael Tschaffon,
Julio I. de Vicente,
Giuseppe Vitagliano,
Harald Weinfurter,
Nikolai Wyderka, 
Xiao-Dong Yu, and
Yuan-Yuan Zhao,
for discussions and collaborations on the subject.

This work has been supported by
the DAAD,
the Deutsche Forschungsgemeinschaft (DFG, German Research Foundation, project numbers 447948357 and 440958198), 
the DFG under Germany’s Excellence Strategy – EXC-2111 – 390814868 (Munich Center for Quantum Science and Technology), 
the Sino-German Center for Research Promotion (Project M-0294), 
the ERC (Consolidator Grant 683107/TempoQ), 
the German Ministry of Education and Research (Project
QuKuK, BMBF Grant No. 16KIS1618K and 16KIS1621), the
National Science Centre
(NCN, Poland) within the Preludium Bis project (Grant No. 2021/43/O/ST2/02679), 
the
Xiamen University Malaysia Research Fund (Grant No.
XMUMRF/2022-C10/IPHY/0002),
the EU (QuantERA eDICT, CHIST-ERA MoDIC),
and the National Research, Development and Innovation Office NKFIH (No.~2019-2.1.7-ERA-NET-2020-00003, 2023-1.2.1-ERA\_NET-2023-00009, and K145927).







\end{document}